\documentclass[twocolumn]{aastex63}
\usepackage{graphicx}
\usepackage{natbib}
\usepackage{amsmath,amsthm,amssymb}
\usepackage{color}
\usepackage{ulem}
\usepackage{epstopdf}
\newcommand{\gaia}{{\it Gaia}}
\newcommand{\PS}{\protect \hbox {Pan-STARRS1}}
\newcommand{\spitzer}{{\it Spitzer}}
\newcommand{\WISE}{{\it WISE}}

\newcommand{\kabstwo}{\ensuremath{{\rm M}_{Ks,{\rm 2MASS}}}}

\newcommand{\ymko}{\ensuremath{Y_{\rm MKO}}}
\newcommand{\jmko}{\ensuremath{J_{\rm MKO}}}
\newcommand{\hmko}{\ensuremath{H_{\rm MKO}}}
\newcommand{\kmko}{\ensuremath{K_{\rm MKO}}}
\newcommand{\mkmko}{\ensuremath{M_{\kmko}}}

\newcommand{\fldg}{\hbox{\sc fld-g}}
\newcommand{\intg}{\hbox{\sc int-g}}
\newcommand{\vlg}{\hbox{\sc vl-g}}

\newcommand{\teff}{\ensuremath{T_{\rm eff}}}
\newcommand{\lbol}{\ensuremath{L_{\rm bol}}}                 
\newcommand{\mbol}{\ensuremath{M_{\rm bol}}}                 
\newcommand{\logg}{\ensuremath{\mathrm{log} g}}                 

\newcommand{\plx}{\ensuremath{\varpi}}
\newcommand{\my}{\protect \hbox {mas yr$^{-1}$}}
\newcommand{\wmss}{\protect \hbox{W\,m$^{-2}$\,s$^{-1}$}}
\newcommand{\mjup}{\ensuremath{M_{\mathrm{Jup}}}}           
\newcommand{\msun}{\ensuremath{M_{\odot}}}                   
\newcommand{\lsun}{\ensuremath{L_{\odot}}}                   
\newcommand{\loglbol}{\protect \hbox{log\,($\lbol/\lsun$)}}       
\newcommand{\mbolsun}{\ensuremath{M_{{\rm bol},\odot}}}       
\newcommand{\feh}{\ensuremath{\rm [Fe/H]}}
\newcommand{\vmax}{\ensuremath{V/V_{\rm max}}}
\newcommand{\exvmax}{\ensuremath{\langle V/V_{\rm max}\rangle}}
\newcommand{\dlim}{\ensuremath{d_{\rm lim}}}
\newcommand{\nlim}{\ensuremath{N_{\rm lim}}}

\newcommand{\varnvollim}{504}
\newcommand{\varncold}{86}
\newcommand{\varnfullplx}{1119}
\newcommand{\varnbinary}{48}
\newcommand{\varnwidecomp}{30}
\newcommand{\varnsingle}{429}

\newcommand{\varcompletedist}{14}
\newcommand{\varcompletedistsig}{16}
\newcommand{\varcompletenum}{143}
\newcommand{\varvmaxfull}{0.39}
\newcommand{\varvmaxfullerr}{0.02}
\newcommand{\varcompletefull}{$\approx$78\%}
\newcommand{\varcompleteearly}{$\approx$92\%}

\newcommand{\vardlim}{\ensuremath{14.3_{-1.4}^{+0.5}}}
\newcommand{\varnlim}{\ensuremath{154_{-39}^{+20}}}

\newcommand{\vardenspois}{$(1.83_{-0.15}^{+0.16})\times10^{-2}$}
\newcommand{\vardensitylt}{$(1.25\pm0.10)\times10^{-2}$}
\newcommand{\vardensityl}{$(5.10_{-0.50}^{+0.53})\times10^{-3}$}
\newcommand{\vardensityt}{$(7.39_{-0.75}^{+0.78})\times10^{-3}$}
\newcommand{\vardensitymbolrange}{$(1.20_{-0.10}^{+0.11})\times10^{-2}$}

\newcommand{\varalphaminchisq}{-0.04}
\newcommand{\varbetaminchisq}{-0.06}
\newcommand{\varmedchisq}{10.8}

\newcommand{\varmedchisqcond}{7.2}

\newcommand{\varalphamaxpdf}{0.16}
\newcommand{\varbetamaxpdf}{-0.12}
\newcommand{\varalphamed}{0.58}
\newcommand{\varbetamed}{-0.44}
\newcommand{\varalphamedconf}{\ensuremath{0.58_{-0.20}^{+0.16}}}
\newcommand{\varbetamedconf}{\ensuremath{-0.44\pm0.14}}
\newcommand{\varalphamaxpdfcond}{1.02}
\newcommand{\varbetamaxpdfcond}{-1.52}
\newcommand{\varalphamedcond}{0.50}
\newcommand{\varbetamedcond}{-0.58}
\newcommand{\varalphamedcondconf}{\ensuremath{0.50\pm0.16}}
\newcommand{\varbetamedcondconf}{\ensuremath{-0.58\pm0.16}}
\newcommand{\varalphamedflat}{-0.06}
\newcommand{\varalphalowerflat}{-0.28}
\newcommand{\varalphaupperflat}{0.26}
\newcommand{\varalphasigmaflat}{2.4}
\newcommand{\varalphasigmaflatkirk}{2.4}
\newcommand{\varbetasigmakirk}{3.1}
\newcommand{\varyoungsinglfrac}{$9.2\%_{-2.6\%}^{+3.0\%}$}

\shorttitle{Substellar Age and Mass Functions}
\shortauthors{Best, W. M. J. et al.}

\begin{document}

\draft{Accepted by {\it The Astrophysical Journal}, 2024 January 11}

\title{A Volume-limited Sample of Ultracool Dwarfs. II. The Substellar Age and
  Mass Functions in the Solar Neighborhood}

\correspondingauthor{William M. J. Best}
\email{wbest@utexas.edu}

\author[0000-0003-0562-1511]{William M. J. Best}
\affiliation{The University of Texas at Austin, Department of Astronomy, 2515 Speedway, C1400, Austin, TX 78712, USA}

\author[0000-0002-1838-4757]{Aniket Sanghi}
\affiliation{The University of Texas at Austin, Department of Astronomy, 2515 Speedway, C1400, Austin, TX 78712, USA}
\affiliation{Cahill Center for Astronomy and Astrophysics, California Institute of Technology, 1200 E. California Boulevard, MC 249-17, Pasadena, CA 91125, USA}
\altaffiliation{NSF Graduate Research Fellow.}

\author[0000-0003-2232-7664]{Michael C. Liu}
\affil{Institute for Astronomy, University of Hawaii, 2680 Woodlawn Drive, Honolulu, HI 96822}

\author[0000-0002-7965-2815]{Eugene A. Magnier}
\affil{Institute for Astronomy, University of Hawaii, 2680 Woodlawn Drive, Honolulu, HI 96822}

\author[0000-0001-9823-1445]{Trent J.~Dupuy}
\affiliation{Institute for Astronomy, University of Edinburgh, Royal Observatory, Blackford Hill, Edinburgh, EH9 3HJ, UK}

\begin{abstract}
  We present the most precise constraints to date for the mass and age
  distributions of single ultracool dwarfs in the solar neighborhood,
  based on an updated volume-limited sample of {\varnvollim} L, T, and
  Y~dwarfs within 25~pc.
  We develop a Monte Carlo approach using the {\exvmax} statistic to
  correct for incompleteness
  and obtain a space density of {\vardenspois}~pc$^{-3}$ for spectral types
  L0--Y2.
  We calculate bolometric luminosities for our sample, 
  using an updated ``super-magnitude'' method for the faintest objects.
  We use our resulting luminosity function
  and a likelihood-based population synthesis approach
  to simultaneously constrain the mass and age distributions.
  We employ the fraction of young L0--L7~dwarfs as a novel input for this
  analysis that is crucial for constraining the age distribution.
  For a power-law mass function $\frac{dN}{dM} \propto M^{-\alpha}$ we find
  $\alpha=\varalphamedconf$, indicating an
  increase in numbers toward lower masses, consistent with measurements in
  nearby star-forming regions.
  For an exponential age distribution $b(t) \propto e^{-\beta t}$ we find
  $\beta=\varbetamedconf$, i.e., a population with fewer old objects than often
  assumed, which may reflect dynamical heating of the Galactic plane as much as
  the historical brown dwarf birthrate.
  We compare our analysis to \citet{Kirkpatrick:2021ik},
  who used a similar volume-limited sample.
  Although our mass function measurements are numerically consistent,
  their assumption of a flat age distribution is disfavored by our analysis, and
  we identify several important methodological differences between our two studies.
  Our calculation of the age distribution of solar neighborhood brown
  dwarfs is the first based on a volume-limited sample.
\end{abstract}


\section{Introduction}
\label{intro}

Brown dwarfs are substellar objects more massive than planets but less massive
than stars \citep[$\approx$13--70 {\mjup};][]{Dupuy:2017ke}.  Given this
intermediate station, studies of brown dwarf formation have sought to determine
whether the objects are extreme low-mass outcomes of standard star formation
processes or high-mass products of planet formation (i.e., in disks around
stars), or have some other origin.  Observations of nearby
($\approx100$--500~pc) star-forming regions indicate that brown dwarfs form like
stars, via turbulent fragmentation and core collapse in molecular clouds
\citep{Luhman:2012bg}, but it is unclear whether other mechanisms such as
massive disk instability \citep{Boss:1997di} or ejection from young multiple
systems \citep{Reipurth:2001go} may contribute significantly to the field
population \citep{Chabrier:2014gg}.

Closer to the Sun (i.e., within a few tens of parsecs), the lack of large and
complete samples has hampered our understanding of the history of brown dwarf
formation.  Most brown dwarfs have been discovered by way of searches in
wide-field sky surveys such as the Sloan Digital Sky Survey
\citep[SDSS;][]{York:2000gn} and the Two Micron All Sky Survey
\citep[2MASS;][]{Skrutskie:2006hl} that are severely magnitude-limited for the
fainter brown dwarf spectral types.  Selection of brown dwarfs from wide-field
surveys has become efficient only with multi-wavelength data
\citep[e.g.][]{Mace:2013jh,Best:2015em}, in particular including the red-optical
bands from the Panoramic Survey Telescope And Rapid Response System (\PS)~3$\pi$
Survey \citep[PS1;][]{Chambers:2020vk} and the mid-infrared bands from the
Wide-Field Infrared Survey Explorer \citep[\WISE;][]{Wright:2010in}.  In
addition, the parallaxes needed to confirm membership in volume-limited samples
are expensive to obtain since most brown dwarfs are too faint to be observed by
{\gaia}, and have only recently become available in sufficient numbers to build
large, complete samples
\citep[e.g.][]{Dupuy:2012bp,Faherty:2012cy,Dahn:2017gu,Smart:2018en,Best:2020jr,Kirkpatrick:2021ik}.

Precise values for the local brown dwarf mass and age distributions would help
to discern the dominant formation mechanism(s), but these quantities have been
difficult to ascertain \citep[e.g.,][]{Marocco:2015iz}.  Brown dwarfs cool
throughout their lifetimes, with more massive brown dwarfs beginning as late-M
dwarfs and progressing through L, T and Y spectral types over billions of years,
while less massive objects reach the cooler spectral types more quickly
\citep[e.g.,][]{Burrows:1997jq,Kirkpatrick:2005cv}.  This continuous cooling
means that a younger, less massive brown dwarf can have the same
directly-observable properties (e.g., {\teff} and luminosity) as an older, more
massive brown dwarf.  Further, there is no evolutionary phase analogous to the
stellar main sequence where {\teff} can serve as a proxy for mass over any given
age range.  The masses and ages of field brown dwarfs are therefore not usually
measurable from observations; exceptions include dynamical masses from orbital
monitoring of binaries and spectroscopic features of unusually old and young
brown dwarfs.  Kinematic distributions have provided statistical measures of age
\citep{Wielen:1977va,Faherty:2009kg}, but have not proven capable of
distinguishing between a constant, evolving, or largely stochastic birth history
\citep[e.g.,][]{Burgasser:2004ed,DayJones:2013hm,Kirkpatrick:2021ik}.

A statistical approach that can overcome the mass-age degeneracy is to
synthesize model populations of brown dwarfs characterized by an initial mass
function (IMF) and age distribution, evolve the populations to present-day
luminosity and {\teff} using an evolutionary model, and compare the synthetic
populations to observations of a well-characterized sample of brown dwarfs.  If
the IMF has not changed over time, and no external influences have altered the
mass distribution of nearby ultracool dwarfs, then the present-day mass function
will be the same as the IMF.  Previous efforts in this vein established the
first estimates of the local low-mass IMF and formation history
\citep[e.g.,][]{Burgasser:2004ed,Allen:2005jf,Deacon:2006ga,Metchev:2008gx,Pinfield:2008jx,Burningham:2010dh,Reyle:2010gq,Kirkpatrick:2012ha,Burningham:2013gt,DayJones:2013hm,Marocco:2015iz}.
These studies used the best space density, luminosity function, binary fraction,
and evolutionary models available at the time, but were unable to place
significant constraints on either the mass or age distribution for ultracool
dwarfs (spectral types M6 and later).  Collectively, they only constrained the
substellar mass function slope to $-1\lesssim\alpha\lesssim1$ (for space density
as a function of object mass given by $\Psi(M)\propto M^{-\alpha}$) and could
only infer broad agreement with a constant formation rate over the history of
the Galaxy.  However, new well-characterized volume-limited samples of brown
dwarfs in the solar neighborhood --- those of \citet[hereinafter
Paper~I]{Best:2021gm} and \citet[hereinafter K21]{Kirkpatrick:2021ik} --- have
breathed new life into the population synthesis approach, enabling precise
measurements of observable distributions such as bolometric luminosity ({\lbol})
or {\teff} needed to constrain the underlying mass and age functions.  K21
presented a full-sky 20~pc volume-limited sample from which they derive a mass
function estimate of $\alpha=0.6\pm0.1$, assuming a uniform age distribution.
In this paper, we present new simultaneous constraints on both the mass and age
distributions of very low-mass stars and brown dwarfs in the solar neighborhood
using bolometric luminosities from a 25-pc volume-limited sample of L, T, and Y
dwarfs.

\section{Volume-Limited Sample}
\label{vlsample}

\subsection{Sample Definition}
\label{vlsample.def}
In Paper~I, we presented a volume-limited sample of 369 L0--T8~dwarfs out to
25~pc defined entirely by parallaxes. Our sample was the first to
comprehensively map the L/T transition (spectral types $\approx$L8--T4), and its
22~young ($\lesssim$200~Myr) members suggested a young-leaning age distribution.
We have now updated our volume-limited sample to include recent nearby brown
dwarf discoveries and parallax measurements from the literature \citep[e.g.,
K21,][]{Zhang:2021jq,GaiaCollaboration:2023jd}.  The boundaries of the sample
remain the same --- declination $-30^\circ \le \delta \le 60^\circ$ (covering
68\% of the sky) out to 25~pc from the Sun --- but we have extended the coolest
spectral type from T8 to Y2, i.e., to include the coldest known
types\footnote{WISE~J085510.83$-$071442.5, the coldest brown dwarf identified to
  date \citep{Luhman:2014jd}, does not yet have a spectroscopically confirmed
  spectral type. We adopt the photometric type of $\ge$Y2 \citep{Leggett:2015dn}
  for convenience, but note the Y4 photometric estimate of
  \citet{Kirkpatrick:2019kt}.}.  We now also require parallax uncertainties to
be less than 1/7 of the parallaxes, rather than 1/5 as in Paper I. This excised
only one object from the Paper~I sample: SDSS J152103.24+013142.7 (spectral type
T3, parallax $43.3\pm6.2$~mas).

We present our updated volume-limited sample in Table~\ref{tbl.sample}. The
sample now contains {\varnvollim} objects, increasing from 369 in Paper~I
primarily due to the inclusion of {\varncold} objects with spectral types T8.5
and cooler.  Nearly all of these were discovered or discussed by K21.  While we
previously required parallaxes and spectroscopic confirmation for inclusion in
our sample, we have allowed 73~objects (15\% of our new sample) lacking one or
both of these measurements that K21 concluded were bona fide brown
dwarfs\footnote{16 have no parallax measurement; 16 have no spectroscopic
  confirmation; 41 have neither.}, so that our volume-limited sample will more
completely represent the coldest members of the solar neighborhood.  All members
of our updated volume-limited sample and associated photometry are tabulated in
the UltracoolSheet\footnote{\url{http://bit.ly/UltracoolSheet}} (which also
identifies the Paper~I volume-limited sample).

\begin{longrotatetable}
\pagestyle{empty}
\begin{deluxetable*}{lCccCCCCCcr}
\centering
\tablecaption{Our volume-limited 25 pc sample of L0--Y2 dwarfs \label{tbl.sample}}
\tabletypesize{\scriptsize}
\setlength{\tabcolsep}{0.025in}
\tablewidth{0pt}
\tablehead{   
  \colhead{Object} &
  \colhead{Parallax\tablenotemark{a}} &
  \colhead{Spectral Type\tablenotemark{a},\tablenotemark{b}} &
  \colhead{Flag\tablenotemark{c}} &
  \colhead{\ymko} &
  \colhead{\jmko} &
  \colhead{\hmko} &
  \colhead{\kmko} &
  \colhead{\mbol} &
  \colhead{Source\tablenotemark{d}} &
  \colhead{References} \\[-3.5mm]
  \colhead{} &
  \colhead{(mas)} &
  \colhead{(Optical/NIR)} &
  \colhead{} &
  \colhead{(mag)} &
  \colhead{(mag)} &
  \colhead{(mag)} &
  \colhead{(mag)} &
  \colhead{(mag)} &
  \colhead{} &
  \colhead{(Disc; \plx; SpT; Flag; Phot)}
}
\startdata
SDSS J000013.54+255418.6 & $70.8\pm1.9$ & T5/T4.5 & \nodata & $15.80\pm0.06$ & $14.73\pm0.03$ & $14.74\pm0.03$ & $14.82\pm0.03$ & $16.75\pm0.18$ & D & 123; 73; 0,32; --; 73,123 \\
WISE J000517.48+373720.5 & $126.9\pm2.1$ & .../T9 & \nodata & $18.48\pm0.02$ & $17.59\pm0.02$ & $17.98\pm0.02$ & $17.99\pm0.02$ & $20.02\pm0.05$ & S & 159; 122; 159; --; 134 \\
2MASS J00132229$-$1143006 & $40.3\pm3.1$ & .../T3pec & \nodata & \nodata & $16.05\pm0.02$ & $15.74\pm0.22$ & $15.76\pm0.22$ & $16.58\pm0.24$ & D & 110; 14; 110; --; 14,15 \\
2MASSW J0015447+351603 & $58.8\pm0.3$ & L2/L1.0 & \nodata & $14.95\pm0.05$ & $13.74\pm0.02$ & $12.96\pm0.02$ & $12.25\pm0.02$ & $14.40\pm0.10$ & D & 114; 89; 0,6; --; 15,14 \\
PSO J004.6359+56.8370 & $46.5\pm3.9$ & .../T4.5 & \nodata & \nodata & $16.22\pm0.02$ & \nodata & \nodata & $17.03\pm0.23$ & J & 14; 14; 14; --; 14 \\
2MASS J00282091+2249050 & $42.9\pm1.6$ & .../L7: & \nodata & $16.82\pm0.05$ & $15.49\pm0.02$ & $14.55\pm0.06$ & $13.77\pm0.06$ & $15.25\pm0.14$ & D & 39; 89; 39; --; 15,14 \\
WISE J003110.04+574936.3 & $71.0\pm3.2$ & .../L9 & BY & $15.92\pm0.05$ & $14.770\pm0.010$ & $13.862\pm0.014$ & $13.21\pm0.03$ & $15.71\pm0.13$ & D & 218; 122; 12; 16,146; 15,78,12 \\
PSO J007.9194+33.5961 & $45.4\pm3.8$ & .../L9 & \nodata & $17.48\pm0.06$ & $16.38\pm0.02$ & $15.46\pm0.05$ & $14.67\pm0.05$ & $16.06\pm0.16$ & D & 13; 14; 13; --; 15,14 \\
\enddata
\end{deluxetable*}
\end{longrotatetable}

\subsection{New Parallaxes}
\label{vlsample.newplx}
We present new $J$-band parallax measurements for four members of the
volume-limited sample, shown in Table~\ref{tbl.newplx} and also included in
Table~\ref{tbl.sample}.  The objects were observed with CFHT/WIRCam as part of
the Hawaii Infrared Parallax Program \citep{Dupuy:2012bp}.  The observation
strategies, data reduction, and astrometric solutions are all as described in
\citet{Dupuy:2012bp} and \citet{Dupuy:2017ke}.

This is the first published parallax for CFBDS~J030135.11$-$161418.0.  K21
published parallaxes for the other three objects, which are generally consistent
with ours. We opted to use our measurements because they are more precise.
\citet{Theissen:2018fy} also published a parallax for 2MASS~J21543318+5942187
using astrometry from {\WISE}. This parallax is formally consistent with ours,
but has a much larger uncertainty.

\floattable
\begin{deluxetable*}{lcCcCCCcCCCCCr}
\centering
\tablecaption{New Parallaxes and Proper Motions \label{tbl.newplx}}
\tabletypesize{\scriptsize}
\setlength{\tabcolsep}{0.065in}
\tablewidth{0pt}
\rotate
\tablehead{   
  \multicolumn{4}{c}{} &
  \multicolumn{3}{c}{Relative} &
  \colhead{} &
  \multicolumn{3}{c}{Absolute} &
  \colhead{} \\
  \cline{5-7}
  \cline{9-11}
  \colhead{Object} &
  \colhead{$\alpha_{\rm J2000}$} &
  \colhead{$\delta_{\rm J2000}$} &
  \colhead{Epoch} &
  \colhead{$\plx_{\rm rel}$} &
  \colhead{$\mu_{\alpha,{\rm rel}}{\rm cos}\,\delta$} &
  \colhead{$\mu_{\delta,{\rm rel}}$} &
  \colhead{} &
  \colhead{$\plx_{\rm abs}$} &
  \colhead{$\mu_{\alpha,{\rm abs}}{\rm cos}\,\delta$} &
  \colhead{$\mu_{\delta,{\rm abs}}$} &
  \colhead{$N_{\rm ep}$} &
  \colhead{$\Delta t$} &
  \colhead{$\chi^2$/dof} \\
  \colhead{} &
  \colhead{(deg)} &
  \colhead{(deg)} &
  \colhead{(MJD)} &
  \colhead{(mas)} &
  \colhead{(\my)} &
  \colhead{(\my)} &
  \colhead{} &
  \colhead{(mas)} &
  \colhead{(\my)} &
  \colhead{(\my)} &
  \colhead{} &
  \colhead{(yr)} &
  \colhead{}
}
\startdata
CFBDS J030135.11$-$161418.0 & 045.3965467 & $-16.2382606$ & 54436.31 & $48.0\pm3.8$ & $297.8\pm3.0$ & $126.1\pm3.1$ &  & $49.1\pm3.8$ & $301.9\pm3.2$ & $124.4\pm3.4$ & 8 & 8.06 & 10.3/11 \\
2MASSI J0512063$-$294954 & 078.0269272 & $-29.8311129$ & 57317.60 & $48.8\pm2.8$ & $-1.0\pm1.6$ & $78.8\pm2.0$ &  & $49.6\pm2.8$ & $0.6\pm1.8$ & $79.9\pm2.3$ & 7 & 2.12 & 10.3/9 \\
WISE J064205.58+410155.5 & 100.5233880 & $+41.0318223$ & 56585.62 & $63.2\pm1.2$ & $-0.5\pm1.2$ & $-374.8\pm1.2$ &  & $64.1\pm1.1$ & $-2.0\pm1.4$ & $-377.4\pm1.5$ & 7 & 2.17 & 10.7/9 \\
2MASS J21543318+5942187 & 328.6376414 & $+59.7040555$ & 55050.45 & $63.3\pm1.5$ & $-160.1\pm0.5$ & $-463.8\pm0.7$ &  & $64.7\pm1.5$ & $-166.0\pm1.1$ & $-469.6\pm0.9$ & 6 & 5.16 & 9.3/7 \\
\enddata
\tablecomments{
  ($\alpha_{\rm J2000}$, $\delta_{\rm J2000}$, Epoch): coordinates and epoch for our first observation of
  that target.}
\end{deluxetable*}

\section{Demographics of L, T, and Y dwarfs}
\label{demo}

\subsection{Lutz-Kelker Bias}
\label{demo.lutzkelker}
Parallax measurements are not exact, so each object in a parallax-defined sample
may in fact lie closer or farther than the distance obtained by inverting the
parallax. At the boundary of a volume-limited sample, the volume of space just
outside the sample is larger than the volume just inside, so typically there
will be more objects with measured distances scattering inward than scattering
outward, artificially inflating the number of objects in the sample. This is a
manifestation of the bias that \citet{Eddington:1913tn} identified for data near
the boundaries of sample bins. \citet{Lutz:1973jv} realized that the same
concept applies to parallax-defined samples: on average, objects are slightly
farther away than measured, and their luminosities are slightly underestimated
by the parallaxes.  In the Lutz-Kelker formulation, the size of the bias depends
on the parallax uncertainty as a fraction of the parallax.  The mean fractional
uncertainty of our sample is 5\%, and the corresponding Lutz-Kelker correction
implies that our parallaxes probe a volume of space $\approx$3\% larger than
their nominal distances indicate.  Since we achieve an uncertainty of
$\approx$10\% for our space density calculation (Section~\ref{demo.density}),
Lutz-Kelker bias is not a significant source of uncertainty, but we nevertheless
account for this bias in our analysis.

\citet{Lutz:1973jv} determined that the distribution of the true parallax
$\varpi$ about the measured parallax $\varpi_0$ is
\begin{equation}
  P(\varpi|\varpi_0) \propto \left(\frac{\varpi}{\varpi_0}\right)^4 \exp\left(-\frac{(\varpi-\varpi_0)^2}{2\sigma^2}\right)
\end{equation}
where $\sigma$ is the standard deviation of $\varpi_0$ (i.e., the measurement
error).
The expected value of $\varpi$ from this distribution is 
\begin{equation}
  \label{lk.corr}
  \varpi = \frac{ \varpi_0+\sqrt{(\varpi_0^2-16\sigma^2) }}{2}
\end{equation}
We corrected the parallaxes in our list to their expected values using
Equation~(\ref{lk.corr}); the median correction was 0.4~mas. We stress that as
the Lutz-Kelker correction is a statistical correction for samples of objects,
we used these corrected parallaxes only for analysis of our volume-limited
sample as a whole and do not quote them for individual objects.  The effect of
the correction was to reduce the membership of our volume-limited sample by
$\approx$3\%, confirming that this bias is not significant relative to our
$\approx$10\% uncertainty on the overall space density.

\subsection{Completeness}
\label{demo.completeness}

We estimated the completeness of our volume-limited sample using the {\vmax}
statistic \citep{Schmidt:1968jc}, employed by both K21 and our Paper~I.
Briefly, $V$ is the volume of space with radius equal to the distance from the
Sun of a given sample member, and $V_{\rm max}$ is the volume of space with
radius equal to the outer limit of the sample (for our full sample, this is
25~pc).  Each object in the sample thus has a {\vmax} value between 0 and 1.  If
a sample has uniform spatial distribution --- a valid assumption for our sample
which sits near the Galactic midplane --- the expectation value will be
$\exvmax=0.5$. Our volume-limited sample is centered on the Sun, so we would
expect any incompleteness to be in the more distant portions of our sample where
objects appear more faint.

\begin{figure*}
  \centering
  \includegraphics[width=2\columnwidth]{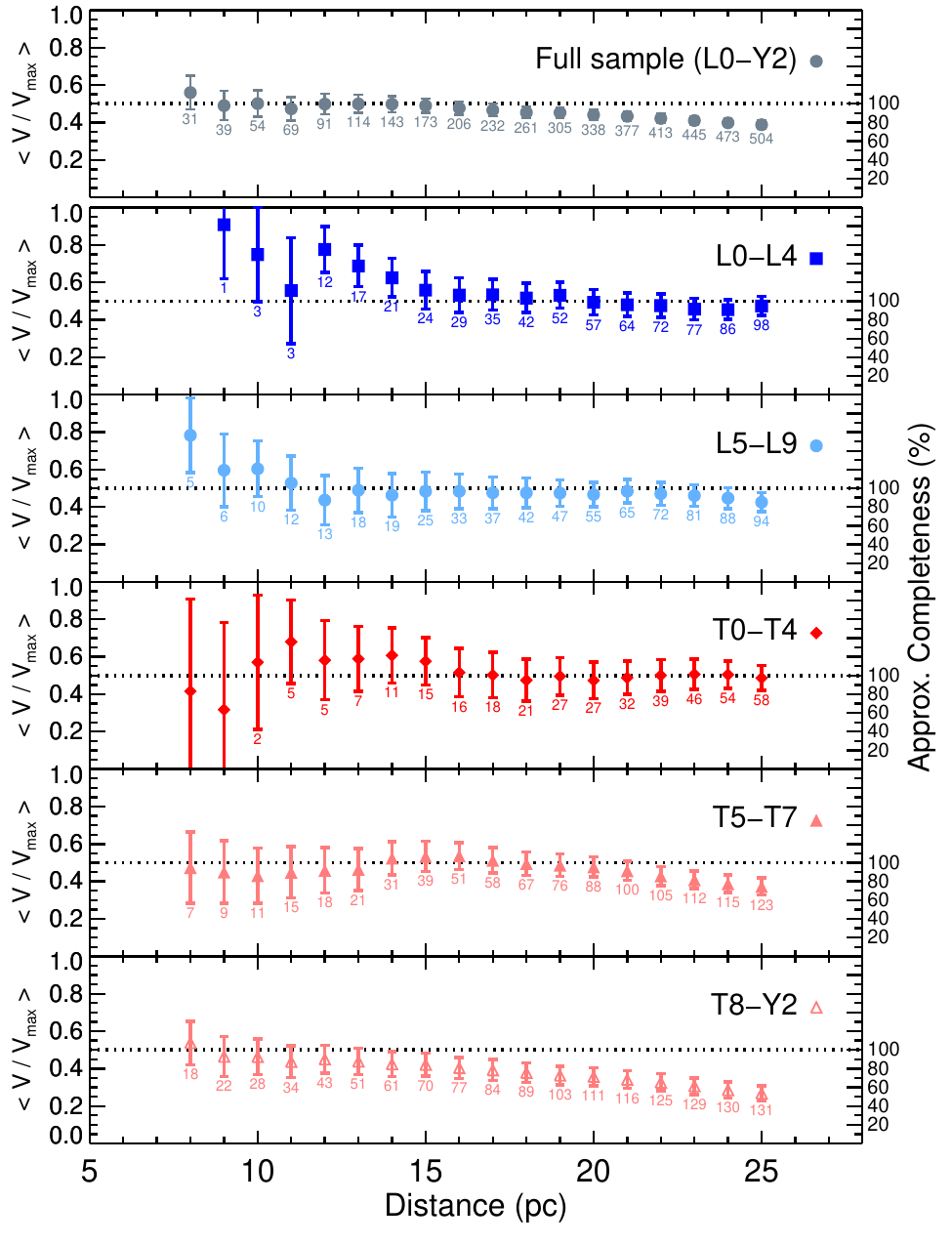}
  \caption{{\exvmax} as a function of distance for our entire volume-limited
    sample (spectral types L0--Y2, top panel) and for five spectral type bins
    (other panels). (This plot updates Figure~3 in Paper~I.) The right-hand axes
    indicate approximate completeness values corresponding to {\exvmax} values
    less than 0.5, estimated as twice the {\exvmax} value.  Our full sample has
    $\exvmax\approx0.5$ to within its uncertainty at {\varcompletedistsig}~pc,
    consistent with completeness, with the declining trend of {\exvmax} implying
    incompleteness beyond this distance.  Most of the incompleteness in our full
    sample is for spectral types T5 and later, which is unsurprising given their
    faintness.  Our sample is {\varcompleteearly} complete for L0--T4.5~dwarfs.}
  \label{fig.vmax}
\end{figure*}

Figure~\ref{fig.vmax} shows {\exvmax} as a function of limiting distance for our
volume-limited sample, for distances 8--25~pc.  In addition, we show {\exvmax}
as a function of distance for five spectral type bins.  Uncertainties for
{\exvmax} were determined using the method described in Paper~I; briefly, we
used Monte Carlo trials to incorporate the parallax uncertainties and
statistical fluctuations (drawn from the binomial distribution) due to our
limited sample size.  Figure~\ref{fig.vmax} makes clear that our volume-limited
sample is close to completeness ({\varcompleteearly} at 25~pc) for spectral
types L0--T4.  Our sample is less complete for cooler spectral types, for which
{\exvmax} steadily decreases at larger distances and is well below 0.5 at 25~pc.
The full sample has $\exvmax=0.50\pm0.04$ at {\varcompletedist}~pc
({\varcompletenum} objects).  {\exvmax} declines beyond this distance to
$\exvmax=\varvmaxfull\pm\varvmaxfullerr$ at 25~pc, indicating {\varcompletefull}
completeness for the full sample.  This is lower than in Paper~I because of the
inclusion of T8.5 and later objects, which are prohibitively faint and difficult
to observe beyond $\approx$15~pc.  Overall, the completeness of our sample is
very similar to that of Paper~I, except for the impact of adding the late-T and
Y~dwarfs from K21.

\subsection{Space Density}
\label{demo.density}

To accurately calculate the space density of L, T, and Y dwarfs in the solar
neighborhood, we needed to correct for incompleteness in our sample.  We also
needed to account for the uncertainties on the parallax measurements, which
affect whether objects observed near the 25~pc boundary are actually inside the
sample.  To address these issues, we used the UltracoolSheet to compile a list
of {\varnfullplx}~L0--Y2 dwarfs from the literature that met the same criteria
as for our volume-limited sample (Section~\ref{vlsample}) but at any distance.
We refer to this as the ``full parallax list''.  It is effectively our
volume-limited sample augmented by L0--Y2 dwarfs with parallax measurements
placing them beyond 25~pc.  Using this list allowed us to incorporate the impact
on our space density measurements of objects appearing to lie beyond 25~pc but
whose uncertainties allowed a significant possibility of membership in our
sample.

From this full parallax list we drew new 25~pc volume-limited samples in a Monte
Carlo fashion, perturbing the Lutz-Kelker-corrected parallaxes according to
their uncertainties (assumed to be normally distributed) and rejecting objects
with \hbox{parallaxes $<40$ mas}.  These perturbations sometimes moved objects
with parallaxes measured near 40~mas from outside to inside the 25~pc distance
limit and vice versa, so the membership varied at the outer edges of the Monte
Carlo samples.

We then needed to correct each Monte Carlo sample for incompleteness.  In
Paper~I, we obtained a rough correction by dividing the number of objects in a
sample by twice its {\exvmax} value (Section~\ref{demo.completeness}).  However,
this approach provides only an estimate because {\exvmax} varies with different
spatial positions of objects within a sample even when the number of objects in
the sample is fixed.  We therefore developed a new method to correct for
incompleteness that accounts for the distribution of objects within the sample
and also gives a statistical estimate of uncertainty.  For each Monte Carlo
trial, we calculated {\exvmax} at multiple distances from 2 to 25~pc (analogous
to Figure~\ref{fig.vmax}), i.e., for a distance $d$, we identified the subsample
of objects with distances less than $d$ and calculated {\exvmax} for that
volume-limited subsample.  We used steps of 0.1~pc, and identified the largest
distance (\dlim) at which a trial subsample had $\exvmax\ge0.5$.  We treated the
Monte Carlo sample as complete at that distance, and calculated the space
density for the sample using the number of objects {\nlim} enclosed in the
volume defined by {\dlim}.  Finally, we calculated {\dlim}, {\nlim}, and the
space density for our volume-limited sample as the median and 68.3\% confidence
limits from all of the Monte Carlo trials.

Table~\ref{tbl.space.density} presents our final space density results and
{\exvmax} for the entire sample, as well as separately for L, T, and Y~dwarfs,
for five spectral type bins spanning the sample, for individual spectral types,
and for single objects, binaries/triples, companions, and young objects.  For
the space density, we report two sets of confidence limits that include noise
from the binomial and Poisson distributions, respectively.  The calculation and
purposes of these confidence limits are described in detail in Paper~I.
Briefly, the binomial uncertainties ($\sigma_{\rm binomial}$) reflect the
uncertainty in our measurement of the space densities within 25~pc of the Sun,
given that our sample only covers 68\% of the associated volume.  The Poisson
uncertainties ($\sigma_{\rm Poisson}$) reflect the uncertainty in our
measurement of the space densities more broadly in our region of our Galaxy, of
which our solar neighborhood is a small part.  As in Paper~I, we adopt the
Poisson uncertainties in order to describe the space density of brown dwarfs in
general and to enable direct comparison with previous estimates.  We note that
the uncertainties for our measurements in Table~\ref{tbl.space.density} have
increased from Paper~I in most categories, with our new values better reflecting
the uncertainty in the completeness of our sample and its subsets.

\floattable
\begin{deluxetable}{lCC|CCCCC}
\centering
\tablecaption{Space Density and {\exvmax} for Our 25 pc Sample of L0--Y2 Dwarfs \label{tbl.space.density}}
\tabletypesize{\footnotesize}
\tablewidth{0pt}
\tablehead{   
  \colhead{} &
  \colhead{} &
  \colhead{} &
  \colhead{} &
  \colhead{} &
  \multicolumn{3}{c}{Space Density} \\
  \colhead{} &
  \colhead{} &
  \colhead{} &
  \colhead{} &
  \colhead{} &
  \multicolumn{3}{c}{(10$^{-3}$ objects pc$^{-3}$)} \\
  \cline{6-8}
  \colhead{Objects} &
  \colhead{$N_\mathrm{25 pc}$} &
  \colhead{\exvmax$_\mathrm{25 pc}$\tablenotemark{a}} &
  \colhead{\dlim} &
  \colhead{\nlim} &
  \colhead{Value} &
  \colhead{$\sigma_{\rm binomial}$} &
  \colhead{$\sigma_{\rm Poisson}$}
}
\startdata
L0 $\le{\rm SpT}<$ L1 & 16 & $0.49\pm0.13$ & $24.9_{-3.2}^{+0.1}$ & $16_{-4}^{+3}$ & 0.36 & $\pm0.05$ & $\pm0.10$ \\
L1 $\le{\rm SpT}<$ L2 & 28 & $0.50\pm0.10$ & $25.0_{-4.4}^{+0.0}$ & $28_{-9}^{+3}$ & 0.63 & $_{-0.09}^{+0.07}$ & $_{-0.13}^{+0.15}$ \\
L2 $\le{\rm SpT}<$ L3 & 19 & $0.46\pm0.11$ & $23.7_{-0.6}^{+0.4}$ & $18_{-2}^{+3}$ & 0.48 & $\pm0.06$ & $_{-0.11}^{+0.12}$ \\
L3 $\le{\rm SpT}<$ L4 & 14 & $0.39\pm0.13$ & $15.2_{-0.1}^{+1.0}$ & $6_{-1}^{+2}$ & 0.60 & $_{-0.10}^{+0.14}$ & $_{-0.21}^{+0.29}$ \\
L4 $\le{\rm SpT}<$ L5 & 21 & $0.48\pm0.11$ & $24.5_{-0.9}^{+0.2}$ & $21_{-4}^{+3}$ & 0.50 & $\pm0.06$ & $_{-0.11}^{+0.13}$ \\
L5 $\le{\rm SpT}<$ L6 & 28 & $0.45\pm0.10$ & $22.3_{-0.8}^{+1.1}$ & $23_{-3}^{+4}$ & 0.72 & $_{-0.09}^{+0.08}$ & $_{-0.14}^{+0.16}$ \\
L6 $\le{\rm SpT}<$ L7 & 15 & $0.42\pm0.13$ & $20.2_{-1.0}^{+1.8}$ & $11_{-2}^{+3}$ & 0.44 & $\pm0.08$ & $_{-0.13}^{+0.15}$ \\
L7 $\le{\rm SpT}<$ L8 & 17 & $0.46\pm0.13$ & $23.8_{-9.1}^{+0.7}$ & $16_{-8}^{+3}$ & 0.41 & $_{-0.07}^{+0.22}$ & $_{-0.12}^{+0.20}$ \\
L8 $\le{\rm SpT}<$ L9 & 17 & $0.40\pm0.12$ & $22.5\pm0.5$ & $16_{-3}^{+2}$ & 0.49 & $\pm0.07$ & $_{-0.12}^{+0.13}$ \\
L9 $\le{\rm SpT}<$ T0 & 17 & $0.40\pm0.13$ & $20.7_{-1.2}^{+1.1}$ & $12\pm2$ & 0.46 & $\pm0.08$ & $_{-0.13}^{+0.15}$ \\
T0 $\le{\rm SpT}<$ T1 & 9 & $0.48\pm0.18$ & $24.6_{-1.1}^{+0.4}$ & $8\pm2$ & 0.20 & $\pm0.04$ & $_{-0.07}^{+0.09}$ \\
T1 $\le{\rm SpT}<$ T2 & 9 & $0.50\pm0.19$ & $25.0_{-1.5}^{+0.0}$ & $8_{-3}^{+2}$ & 0.18 & $\pm0.04$ & $\pm0.07$ \\
T2 $\le{\rm SpT}<$ T3 & 15 & $0.43\pm0.14$ & $21.1_{-6.5}^{+2.6}$ & $11_{-6}^{+4}$ & 0.41 & $_{-0.08}^{+0.13}$ & $_{-0.13}^{+0.19}$ \\
T3 $\le{\rm SpT}<$ T4 & 11 & $0.64\pm0.16$ & $25.0\pm0.0$ & $9_{-3}^{+2}$ & 0.20 & $_{-0.07}^{+0.04}$ & $\pm0.07$ \\
T4 $\le{\rm SpT}<$ T5 & 14 & $0.44\pm0.14$ & $20.6_{-1.6}^{+3.7}$ & $10_{-3}^{+4}$ & 0.40 & $\pm0.08$ & $_{-0.12}^{+0.14}$ \\
T5 $\le{\rm SpT}<$ T6 & 40 & $0.41\pm0.08$ & $21.2_{-0.5}^{+0.8}$ & $29_{-3}^{+4}$ & 1.06 & $\pm0.12$ & $_{-0.19}^{+0.21}$ \\
T6 $\le{\rm SpT}<$ T7 & 34 & $0.35\pm0.09$ & $17.7_{-1.0}^{+1.5}$ & $21_{-4}^{+5}$ & 1.28 & $\pm0.17$ & $_{-0.28}^{+0.30}$ \\
T7 $\le{\rm SpT}<$ T8 & 49 & $0.35\pm0.07$ & $14.9_{-0.6}^{+1.3}$ & $19_{-3}^{+5}$ & 1.98 & $_{-0.27}^{+0.26}$ & $_{-0.43}^{+0.49}$ \\
T8 $\le{\rm SpT}<$ T9 & 67 & $0.32\pm0.06$ & $14.5_{-0.5}^{+0.7}$ & $29_{-5}^{+4}$ & 3.35 & $_{-0.36}^{+0.39}$ & $_{-0.61}^{+0.63}$ \\
T9 $\le{\rm SpT}<$ Y0 & 35 & $0.31\pm0.08$ & $14.3_{-1.8}^{+1.5}$ & $15_{-4}^{+6}$ & 1.81 & $_{-0.29}^{+0.32}$ & $_{-0.46}^{+0.57}$ \\
Y0 $\le{\rm SpT}<$ Y1 & 16 & $0.19\pm0.11$ & $9.3_{-0.1}^{+0.2}$ & $8\pm2$ & 3.45 & $_{-0.62}^{+0.73}$ & $_{-1.03}^{+1.35}$ \\
Y1 $\le{\rm SpT}<$ Y2 & 11 & $0.17\pm0.12$ & $11.6_{-1.1}^{+2.2}$ & $6\pm2$ & 1.31 & $_{-0.35}^{+0.41}$ & $_{-0.51}^{+0.69}$ \\
${\rm SpT}=$ Y2 & 2 & $0.03\pm0.13$ & $2.8_{-0.0}^{+7.2}$ & $1\pm1$ & 15.92 & $_{-1.02}^{+14.91}$ & $_{-1.80}^{+30.05}$ \\
\hline
L0 $\le{\rm SpT}<$ L5 & 98 & $0.47\pm0.05$ & $20.1\pm0.4$ & $58\pm5$ & 2.51 & $_{-0.18}^{+0.19}$ & $_{-0.33}^{+0.36}$ \\
L5 $\le{\rm SpT}<$ T0 & 94 & $0.43\pm0.05$ & $18.6\pm2.7$ & $48_{-17}^{+21}$ & 2.63 & $_{-0.22}^{+0.29}$ & $_{-0.37}^{+0.44}$ \\
T0 $\le{\rm SpT}<$ T5 & 58 & $0.49\pm0.07$ & $24.6_{-1.2}^{+0.4}$ & $53_{-8}^{+6}$ & 1.25 & $\pm0.11$ & $_{-0.17}^{+0.20}$ \\
T5 $\le{\rm SpT}\le$ T8 & 168 & $0.37\pm0.05$ & $16.4_{-0.4}^{+0.5}$ & $77\pm8$ & 6.08 & $_{-0.41}^{+0.40}$ & $\pm0.71$ \\
T8.5 $\le{\rm SpT}\le$ Y2 & 86 & $0.28\pm0.04$ & $10.4\pm0.2$ & $27_{-5}^{+3}$ & 8.23 & $_{-0.98}^{+0.96}$ & $_{-1.57}^{+1.66}$ \\
\hline
L0 $\le{\rm SpT}<$ T0 & 192 & $0.45\pm0.04$ & $19.0_{-0.3}^{+0.5}$ & $101_{-8}^{+12}$ & 5.10 & $\pm0.29$ & $_{-0.50}^{+0.53}$ \\
T0 $\le{\rm SpT}\le$ T8 & 226 & $0.40\pm0.04$ & $16.4\pm0.4$ & $93_{-8}^{+9}$ & 7.39 & $_{-0.45}^{+0.44}$ & $_{-0.75}^{+0.78}$ \\
T0 $\le{\rm SpT}<$ Y0 & 283 & $0.37\pm0.03$ & $15.8_{-0.6}^{+0.4}$ & $116_{-13}^{+11}$ & 10.15 & $_{-0.58}^{+0.55}$ & $_{-0.96}^{+0.98}$ \\
Y0 $\le{\rm SpT}\le$ Y2 & 29 & $0.17\pm0.07$ & $9.3_{-1.0}^{+0.8}$ & $13_{-4}^{+3}$ & 5.57 & $_{-0.97}^{+0.95}$ & $_{-1.52}^{+1.73}$ \\
\hline
Single & 429 & $0.39\pm0.02$ & $14.7_{-1.0}^{+0.6}$ & $143_{-27}^{+19}$ & 15.77 & $_{-0.79}^{+0.81}$ & $_{-1.33}^{+1.42}$ \\
Binary/triple\tablenotemark{b} & 48 & $0.48\pm0.08$ & $24.5_{-1.3}^{+0.4}$ & $45_{-6}^{+5}$ & 1.09 & $_{-0.10}^{+0.09}$ & $_{-0.16}^{+0.17}$ \\
Companion\tablenotemark{c} & 30 & $0.35\pm0.09$ & $18.7_{-0.0}^{+0.4}$ & $20_{-2}^{+3}$ & 1.07 & $_{-0.15}^{+0.13}$ & $_{-0.21}^{+0.27}$ \\
Young & 20 & $0.41\pm0.11$ & $23.2_{-0.5}^{+0.2}$ & $20_{-3}^{+2}$ & 0.56 & $_{-0.08}^{+0.07}$ & $_{-0.12}^{+0.13}$ \\
\hline
All L0 $\le{\rm SpT}\le$ T8 & 418 & $0.42\pm0.03$ & $16.4_{-0.3}^{+0.7}$ & $161_{-14}^{+20}$ & 12.46 & $_{-0.57}^{+0.56}$ & $_{-1.01}^{+0.99}$ \\
All & 504 & $0.39\pm0.02$ & $14.3_{-1.4}^{+0.5}$ & $154_{-39}^{+20}$ & 18.33 & $_{-0.92}^{+0.91}$ & $_{-1.53}^{+1.62}$ \\
\enddata
\tablecomments{
This table expands Table~2 from Paper~I by adding spectral types T8.5--Y2,
  and adopts a new correction for incompleteness (Section~\ref{demo.completeness}).
  $N_\mathrm{25 pc}$: Number of objects in our full volume-limited sample.
  \exvmax$_\mathrm{25 pc}$: Calculated for our full volume-limited sample. A sample with uniform
  spatial distribution will have $\exvmax=0.5$.
  {\dlim}: Median and 68\% confidence limits for the largest distance at
  which $\exvmax\ge0.5$ (implying a complete sample) from our Monte Carlo trials.
  {\nlim}: Median and 68\% confidence limits for the number of objects in
  the sample out to {\dlim}.
  \emph{Space Density}: Median and 68\% confidence limits for {\nlim}
  divided by the volume of the sample at {\dlim} from our Monte Carlo
  trials. $\sigma_{\rm binomial}$ describes how precisely
  our space density measurements represent the full 25~pc volume around the
  Sun. $\sigma_{\rm Poisson}$ describes how precisely our space density
  measurements represent brown dwarfs in our general neighborhood of the Galaxy.
  The calculation of $\sigma_{\rm binomial}$ and $\sigma_{\rm Poisson}$ is
  described in Appendix~B of Paper~I.
}
\tablenotetext{a}{Mean and standard deviation from Monte Carlo trials that
  resample the parallaxes from their errors and incorporate binomial
  uncertainties to account for statistical fluctuations in our sample.}
\tablenotetext{b}{Close binaries and triples are counted as single objects with
  unresolved spectral types.}
\tablenotetext{c}{Three companions are themselves binaries (see Paper~I for
  details) and are also included in the binary/triple bin.}
\end{deluxetable}

For our full 25~pc sample of L0--Y2~dwarfs, we find $\dlim=\vardlim$~pc,
$\nlim=\varnlim$~objects, and a space density of {\vardenspois}~pc$^{-3}$.
However, proceeding through our sample toward colder spectral types, {\dlim}
drops steadily beyond T3, while the number objects per spectral type peaks at T8
and then drops rapidly.  It is unclear to what extent this latter decline is due
to an actually smaller number of T9 and Y~dwarfs in the solar neighborhood, but
the decreasing {\dlim} and {\exvmax} imply that our sample is significantly
incomplete and our space density estimates may be less accurate for these cold,
faint spectral types.  Table~\ref{tbl.space.density} therefore includes several
sub-samples with spectral types down to only T8 (not including T8.5) to
facilitate comparison with our Paper~I sample, and also to emphasize that our
results for sub-samples including T8.5 and later-type dwarfs should be used with
caution.

For L0--T8~dwarfs, our space density estimate of \vardensitylt~pc$^{-3}$ has
increased by 27\% from Paper~I, more than the 13\% increase in the number of
objects with these spectral types.  Inspecting the sub-samples in
Table~\ref{tbl.space.density}, we note that for sub-samples for which {\dlim} is
close to 25~pc (implying completeness), we obtain space densities consistent
with Paper~I when accounting for the larger number of members of our updated
sample.  Where {\dlim} is significantly less than 25~pc, we systematically find
larger space densities than in Paper~I, which indicates that the method we used
to correct for incompleteness in Paper~I failed to account for the spatial
distribution of objects within the sub-samples and underestimated the space
densities of incomplete sub-samples.

Our updated \vardensityl~pc$^{-3}$ space density for L~dwarfs, as well as our
values for the L~dwarf subclasses, are systematically $\approx$20--80\% larger
than many previous estimates \citep{Cruz:2007kb,Reyle:2010gq,Marocco:2015iz},
although they are formally different only at the $\sim$1--2$\sigma$ level due to
large uncertainties.  \citet{BardalezGagliuffi:2019gn} have also recently
published space densities for M7--L5 dwarfs that are larger than previous
estimates, but our early-L measurements are $\approx$70\% smaller than
theirs. We addressed this discrepancy in Paper~I.  Our L~dwarf space density is
consistent with the recent measurements of K21 based on their 20~pc all-sky
volume-limited sample, which is somewhat surprising given that they treat the
binary components as separate objects whereas we do not
(Section~\ref{demo.luminosity}); however, assuming a $\approx$15\% binary
fraction for our sample, our space densities would differ by $\approx$10\%,
i.e., consistent within uncertainties.  For T0--T8~dwarfs, a similar trend is
clear: our updated \vardensityt~pc$^{-3}$ space density is larger than some
previous estimates \citep{Burningham:2013gt,Marocco:2015iz} but consistent with
those of \citet{Metchev:2008gx}, \citet{Reyle:2010gq}, and K21.

\subsection{Bolometric Magnitudes}
\label{demo.mbol}
Bolometric luminosities (\lbol) are fundamental, observable physical quantities
of ultracool dwarfs that can be directly compared to evolutionary models,
providing constraints on ages and masses.  To determine these constraints for L,
T, and Y~dwarfs in the solar neighborhood, we calculated the bolometric
luminosity (\lbol) of each member of our volume-limited sample.  We used the
{\loglbol} vs. {\mkmko} polynomial of \citet{Dupuy:2017ke}, derived from {\lbol}
calculated by \citet{Filippazzo:2015dv}, for objects with
$9.1\le\mkmko\le17.8$~mag, the magnitude range over which this polynomial is
valid.  This range included the brightest objects in our sample and provided
{\lbol} for 370~objects, but did not include 133~objects with $\mkmko>17.8$~mag
or which lacked $K$-band photometry.

For the remaining objects (mostly having spectral types $\ge$T8), we created an
updated version of the ``super-magnitude'' method developed by
\citet{Dupuy:2013ks}, applicable to objects with spectral types L4 and later.
Our updated method is described in detail in Appendix~\ref{appendix.supermag}.
Briefly, we used the Sonora-Bobcat model atmospheres \citep{Marley:2021ba} to
define polynomials that convert the combined absolute flux in several bands
(usually {\jmko}, {\hmko}, and {\spitzer}/IRAC [3.6] and [4.5], or {\jmko},
{\hmko}, and AllWISE $W1$ and $W2$) into {\lbol}.  The Sonora-Bobcat models
cover {\teff} down to 200~K, below the coldest objects in our sample, so using
this super-magnitude method we were able to calculate {\lbol} for 105 of the
remaining 133~objects, many of which had no previous determination of {\lbol}.

We used the {\kmko} (when possible) or {\jmko} bolometric corrections of
\citet{Liu:2010cw} to determine {\lbol} for 26 of the remaining objects, leaving
two for which we could not determine a value of {\lbol} because they lack the
photometry needed for all of the above methods.  These two objects,
ULAS~J074502.79+233240.3 and GJ~758B, were therefore not included in our
volume-limited sample.

We converted {\lbol} to bolometric magnitude (\mbol) using $\mbolsun=4.74$~mag.
For the super-magnitude-derived {\lbol}, we propagated the uncertainties from
the objects' parallaxes and photometry through Monte Carlo trials to obtain
uncertainties for the {\mbol}.  We present our {\mbol} results in
Table~\ref{tbl.sample}.  We caution that for unresolved binaries our {\mbol}
calculation should be treated as less reliable, particularly for pairs in which
the secondary component contributes a significant amount of the total flux.

\subsection{Luminosity Function for Single Objects}
\label{demo.luminosity}
As discussed at length in Paper~I, there is evidence suggesting that the
components of binaries and companions to high-mass primaries may have different
mass distributions and formation histories from single objects, e.g., the
spectral type distribution of wide companions notably favors T~dwarfs more than
in the rest of our volume-limited sample.  The numbers of binaries and
companions in our volume-limited sample may also be impacted by different
selection effects.  These related but distinct populations therefore need to be
treated as such.  In addition, a proper analysis of binaries needs {\lbol} to be
calculated separately for each component, requiring resolved absolute photometry
which is often not available, in particular in the mid-infrared {\spitzer}/IRAC
and WISE bands.  We therefore focus our remaining analysis on the single objects
in our volume-limited sample.

We used The UltracoolSheet and our own high-angular resolution imaging survey
(W. Best et al., in preparation, described briefly in Paper~I) to identify
\varnbinary~binaries and multiples and \varnwidecomp~wide companions
(separations typically $>$10\arcsec) to main-sequence primaries in our sample,
which we removed to create a volume-limited sample of \varnsingle~single
objects.  Similarly, we removed binaries and wide companions from the full
parallax list to create a list of single L0--Y2~dwarfs in our volume-limited
sample and beyond.  Using this list, we determined the bolometric luminosity
function of the single objects in our volume-limited sample, correcting for
incompleteness and incorporating Poisson uncertainties in the manner described
in Section~\ref{demo.density}.  We used bins of 1~mag in {\mbol} and Monte Carlo
trials to calculate uncertainties.  The mean effect of our Lutz-Kelker
correction (Section~\ref{demo.lutzkelker}) on our luminosity bins was
\hbox{$-0.03\times10^{-3}$ pc$^{-3}$} per bin. This is less than the
uncertainties on the space densities in those bins, confirming that our
luminosity function was not significantly impacted by the correction.

\begin{figure*}
  \centering
  \includegraphics[width=2\columnwidth, trim=32 0 0 0mm, clip]{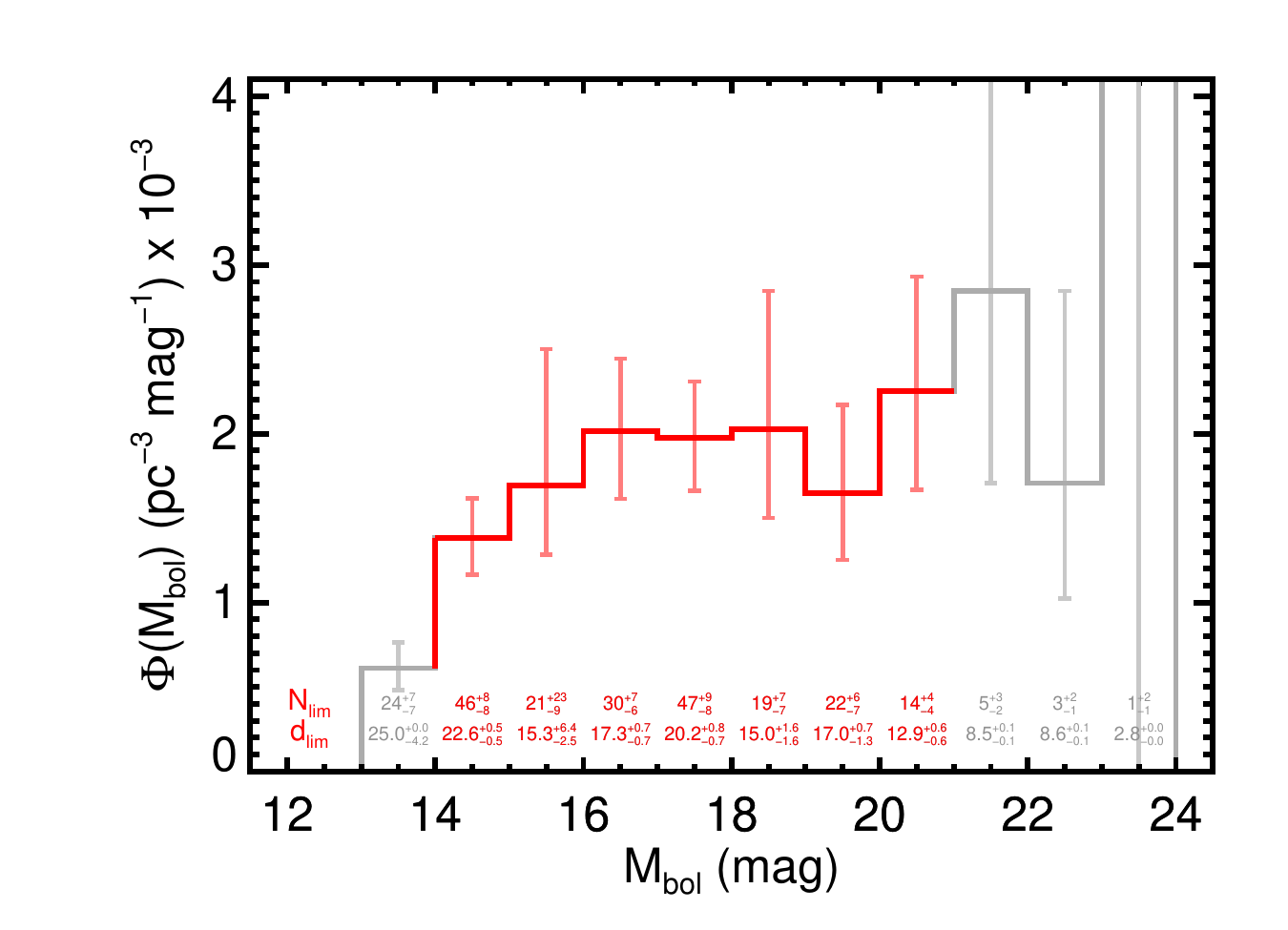}
  \caption{ The bolometric luminosity function for single objects in our
    volume-limited sample, plotted as the space density of bins of 1~bolometric
    magnitude. Along the bottom we print the number of objects within the
    volume-complete distance (\nlim) and the volume-complete distance (\dlim)
    for each bin from Table~\ref{tbl.lum.func}; we use these numbers to
    calculate the space densities.  Bins plotted in red have robustly-measured
    space densities, which we use to infer population parameters in
    Section~\ref{popsynth}.  Bins excluded from our analysis are plotted in
    gray: the brightest bin (13--14~mag) is incomplete due to our exclusion of
    M~dwarfs from our sample, while the three faintest bins contain fewer than
    10 objects within their {\dlim}. The luminosity function is consistent with
    a flat distribution for $\mbol\ge15$~mag, but the overall shape suggests an
    increase toward fainter magnitudes.}
  \label{fig.lum.func}
\end{figure*}

We present our completeness-corrected luminosity function in
Table~\ref{tbl.lum.func}, and plot it in Figure~\ref{fig.lum.func}.  This is the
first bolometric luminosity function calculated for single brown dwarfs in the
solar neighborhood.  The function spans $13\le\mbol\le24$~mag, and appears
consistent with a flat distribution for $\mbol\ge15$~mag, although there is a
suggestion of gradual increase in space density toward fainter luminosities.
The brightest bin ($\mbol=13$--14~mag) corresponds to $\approx$M8--L1~dwarfs
\citep{Zhang:2020jn} and is thus significantly incomplete due to our exclusion
of M~ dwarfs from our sample, so we do not include this bin in our subsequent
population synthesis analysis.  On the faint end, the three bins with
$\mbol\ge21$~mag have large uncertainties due to fewer objects and
$\dlim<10$~pc, so we likewise exclude these bins from our analysis.

\floattable
\begin{deluxetable}{LCC|CCCCC}
\centering
\tablecaption{Luminosity Function for Our 25~pc Volume-Limited Sample of Single L0--Y2 Dwarfs\label{tbl.lum.func}}
\setlength{\tabcolsep}{0.10in}
\tablewidth{0pt}
\tablehead{   
  \colhead{} &
  \colhead{} &
  \colhead{} &
  \colhead{} &
  \colhead{} &
  \multicolumn{3}{c}{Space Density} \\
  \colhead{} &
  \colhead{} &
  \colhead{} &
  \colhead{} &
  \colhead{} &
  \multicolumn{3}{c}{(10$^{-3}$ objects pc$^{-3}$)} \\
  \cline{6-8}
  \colhead{Objects} &
  \colhead{$N_\mathrm{25 pc}$} &
  \colhead{\exvmax$_\mathrm{25 pc}$} &
  \colhead{\dlim} &
  \colhead{\nlim} &
  \colhead{Value} &
  \colhead{$\sigma_{\rm binomial}$} &
  \colhead{$\sigma_{\rm Poisson}$}
}
\startdata
13.0\le\mbol<14.0\tablenotemark{a} & 26 & 0.50\pm0.10 & 25.0_{-4.2}^{+0.0} & 25_{-7}^{+4} & 0.60 & _{-0.09}^{+0.11} & _{-0.13}^{+0.15} \\
14.0\le\mbol<15.0 & 56 & 0.46\pm0.07 & 22.6\pm0.5 & 45\pm5 & 1.37 & _{-0.14}^{+0.15} & _{-0.22}^{+0.23} \\
15.0\le\mbol<16.0 & 55 & 0.44\pm0.07 & 15.3_{-2.5}^{+6.4} & 19_{-8}^{+23} & 1.82 & _{-0.34}^{+0.65} & _{-0.41}^{+0.81} \\
16.0\le\mbol<17.0 & 58 & 0.42\pm0.07 & 17.3\pm0.7 & 30\pm5 & 1.99 & \pm0.27 & _{-0.40}^{+0.43} \\
17.0\le\mbol<18.0 & 70 & 0.39\pm0.07 & 20.2_{-0.7}^{+0.8} & 47_{-7}^{+6} & 1.96 & \pm0.21 & _{-0.31}^{+0.33} \\
18.0\le\mbol<19.0 & 48 & 0.35\pm0.08 & 15.0\pm1.6 & 19_{-7}^{+6} & 1.99 & _{-0.39}^{+0.60} & _{-0.53}^{+0.82} \\
19.0\le\mbol<20.0 & 44 & 0.31\pm0.08 & 17.0_{-1.3}^{+0.7} & 22_{-6}^{+4} & 1.62 & _{-0.29}^{+0.37} & _{-0.40}^{+0.52} \\
20.0\le\mbol<21.0 & 41 & 0.15\pm0.08 & 12.9\pm0.6 & 14_{-2}^{+3} & 2.22 & _{-0.39}^{+0.43} & _{-0.59}^{+0.68} \\
21.0\le\mbol<22.0\tablenotemark{b} & 22 & 0.07\pm0.09 & 8.5\pm0.1 & 5\pm1 & 2.85 & _{-0.65}^{+0.57} & _{-1.14}^{+1.71} \\
22.0\le\mbol<23.0\tablenotemark{b} & 7 & 0.02\pm0.08 & 8.6\pm0.1 & 3\pm1 & 1.65 & \pm0.53 & _{-0.68}^{+1.14} \\
23.0\le\mbol<24.0\tablenotemark{b} & 1 & 0.00\pm0.02 & 2.8\pm0.0 & 1_{-1}^{+0} & 15.92 & _{-15.92}^{+0.00} & _{-15.92}^{+31.85} \\
\enddata
\tablecomments{The columns are the same as in Table~\ref{tbl.space.density},
  presented here for bins of 1~bolometric magnitude. {\dlim}, {\nlim}, 
  and space densities were calculated and corrected for incompleteness using 
  the same method as for the spectral type bins in Table~\ref{tbl.space.density}.
  The space densities are plotted with Poisson uncertainties in
  Figure~\ref{fig.lum.func}.
}
\tablenotetext{a}{This bin is incomplete due to our exclusion of M~dwarfs from 
  our sample. We do not use this {\mbol} bin in our comparison with synthetic 
  populations (Section~\ref{popsynth.construct}).}
\tablenotetext{b}{Space density is based on $\nlim<10$ and has large uncertainties.
  We do not use this {\mbol} bin in our comparison with synthetic populations
  (Section~\ref{popsynth.construct}).}
\end{deluxetable}

\section{Population Synthesis}
\label{popsynth}

\subsection{Overview}
\label{popsynth.overview}
To constrain the mass and age distributions underlying the observed luminosity
function of ultracool dwarfs, we employed population synthesis, i.e., forward
modeling of our volume-limited sample.  Such modeling adopts parametrized
distributions of masses and ages of objects, from which synthetic objects are
drawn.  Each synthetic object's mass and age can then be used to derive {\lbol},
{\teff}, and other physical properties using evolutionary models.  We can then
compare luminosity functions for our synthetic populations to that of our
volume-limited sample to constrain the underlying age and mass parameters.  This
comparison enables us to simultaneously constrain the mass and age distributions
of nearby brown dwarfs.

\subsection{Initial Mass Function}
\label{popsynth.imf}
\begin{figure*}
  \includegraphics[width=1\columnwidth, trim = 10mm 0 10mm 0]{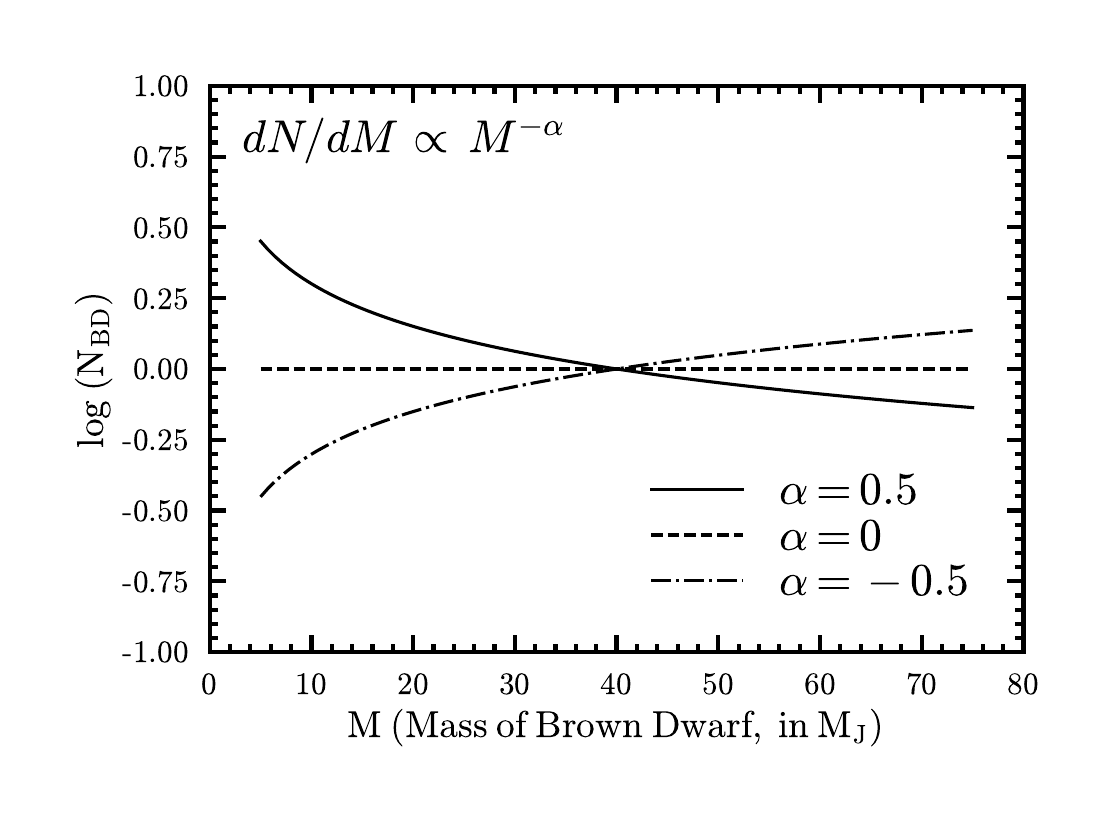}
  \includegraphics[width=1\columnwidth, trim = 0mm 0 20mm 0]{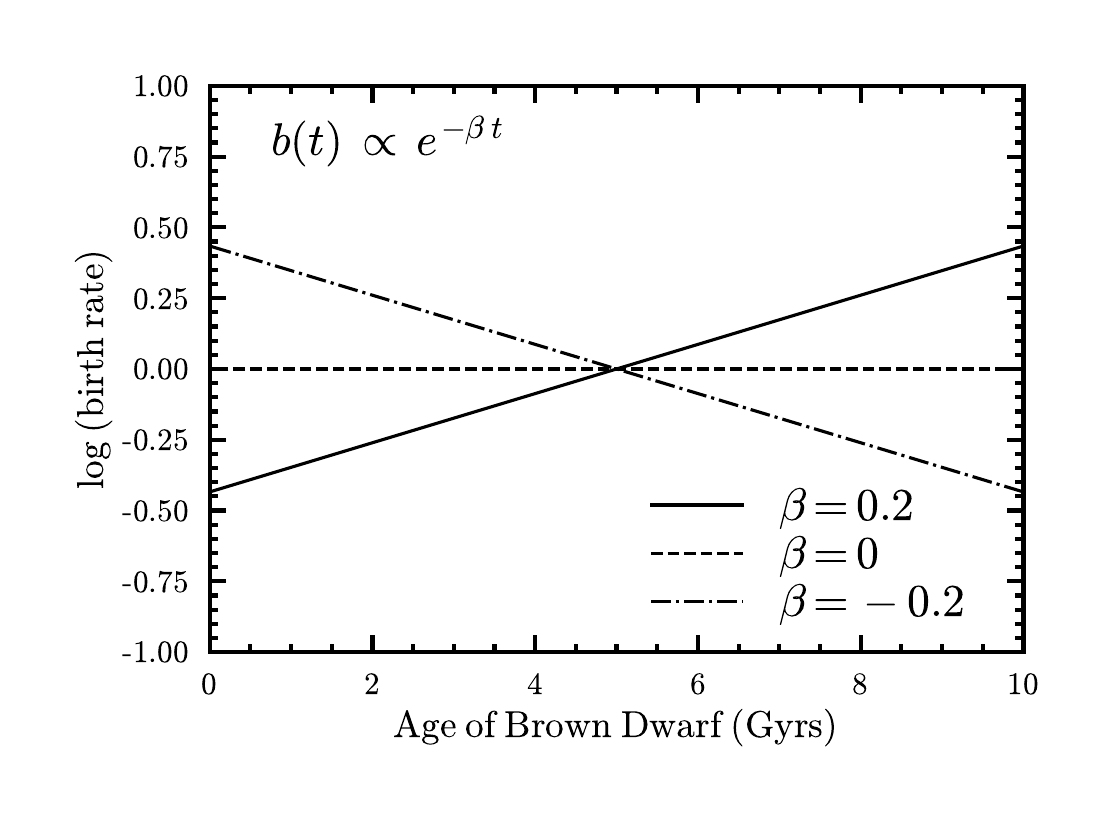}
  \caption{Left: Examples of mass distributions drawn from Kroupa’s power law
    form of the IMF for $\alpha=\{-0.5, 0, +0.5\}$.  Right: Examples of
    exponential age distributions with $\beta=\{-0.2, 0, +0.2\}$.  Both vertical
    axis scales are arbitrary.  Previous ultracool dwarf studies have found
    estimates spanning $-1\lesssim\alpha\lesssim1$ with $\alpha=0$ corresponding
    to a uniform mass function, and have typically assumed a uniform age
    distribution ($\beta=0$).}
\label{fig.distrib.examples}
\end{figure*}

Studies of the stellar IMF typically adopt a power law with form
\hbox{$\xi(M)\equiv dN/d(\log M)\propto M^{-\Gamma}$},
where $M$ is stellar mass and $N$ is the occurrence rate.
This is the form used in the seminal work by \citet{Salpeter:1955hz}, who
determined that occurrence decreases toward larger masses, with $\Gamma=1.35$
for stars with \hbox{${\rm masses}\ge1\msun$} in nearby clusters.  Previous work
on ultracool dwarfs has commonly used the alternate power-law form
\citep{Kroupa:2001ki}
\begin{equation}
  \label{eqn.imf}
  \Psi(M) \equiv \frac{dN}{dM} \propto M^{-\alpha}
\end{equation}
with $\Psi$ representing space density,
so we also used this form,
which is related to the Salpeter form by
$\alpha=\Gamma+1$ (so $\alpha=2.35$ for $\ge1\msun$ stars).
Figure~\ref{fig.distrib.examples} shows three examples of the brown dwarf mass
distributions drawn from different choices of $\alpha$, where $\alpha=0$ means a
uniform distribution of masses.
\citet{Chabrier:2003ki} has also shown that an IMF which takes the form of a
power law above 1~\msun\ and a lognormal distribution below 1~\msun\ (defined by
a peak mass and a characteristic width) is consistent with data available at the
time.

\subsection{Age Distribution}
\label{popsynth.age}
For age distributions, we used the exponential form
\begin{equation}
  \label{eqn.birthrate}
  b(t) \propto e^{-\beta t}
\end{equation}
where $b$ is the birthrate and $t\in[0,10]$~Gyr is the time since the formation
of the Galaxy; hence, the present-day age of an object born at time $t$ is
($10-t$)~Gyr.  This is the most commonly used form in previous work that does
not simply assume a flat age distribution
\citep[e.g.,][]{Allen:2005jf,Deacon:2006ga,DayJones:2013hm}.
Figure~\ref{fig.distrib.examples} shows three examples of age distributions
resulting from different choices of $\beta$.  Note that $\beta=0$ is equivalent
to a flat age distribution, which is consistent with many estimates for the
formation history of nearby stars \citep[e.g.,][]{Soderblom:1991do,Gizis:2002ej}
but has been called into question by more recent work
\citep[e.g.,][]{Mor:2019ge,Fantin:2019fa}.

\subsection{Evolutionary Models}
\label{popsynth.evomodel}
We used two sets of evolutionary models to generate {\lbol} and {\teff} for
members of our synthetic ultracool dwarf populations
(Section~\ref{popsynth.construct}).
 
(1) The solar-metallicity ``hybrid'' evolutionary models of \citet[hereinafter
SM08]{Saumon:2008im} have to date provided the best matches to measured masses
and luminosities for L through mid-T dwarfs
\citep{Dupuy:2015gl,Dupuy:2017ke,Chen:2022hl}.  The SM08 models assume a gradual
loss of cloud opacity as objects cool through the L/T transition, coupling
cloudy models at 1400~K with cloudless models at 1200~K by linearly
interpolating the surface boundary condition in {\teff} for each surface gravity
in the model grid.  The models are conjectural in that they do not derive from a
physical explanation for the cloud-clearing, but they are coupled to the cooling
of the brown dwarf interiors.  The SM08 models span a mass range of
0.002~{\msun} to 0.085~{\msun} and an age range of 3~Myr to 10~Gyr, so these are
the boundaries of our synthetic populations.

(2) The AMES-COND evolutionary models \citep[hereinafter COND]{Baraffe:2003bj}
underlie the widely-used BT-Settl model atmosphere grids \citep{Allard:2012jx}.
BT-Settl includes clouds for L~dwarfs and a physical prescription for the
transition to clear-photosphere T~dwarfs, making these models well-suited for
studying both L and T~dwarf atmospheres as well as the L/T transition.  However,
the atmospheric component of the BT-Settl models is not coupled with the
evolutionary component and does not impact the {\lbol} and {\teff} in the model
grid.  The evolutionary components of the BT-Settl models are thus identical to
each other and to the original COND models, except that the BT-Settl versions
have more densely sampled age grids and extend to higher masses (beyond the
upper mass limit of our analysis).  We therefore refer to this evolutionary
model grid as simply COND, although we use the solar-metallicity grid published
with BT-Settl.  The COND models have a wider mass range of 0.0005~{\msun} to
0.1~{\msun} and age range of 1~Myr to 10~Gyr, but for consistency we limit our
synthetic populations to the boundaries required by the SM08 models.

We note that more recent evolutionary models are available, but they sample
less of the brown dwarf parameter space or do not incorporate clouds. 
BHAC15 \citep{Baraffe:2015fw} has a lower mass limit of 0.01~{\msun} and a
lower {\teff} limit of 1200~K, and thus excludes most T~dwarfs and all Y~dwarfs.
ATMO2020 \citep{Phillips:2020ea} has an upper mass limit of 0.075~{\msun}, and
therefore does not include objects with $\mbol\le14$~mag at ages
$\gtrsim$1~Gyr, thus excluding synthetic analogs of the low-mass stars in our
volume-limited sample.
In addition, ATMO2020 does not incorporate the impact of atmospheric clouds on
evolution.
The Sonora Bobcat models \citep{Marley:2021ba} have masses spanning
0.0005~{\msun} to 0.1~{\msun} and age range of 0.01~Myr to 20~Gyr but do not
incorporate clouds.

\subsection{Construction and Comparison of Synthetic Luminosity Functions}
\label{popsynth.construct}
\begin{figure*}
\fig{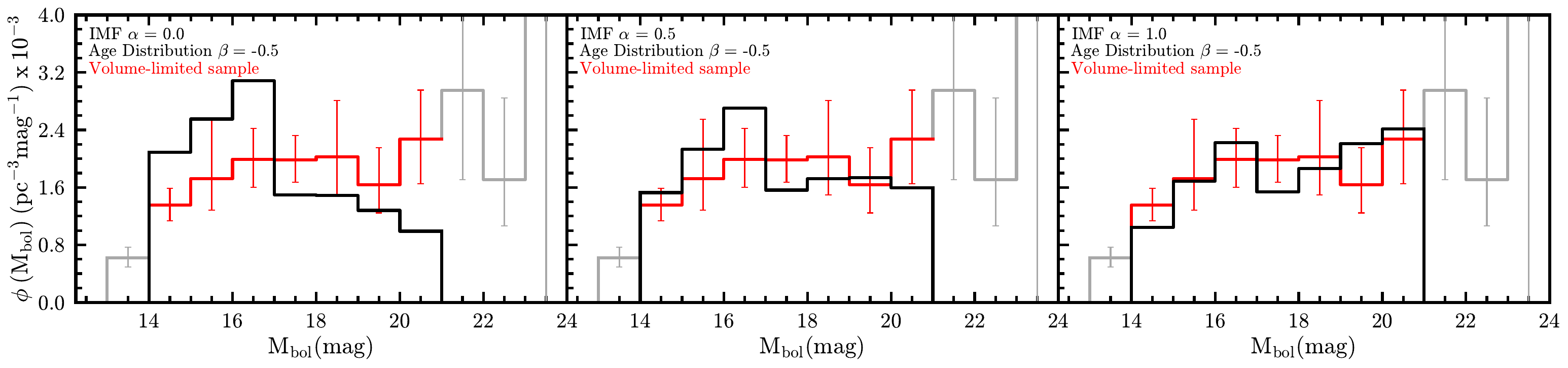}{1\textwidth}{}
\caption{ Synthetic bolometric luminosity functions based on power-law IMFs with
  three representative values of $\alpha$ (black) compared with our
  volume-limited sample's luminosity function (red/gray, from
  Figure~\ref{fig.lum.func}).  The synthetic {\mbol} were derived from the SM08
  models and assume an exponential age distribution with $\beta=-0.5$.
  Increasing $\alpha$ (left to right) leads to populations with more faint
  objects because larger $\alpha$ favors the creation of more low-mass objects.
}
\label{fig.alpha.examples}
\end{figure*}

\begin{figure*}
\fig{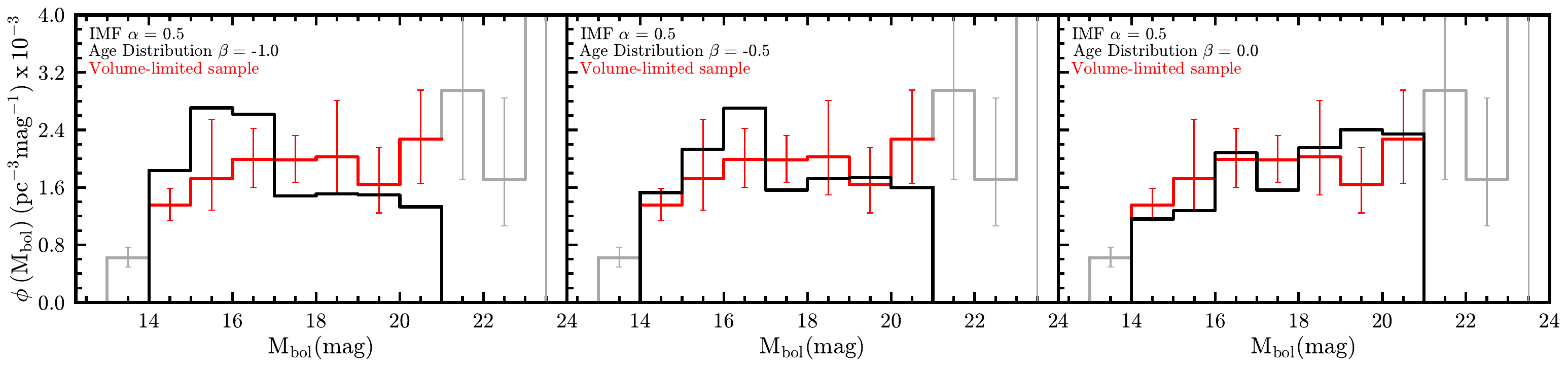}{1\textwidth}{}
\caption{ Synthetic bolometric luminosity functions based on exponential age
  distributions with three representative values of $\beta$ (black) compared
  with the luminosity function for our volume-limited sample (red/gray, from
  Figure~\ref{fig.lum.func}).  The synthetic {\mbol} were derived from the SM08
  models and assume a power-law IMF with $\alpha=0.5$.  Increasing $\beta$ (left
  to right) leads to populations with more faint objects because larger $\beta$
  favors older objects which have cooled to fainter luminosities.  }
\label{fig.beta.examples}
\end{figure*}

We constructed synthetic populations with the same volume and space density as
our corrected 25~pc sample (Table~\ref{tbl.lum.func}) of single objects,
oversampled by a factor of 1000.
We drew masses from the power-law distribution of Equation~(\ref{eqn.imf})
and ages from the exponential distribution of Equation~(\ref{eqn.birthrate}),
and used the evolutionary models to interpolate {\lbol} and {\teff} for each object.
We converted the synthetic {\lbol} to {\mbol} and binned synthetic objects into
the same 1~mag \mbol\ bins we used for our volume-limited sample
(Table~\ref{tbl.lum.func}).
Figure~\ref{fig.alpha.examples} shows examples of these synthetic luminosity
functions and illustrates the impact of varying the mass distribution
($\alpha$) on synthetic luminosity functions with the age distribution
($\beta$) held constant.
Larger values of $\alpha$ generate populations with more
low-luminosity objects, since larger $\alpha$ favors lower masses
(Figure~\ref{fig.distrib.examples}).
Figure~\ref{fig.beta.examples} shows the impact of varying $\beta$ with $\alpha$
held constant: Larger $\beta$ similarly results in populations with more
low-luminosity objects, since those populations have larger numbers of older
objects.

When comparing the synthetic luminosity functions to our volume-limited sample,
we did not use the \hbox{$13\le\mbol<14$ mag} bin, which is significantly
incomplete in our volume-limited sample due to our exclusion of M~dwarfs.  We
also did not use any objects with $\mbol\ge21$~mag because our volume-limited
sample contains fewer than 10~objects within the completeness distance (\dlim)
in each of those faint magnitude bins, so the completeness-corrected space
densities in those bins have very large uncertainties.  The total space density
of the seven bins spanning $14\le\mbol<21$~mag is
{\vardensitymbolrange}~pc$^{-3}$. We normalized the luminosity functions of our
synthetic populations to match this.

The SM08 models include masses 0.002~{\msun} to 0.085~{\msun}
and ages 3~Myr to 10~Gyr,
so we could only assign masses and ages to our synthetic objects within these
limits.
We explored whether this impacted the distribution of $14\le\mbol<21$~mag
objects in our synthetic populations.
The SM08 model grid includes objects with $\mbol<14$~mag at all ages,
so the upper mass limit of the models does not encroach on the $\mbol\ge14$~mag
cutoff for our analysis.
Objects with ${\rm mass}=0.002$~{\msun} have synthetic $\mbol\ge21$~mag for
${\rm ages}\ge100$~Myr,
so the lower mass limit of the models only impacts the faint end of our analysis 
at young $<$100~Myr ages.
The COND models have wider mass and age ranges
(Section~\ref{popsynth.evomodel}),
so for consistency with our SM08 population synthesis, we only draw masses and
ages from the same ranges.
Limitations in model parameter space are not a significant issue in our analysis
given our wide {\mbol} range.

We constructed a grid of synthetic populations using $-2\le\alpha\le2$ in steps
of 0.02 and $-5\le\beta\le1.5$ in steps of 0.02.
We initially calculated $\chi^2$ for the fit of our volume-limited sample's
luminosity function to that of each synthetic population
using our mean and rms for each {\mbol} bin,
and we identified a single $\chi^2$ minimum at
$(\alpha,\beta)=(\varalphaminchisq, \varbetaminchisq)$.
However, we also noted significant asymmetry in some of the uncertainties on
our volume-limited sample's 
{\mbol} bins (Table~\ref{tbl.lum.func}; Figure~\ref{fig.lum.func}).
We therefore sought a method to quantitatively compare our synthetic luminosity
functions that did not assume a Gaussian distribution of luminosities in each
bin of our volume-limited sample, as is inherent to the $\chi^2$ statistic.

For each bin of {\mbol}, we determined the space density histogram from our
Monte Carlo trials (Section~\ref{demo.luminosity}) in bins of
$10^{-5}$~pc$^{-3}$, and we smoothed these histograms using a boxcar of width
$2.5\times10^{-4}$~pc$^{-3}$ ($\approx$10\% of the extent of the 95\% confidence
intervals for the {\mbol} bins).  We normalized each smoothed histogram to a
total of 1 and treated these as probability distribution functions (PDFs) for
the space densities of the {\mbol} bins in our volume-limited sample.  To
compare each synthetic luminosity function to our volume-limited sample, we
assigned the probability from the corresponding PDF to the synthetic space
density in each {\mbol} bin.  We took the product of these probabilities for
each synthetic luminosity function to be its overall likelihood.

\begin{figure}
  \centering
  \includegraphics[width=1\columnwidth, trim = 20mm 0 10mm 0]{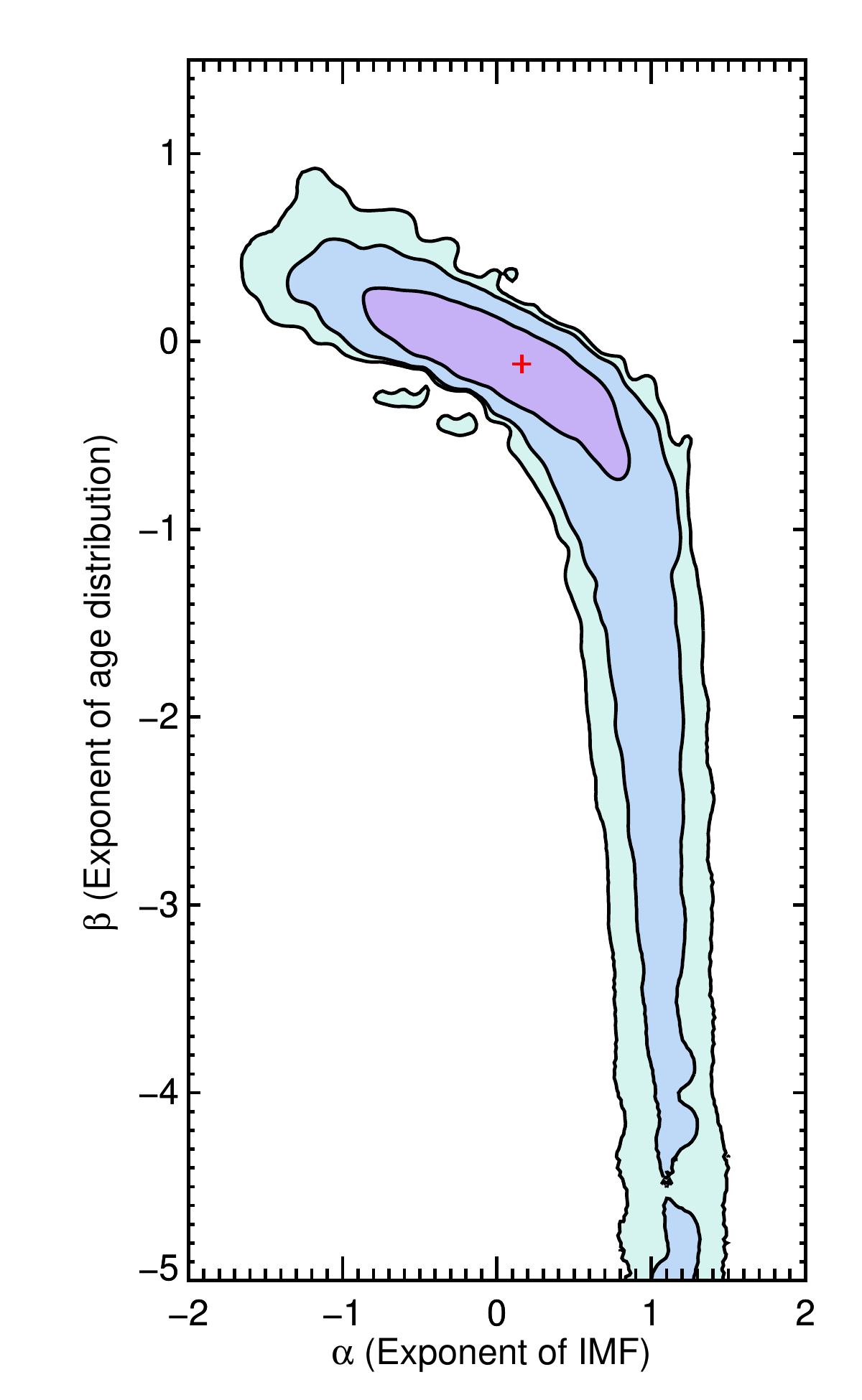}
  \caption{Contour plot (smoothed) of likelihoods for the fits of our
    volume-limited sample's luminosity function to those from synthetic
    populations based on the $\alpha$ (IMF) and $\beta$ (age distribution)
    parameters and the SM08 evolutionary models.  The red cross at
    $\alpha=\varalphamaxpdf$ and $\beta=\varbetamaxpdf$ indicates the maximum
    likelihood.  Contours trace the 68.3\%, 95.5\%, and 99.7\% confidence
    limits.  The mass function is constrained only to a broad
    $-1\lesssim\alpha\lesssim1$ range.  $\beta$ is only constrained toward
    negative values, and plausible (95.5\% confidence) fits extend to $\beta<-5$
    where populations are unrealistically young.  Comparing our luminosity
    function to synthetic populations cannot by itself produce meaningful
    constraints for the mass and age distributions.}
  \label{fig.alpha.beta.prob.sm08}
\end{figure}

Figure~\ref{fig.alpha.beta.prob.sm08} shows the distribution of likelihoods over
the ($\alpha$, $\beta$) parameter grid for the synthetic populations based on
the SM08 models.  We identified a maximum likelihood at
$(\alpha,\beta)=(\varalphamaxpdf, \varbetamaxpdf)$ (similar to the $\chi^2$
minimum), but also a large range of $(\alpha,\beta)$ having likelihoods near the
maximum.
For synthetic populations based on the COND models, we obtained a broadly
similar distribution but with a maximum likelihood at
$(\alpha,\beta) = (\varalphamaxpdfcond,\varbetamaxpdfcond)$.  The confidence
limits for both sets of evolutionary models show that the mass function is
constrained only to a broad $-1\lesssim\alpha\lesssim1$, which encompasses all
previous literature estimates.  $\beta$ is constrained to $\lesssim0.5$, but can
extend to large negative values, implying that our volume-limited sample's
luminosity function is consistent with a wide range of age distributions,
excluding only very old populations.  Plausible fits to the volume-limited
sample within the 95\% confidence limits extend to $\beta<-5$ where populations
have unrealistically young age distributions, i.e, $>$40\% of objects in the
volume-limited sample would have ages less than 200~Myr, and there would be no
objects older than $\approx$2~Gyr.  Such populations are clearly in conflict
with even the limited age constraints from lithium depletion, binaries with
measured dynamical masses, and kinematics
\citep[e.g.,][]{Kirkpatrick:2008ec,Dupuy:2017ke,Hsu:2021kr}.

\subsubsection{A Constraint from the Young L0--L7 Fraction}
\label{popsynth.construct.youngl}
The lack of strong constraints on $\beta$ from this luminosity function analysis
indicates the need for another source of constraint on the age distribution of 
our volume-limited sample.
In Paper~I, we identified young objects based on spectroscopic
indicators of youth or kinematic association with a young moving group of
stars. 
Here, we focus specifically on L0--L7~dwarfs in our sample, for which
red-optical and near-IR spectroscopic indicators of low surface gravity 
are available for ages $\lesssim$200~Myr
\citep{Cruz:2009gs,Allers:2013hk,Gagne:2015dc,Liu:2016co}.
Our volume-limited sample contains 12 single L0--L7~dwarfs having such
indicators of youth,
After correcting for incompleteness and Lutz-Kelker bias,
12 young out of 127 single L0--L7~dwarfs
represents a fraction of
{\varyoungsinglfrac} in our volume-limited sample.
For our analysis, we adopt an age of 0--200~Myr for these objects as suggested
by \citet{Allers:2013hk} and \citet{Liu:2016co}, with the caveat that no
well-calibrated age scale has been established for the spectroscopic low gravity
indicators.
(We discuss plausible age ranges for young L0--L7~dwarfs and their
  impact on our results in Section~\ref{results.youngl}).
We note that our young L0--L7~dwarf fraction is consistent with the
$7.6\%\pm1.6\%$ fraction of L~dwarfs with ages~$<100$~Myr found by
\citet{Kirkpatrick:2008ec} based on the presence of lithium absorption.

\begin{figure}
  \centering
  \includegraphics[width=1\columnwidth, trim = 20mm 0 10mm 0]{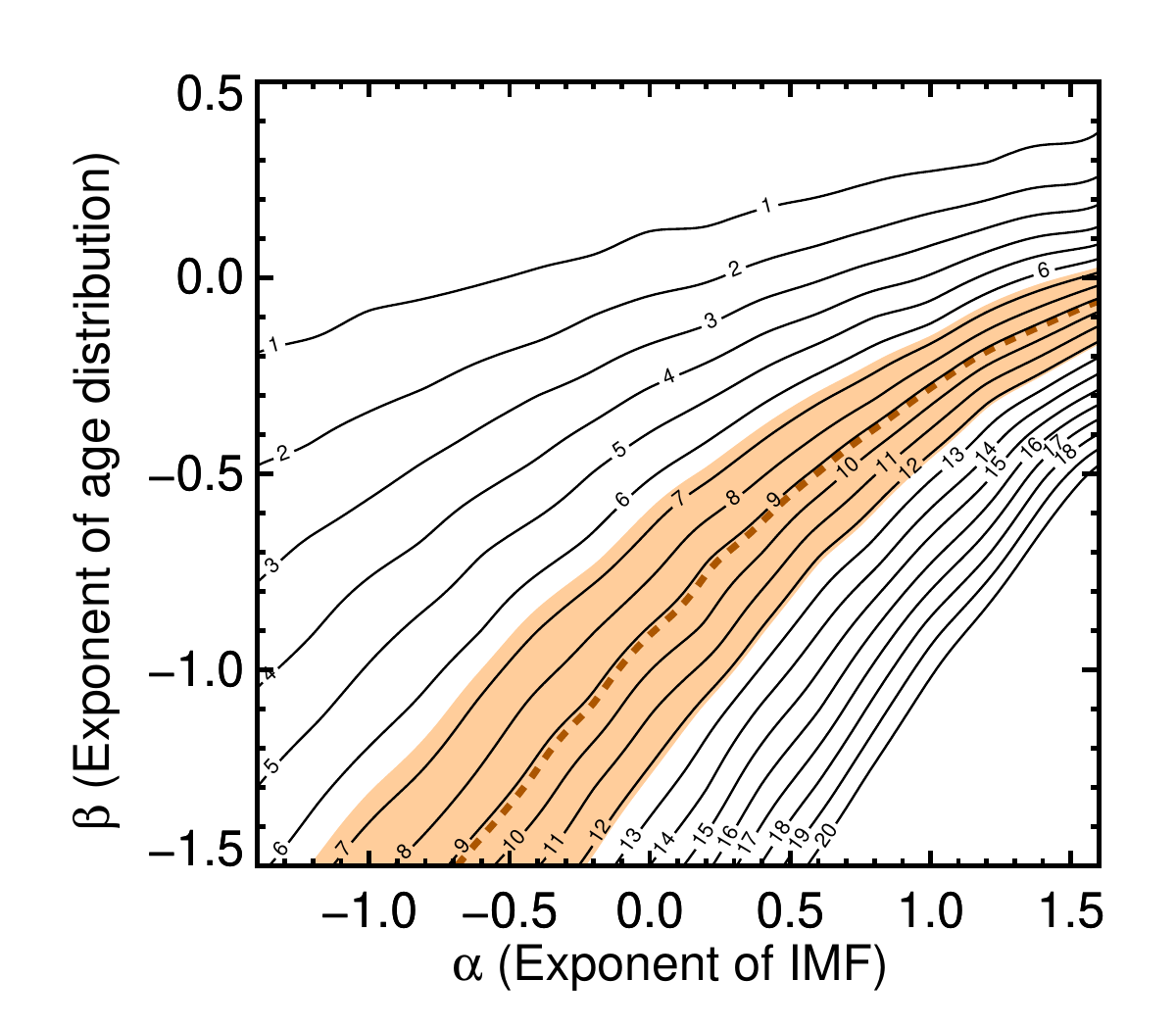}
  \caption{Smoothed contours showing the percentage of single L0--L7~dwarfs in
    our synthetic populations that are young ($<200$~Myr), as a function of the
    $\alpha$ (mass function) and $\beta$ (age distribution) parameters.
    Synthetic populations with a larger percentage of young L0--L7~dwarfs occur
    for higher $\alpha$ (more lower-mass objects) and lower $\beta$ (more young
    objects).  The brown dashed contour and shaded region indicate where
    {\varyoungsinglfrac} of L0--L7~dwarfs in the synthetic populations are
    young, corresponding to the percentage measured in our volume-limited
    sample.}
  \label{fig.youngfrac.chisq}
\end{figure}

We converted the {\teff} values for our synthetic population members to spectral
types by inverting the SpT--{\teff} relation of \citet{Stephens:2009cc},
identifying objects with synthetic spectral types L0--L7 (specifically, later
than M9.75 and earlier than L7.25).
We then calculated the percentage of L0--L7~dwarfs with ages less than 200~Myr
in each of our synthetic populations.
Figure~\ref{fig.youngfrac.chisq} shows these percentages as a contour plot in
the same $\beta$ vs. $\alpha$ space used in
Figure~\ref{fig.alpha.beta.prob.sm08},
but for a narrower, more realistic range of $\alpha$ and $\beta$ --- in
particular, setting $-1.5\le\beta\le0.5$ to exclude implausibly young age distributions.
Within this more realistic range,
synthetic populations with lower values of $\beta$ have more young objects and
thus naturally have a higher percentage of young L0--L7~dwarfs.
Populations with higher values of $\alpha$ also have a higher percentage of 
young L0--L7~dwarfs because they have more low-mass objects which cool quickly
to later spectral types, so the warmer L0--L7~dwarfs that are in the sample are more
likely to be young.
Figure~\ref{fig.youngfrac.chisq} illustrates that the constraints in ($\alpha$,
$\beta$) from the young L0--L7 population are complementary to those from the
luminosity function analysis (Figure~\ref{fig.alpha.beta.prob.sm08}).

\begin{figure*}
  \centering
  \begin{minipage}[t]{0.48\textwidth}
    \includegraphics[width=1\columnwidth, trim = 20mm 0 10mm 0]{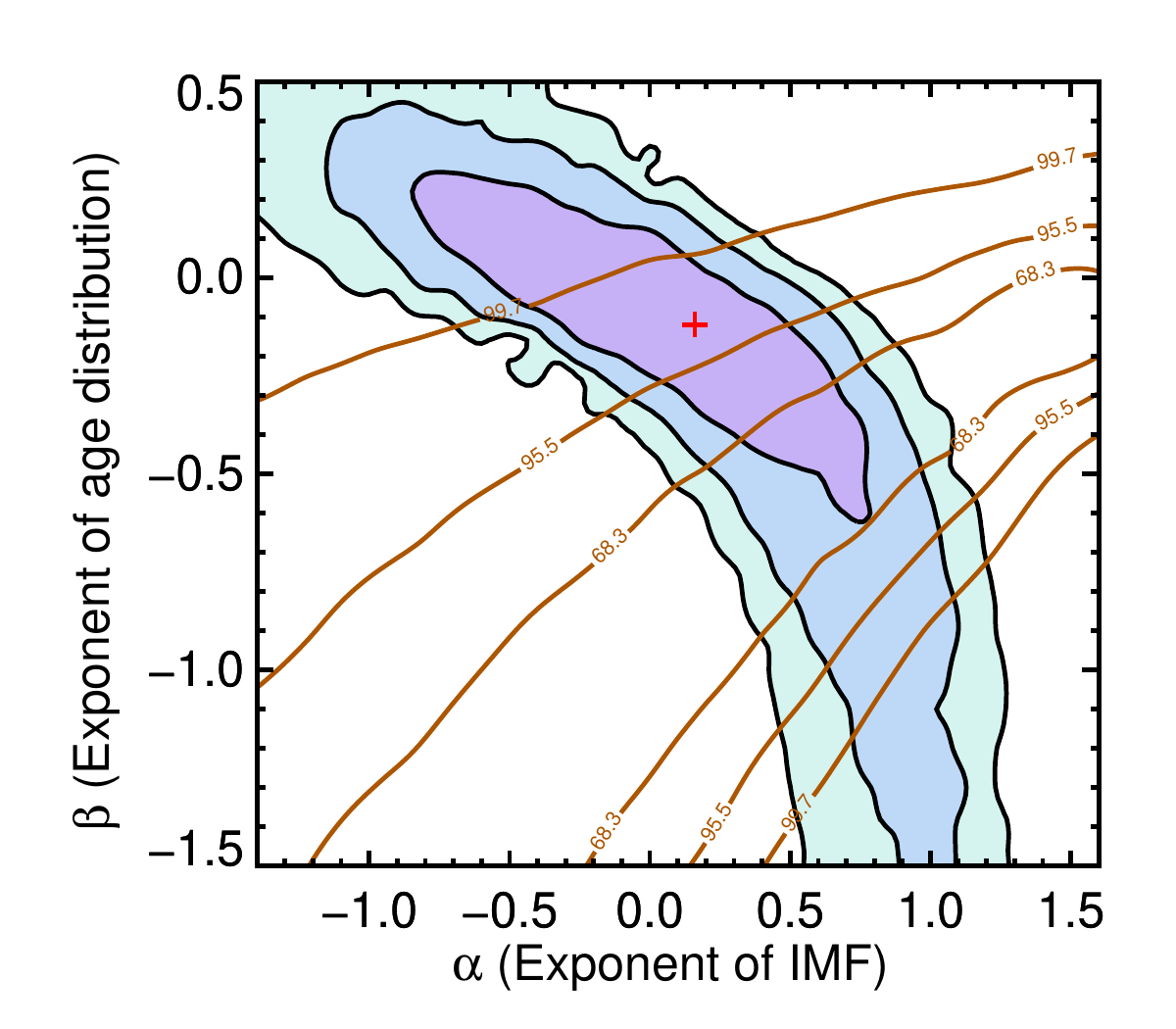}
  \end{minipage}
  \hfill
  \begin{minipage}[t]{0.48\textwidth}
    \includegraphics[width=1\columnwidth, trim = 20mm 0 10mm 0]{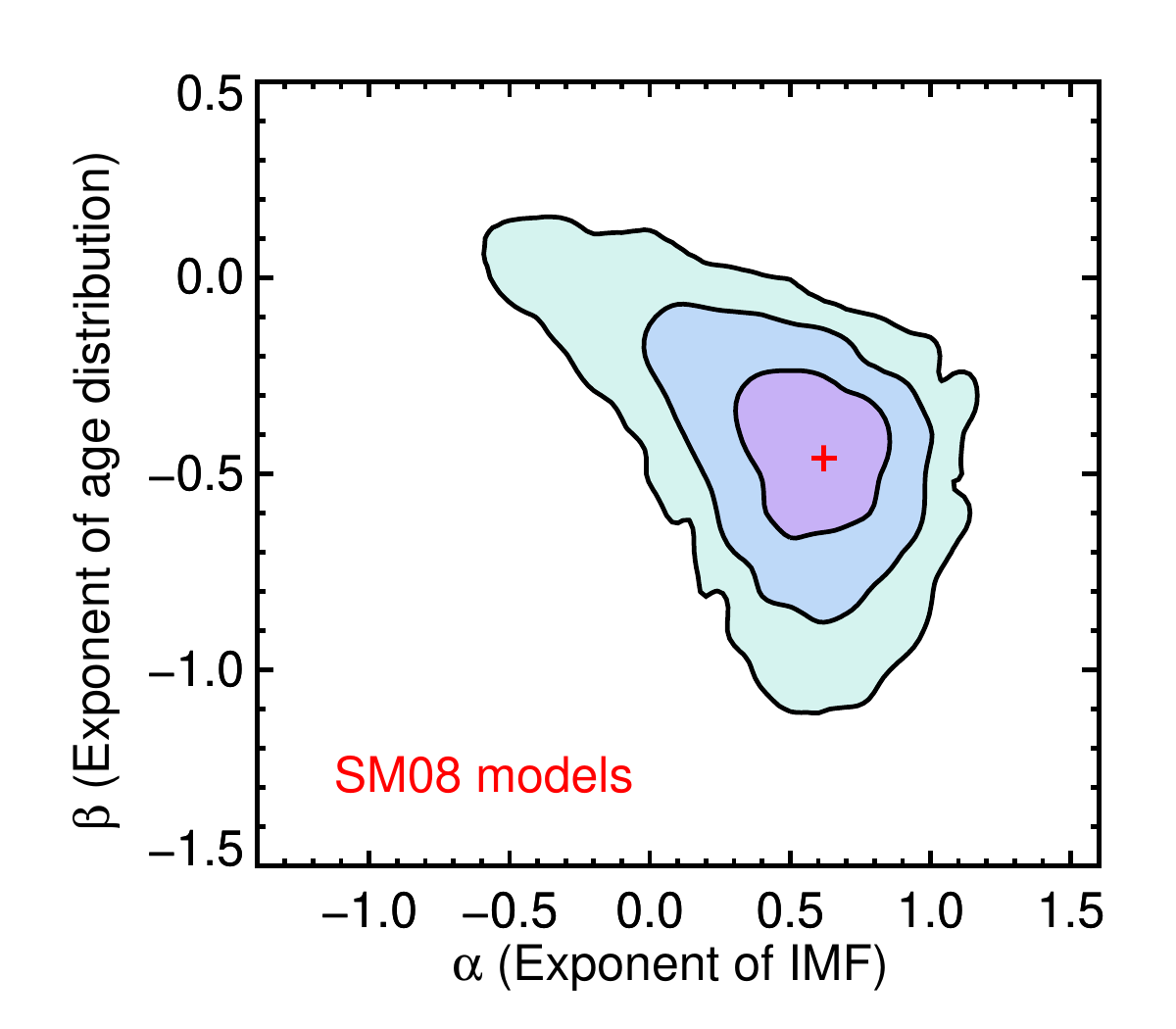}
  \end{minipage}
  \caption{{\it Left}: Constraints on the $\alpha$ (mass function) and $\beta$
    (age distribution) parameters from our luminosity function analysis using
    the SM08 evolutionary models, calculated for the same ($\alpha$, $\beta$)
    grid as in Figure~\ref{fig.youngfrac.chisq}. The red cross indicates the
    likelihood maximum.  The overlaid brown 68.3\% ($1\sigma$), 95.5\%
    ($2\sigma$), and 99.7\% ($3\sigma$) contours represent the
    {\varyoungsinglfrac} fraction of single young L0--L7~dwarfs in our volume
    limited sample (shown in Figure~\ref{fig.youngfrac.chisq}).  {\it Right}:
    Product of the probability distributions from the luminosity function and
    young L0--L7 fraction analyses, with contours showing the 68.3\%, 95.5\%,
    and 99.7\% confidence limits, and a maximum likelihood at
    $\alpha=\varalphamedconf$ and $\beta=\varbetamedconf$.  The intersection of
    the luminosity function and young L0--L7 fraction constraints provides a
    much stronger joint constraint on $\alpha$ and $\beta$ than either analysis
    alone, thereby yielding the strongest statistical constraints to date on the
    mass and age distributions of nearby L0--Y2~dwarfs.}
\label{fig.product.contours.sm08}
\end{figure*}

Figure~\ref{fig.product.contours.sm08} combines the ($\alpha,\beta$) constraints
from the luminosity function analysis,
recalculated for the grid used in Figure~\ref{fig.youngfrac.chisq},
with the constraints from the percentage of young L0--L7~dwarfs in our
volume-limited sample.
The two sets of contours are roughly perpendicular in the ($\alpha,\beta$)
plane.  We multiplied the probability grids from each analysis to combine these
two constraints, obtaining the final ($\alpha,\beta$) constraints shown in
Figure~\ref{fig.product.contours.sm08}.
This combined analysis establishes strong constraints on both $\alpha$
and $\beta$, which were not possible based on either analysis separately.
We then marginalized the combined probability distribution over the
($\alpha$,$\beta$) grid for each parameter to obtain their median values and
confidence limits.

\section{Mass and Age Distribution Constraints}
\label{results}
\begin{figure}   
  \centering
  \includegraphics[width=1\columnwidth, trim = 20mm 0 10mm 0]{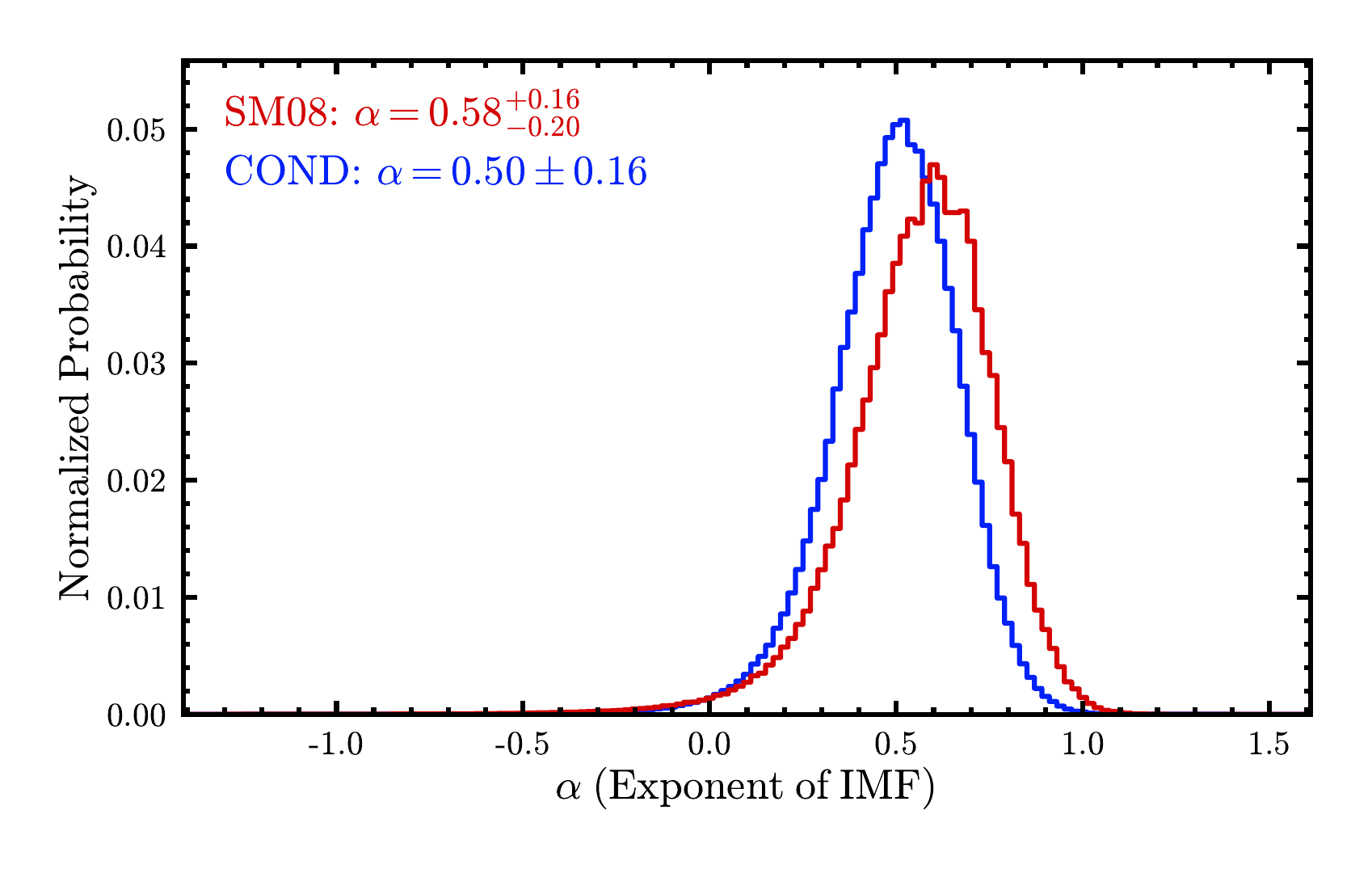}
  \includegraphics[width=1\columnwidth, trim = 20mm 0 10mm 0]{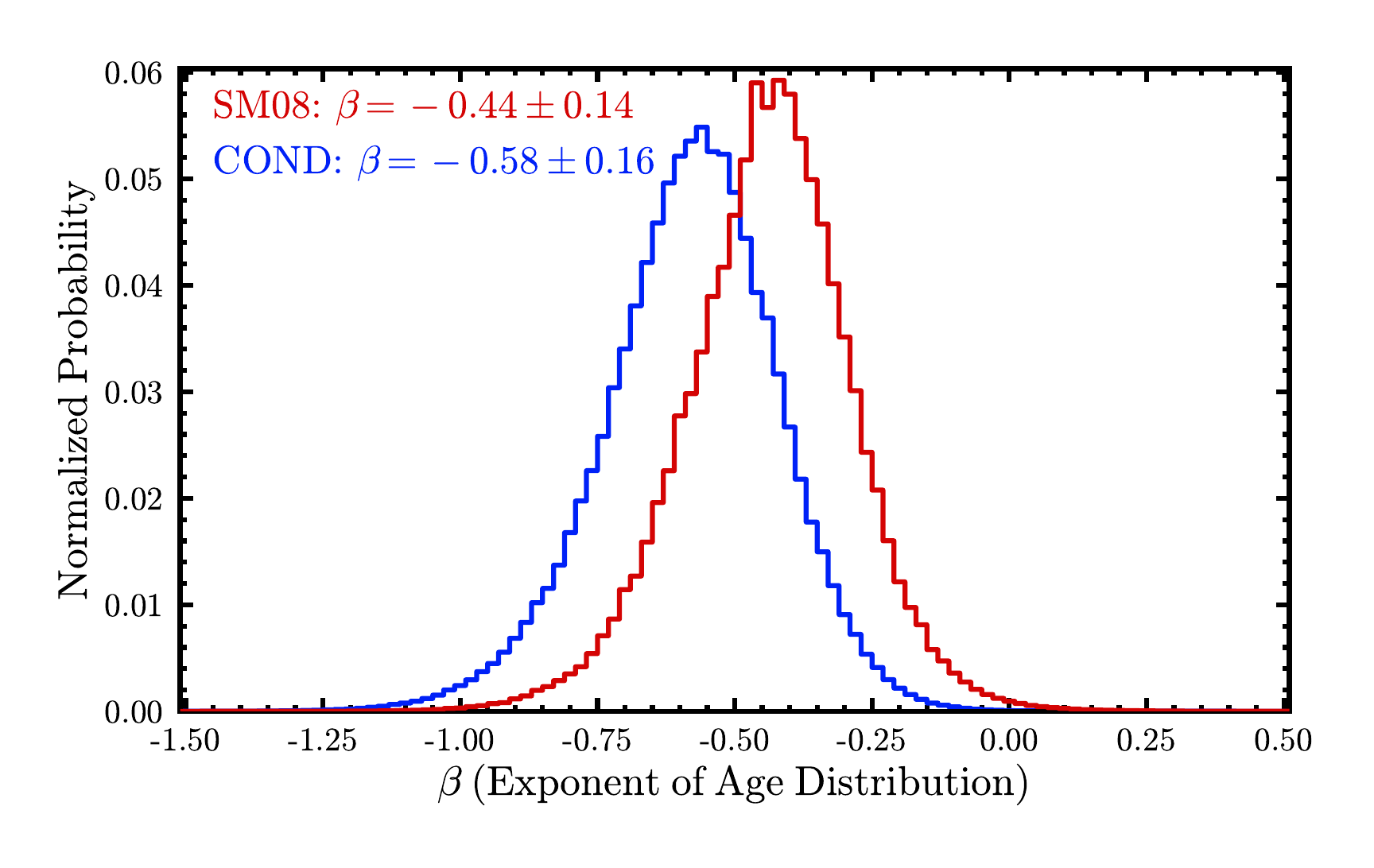}
  \caption{Probability distribution functions (PDFs) for the $\alpha$ (mass
    function, {\it top}) and $\beta$ (age distribution, {\it bottom}) parameters
    marginalized from the combined probability distributions shown in
    Figures~\ref{fig.product.contours.sm08} and~\ref{fig.product.contours.cond},
    using the SM08 (red) and COND (blue) evolutionary models, respectively. The
    median and 68\% confidence limits are shown at upper left in both
    plots. Both parameters are well constrained, and the significant overlap in
    the PDFs for each parameter confirm that the choice of evolutionary model
    does not significantly impact our results.  }
  \label{fig.pdf.imf.sfh}
\end{figure}

For the SM08 evolutionary models, we obtained $\alpha=\varalphamedconf$ and
$\beta=\varbetamedconf$ (median values with 68\% confidence limits) with
nearly-symmetric posterior distributions (Figures
\ref{fig.product.contours.sm08} and~\ref{fig.pdf.imf.sfh}).  These are more
precise statistical constraints on both parameters than from any previous
analysis.  The fit of the luminosity function for our volume-limited sample with
that of the synthetic population generated using $\alpha=\varalphamed$ and
$\beta=\varbetamed$ yields $\chi^2=\varmedchisq$ for 4 degrees of freedom.

\begin{figure*}
  \centering
  \begin{minipage}[t]{0.48\textwidth}
    \includegraphics[width=1\columnwidth, trim = 20mm 0 10mm 0]{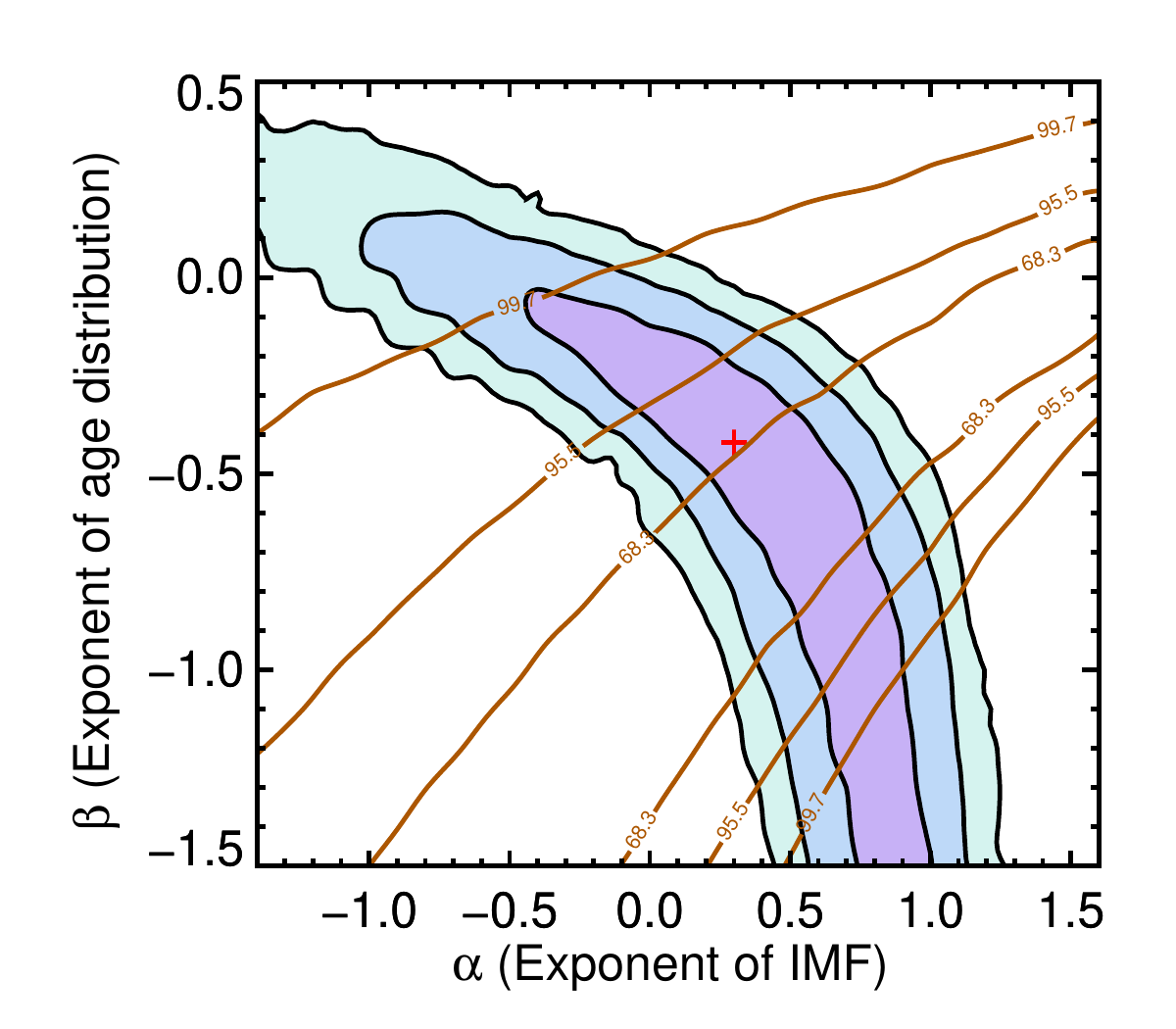}
  \end{minipage}
  \hfill
  \begin{minipage}[t]{0.48\textwidth}
    \includegraphics[width=1\columnwidth, trim = 20mm 0 10mm 0]{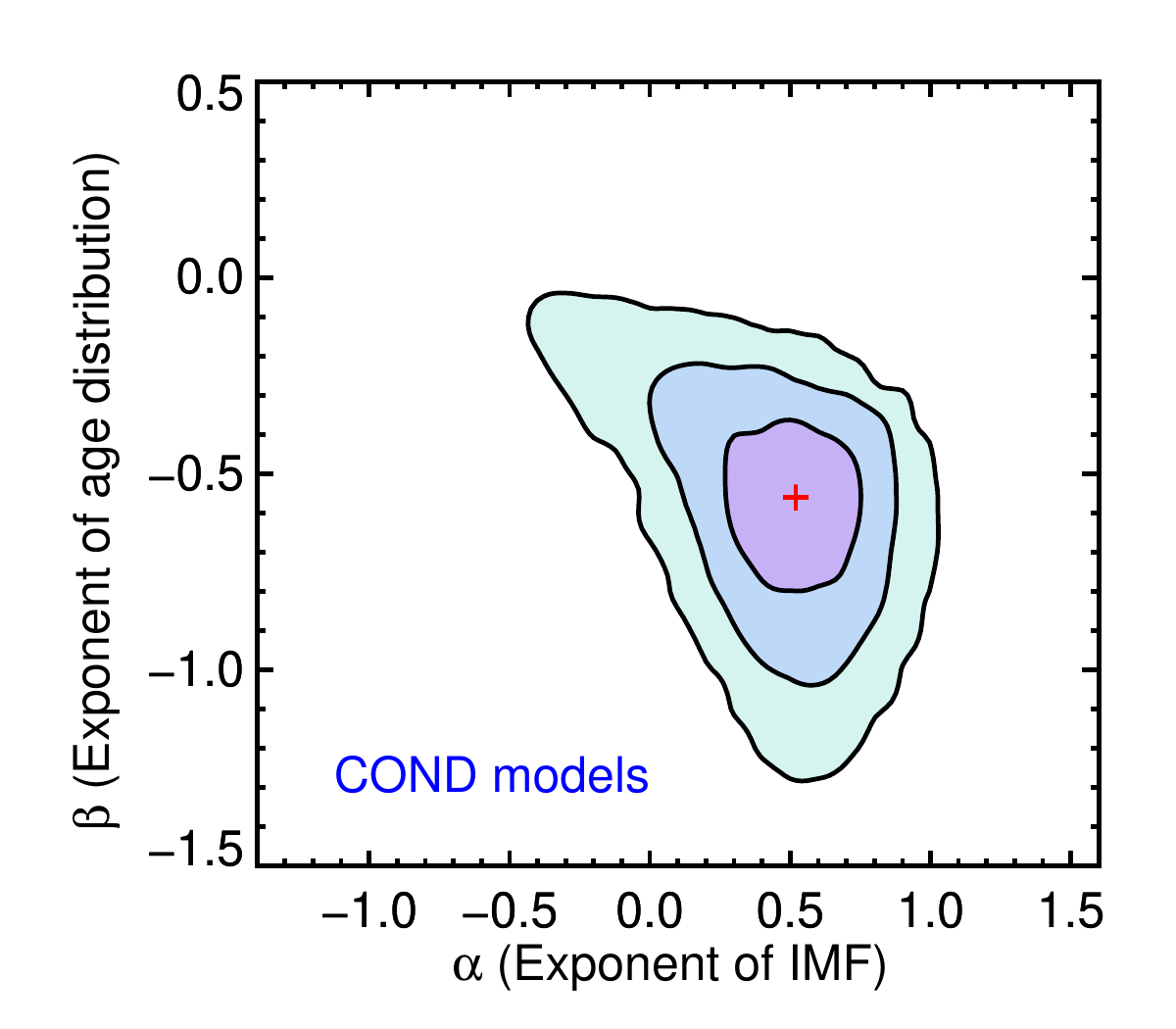}
  \end{minipage}
  \caption{Same as 
    Figure~\ref{fig.product.contours.sm08}, but using the COND
    evolutionary models, resulting in constraints of 
    $\alpha=\varalphamedcondconf$ and
    $\beta=\varbetamedcondconf$.
    The contours and median parameter values are consistent with those found
    using the SM08 models, indicating that the choice of evolutionary model does
    not significantly impact our analysis.}
  \label{fig.product.contours.cond}
\end{figure*}

Repeating the steps outlined in Section~\ref{popsynth.construct} but using the
COND evolutionary models instead of SM08, we obtained
$\alpha=\varalphamedcondconf$ and $\beta=\varbetamedcondconf$, again from
near-symmetric posteriors (Figures \ref{fig.pdf.imf.sfh}
and~\ref{fig.product.contours.cond}).  These values are consistent with those
found using the SM08 models, with the corresponding synthetic luminosity
function providing a fit with $\chi^2=\varmedchisqcond$ (4 degrees of freedom).
The similarity in the $\alpha$ and $\beta$ posteriors indicates that the choice
of evolutionary model does not significantly impact our analysis, as the young
L0--L7 fraction constraint exerts a strong influence.

Our sample includes 57 objects that have only photometric distances from K21 (no
parallaxes; Section~\ref{vlsample.def}). We explored whether the bolometric
luminosities we calculated for these objects could be systematically impacting
our population synthesis results. We repeated our analysis using the SM08 models
but excluding these objects, and obtained nearly identical contours, medians,
and confidence limits for $\alpha$ and $\beta$. This perhaps surprising result
is a consequence of our method of correcting for incompleteness
(Section~\ref{demo.density}, which effectively discards objects beyond the
largest distance at which a sample or subsample is complete ({\dlim}) from our
population analysis. These 57 objects without parallaxes are almost all farther
away than the completeness distances for their respective bolometric magnitude
bins, so removing them from our sample had little impact on our calculations.

\subsection{Best-fit Synthetic Luminosity Functions}
\begin{figure}
  \centering
  \includegraphics[width=1\columnwidth, trim = 20mm 0 10mm 0]{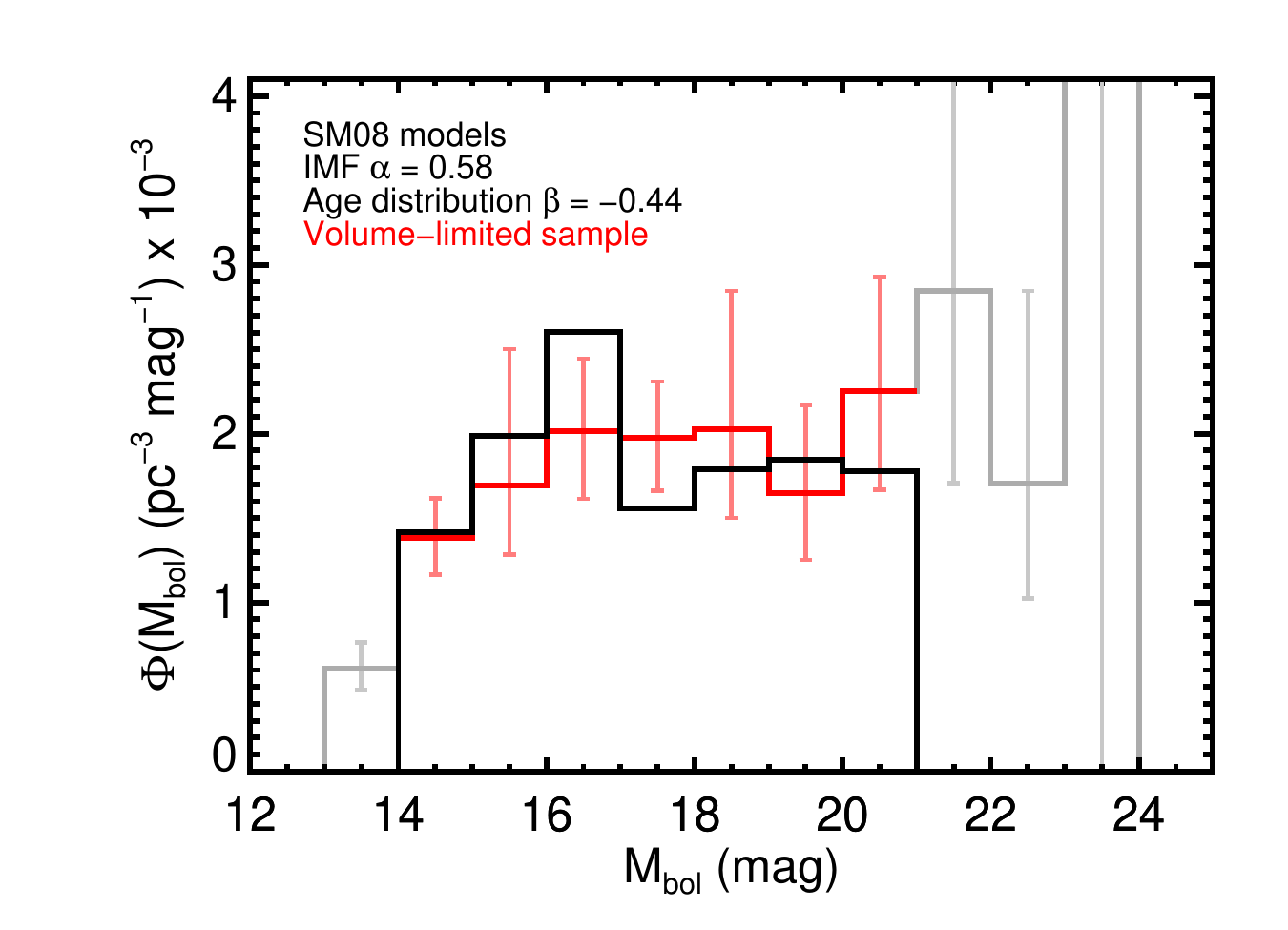}
  \includegraphics[width=1\columnwidth, trim = 20mm 0 10mm 0]{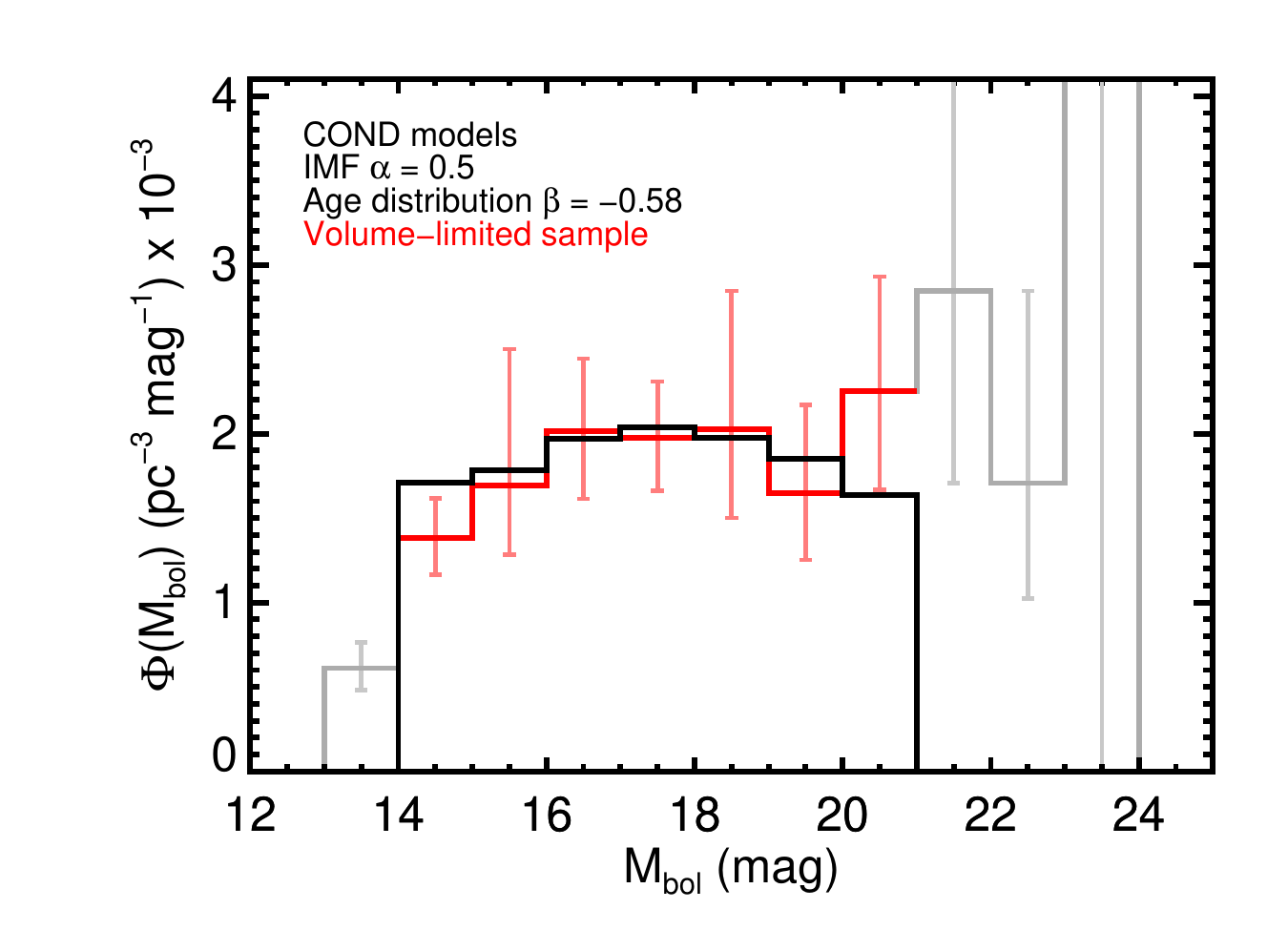}
  \caption{{\it Top}: Our 25~pc luminosity function (red/gray) overlaid with the
    best-fit luminosity function (black) from a synthetic population
    based on the SM08 hybrid models
    with IMF \hbox{$dN/dM \propto M^{\varalphamed}$}
    and age distribution $b(t) \propto e^{\varbetamed t}$.
    The bins with $\mbol<14$~mag and $\mbol>21$~mag (gray) were not used
    in our analysis. 
    {\it Bottom}: Same as top, but showing the best-fit luminosity
    function from a synthetic population
    based on the COND models,
    with IMF \hbox{$dN/dM \propto M^{\varalphamedcond}$}
    and age distribution $b(t) \propto e^{\varbetamedcond t}$.
    The COND-based synthetic luminosity functions match that of
    our volume-limited sample better overall, in particular across the L/T
    transition ($\mbol\approx15$--17~mag).}
  \label{best.synth.lum}
\end{figure}

Figure~\ref{best.synth.lum} shows the synthetic luminosity function based on the
SM08 and COND models for our respective best-fit $\alpha$ and $\beta$.  The SM08
luminosity function is reasonably consistent with that of our volume-limited
sample, but we note one apparent difference.  The models predict a peak in the
luminosity function at $\mbol=$~16--17~mag and a deficit at 17--18~mag.  SM08
ascribe this feature to a release of entropy as brown dwarfs lose the clouds
from their photospheres in the transition from L to T dwarfs, causing the
objects to temporarily slow their rate of cooling at $\mbol\approx$~16--17~mag
followed by a more rapid decline.  This scenario is supported by binaries with
dynamically-determined masses, whose mass-luminosity relation is better matched
to the SM08 ``hybrid'' models than to cloudless models
\citep{Dupuy:2015gl,Dupuy:2017ke}.  Our volume-limited sample's luminosity
function is essentially flat across $\mbol=$~15--21~mag and does not appear to
reflect this predicted slowdown in cooling, although the difference is not
strongly significant; our {\mbol} bins differ from the SM08 luminosity function
by $1.5\sigma$ and $1.4\sigma$ in the $\mbol=$~16--17~mag and 17--18~mag bins,
respectively.
One possible explanation for the difference is that the transition from cloudy
to cloudless objects occurs at lower temperatures in younger, lower-gravity
brown dwarfs;
this could wash out the luminosity peak seen in the SM08 models
\citep[e.g.,][]{Metchev:2006bq,Liu:2016co}, in which the transition occurs over
the same {\teff} range for all objects.

The COND-based synthetic luminosity function (Figure~\ref{best.synth.lum}) appears
more closely matched to our volume-limited sample, especially over the
$\mbol=$~15--20~mag range,
and does not show a peak/deficit or other feature that indicates an impact from
the L/T transition cloud clearing on bolometric luminosity.
This is expected from the cloudless COND models, but is more surprising in our
volume-limited sample that contains many L/T transition objects
which are expected to be in various stages of cloud-clearing.
Future evolutionary models that provide a better accounting for the effects of
clouds may shed more light on the impact of
L/T transition cloud-clearing on the luminosity function.
Alternatives to a single power-law mass function and/or an exponential age distribution
may also better represent the local ultracool dwarf population.

For subsequent discussion, we adopt the values of $\alpha$ and $\beta$ we
obtained using the SM08 models
because those models better represent the properties of our L0--Y2 sample due
to the inclusion of clouds for L~dwarfs.

\subsection{Ages of Young L0--L7 Dwarfs}
\label{results.youngl}
In our analysis, the choice of evolutionary models does not have a major impact
on our results because the fraction of young L0--L7~dwarfs 
provides a powerful and complementary constraint for the mass and age
distributions.
However, the usefulness of the young L0--L7~dwarf fraction is determined by
our ability to identify such objects in our volume-limited sample,
estimate their ages,
and understand the completeness of our young single L0--L7~dwarf sample.
The objects in our sample have mostly been vetted for low-gravity
spectroscopic signatures in the literature:
Of the 127 single L0--L7 dwarfs (optical or NIR spectral types) in our
volume-limited sample, 109 (86\%) have an optical or NIR spectroscopic surface
gravity classification.
None of the remaining 18 have other indications of youth, e.g., kinematic
association with a young moving group, so our young single L0--L7~dwarf sample
is likely to be complete.
We assume these 18 objects are not young, so our {\varyoungsinglfrac} young
single L0--L7~dwarf fraction (Section~\ref{popsynth.construct.youngl}) is
technically a lower limit.
The identification of more young single L0--L7~dwarfs in our sample would lead
to our population analysis finding a more positive $\alpha$ and more negative
$\beta$, further reinforcing the major trends of our results.

However, the ages corresponding to such low-gravity signatures are not well
established\footnote{We note also that two objects we identify as young in our
  volume-limited sample, 2MASS~J00332386$-$1521309 and 2MASS J10224821+5825453,
  have optical gravity class $\beta$ indicating moderately low gravity
  \citep{Cruz:2009gs} but kinematics indicating possible membership in the
  Galactic thick disk \citep{Gonzales:2019ia}, which suggests a much older
  age.}.  The $\lesssim$200~Myr age limit for objects with low-gravity
signatures quoted by \citet{Allers:2013hk} and \citet{Liu:2016co} derives from
the ages of young moving groups, which have significant uncertainties at the
older end where low-gravity signatures are weak and fewer such groups have been
identified \citep[e.g.,][]{Gagne:2018jj}. We explored the uncertainty in this
200~Myr limit by considering the ages of moving groups whose members have weaker
low-gravity features.  For example, the majority of members of the AB Doradus
Moving Group \citep[ABDMG;][]{Zuckerman:2004ds} with spectroscopic gravity
classifications have intermediate {\intg} spectra \citep[e.g.,][]{Liu:2016co},
and the $149^{+51}_{-19}$~Myr age of ABDMG \citep{Bell:2015gw} has 3$\sigma$
limits spanning $\approx$100--300~Myr.  The Carina-Near Moving Group
\citep[CARN;][]{Zuckerman:2006ke}, age $200\pm50$~Myr, has L0--L7~dwarfs with
both {\intg} and {\fldg} (field-age gravity) but no {\vlg} (very low-gravity)
classifications (UltracoolSheet, version 2.0.0, in preparation), and its
2$\sigma$ age range spans 100--300~Myr.

We therefore re-ran our analysis using age limits spanning 100--300~Myr for the
young L0--L7~dwarfs.
This yielded parameters in the ranges
$0.36\le\alpha\le0.88$ for the mass function and
$-0.98\le\beta\le-0.26$ for the age distribution,
with younger age limits corresponding to higher values
of $\alpha$ and lower values of $\beta$,
i.e., lower-mass and younger populations.
These ranges are somewhat broader than the 68\% confidence limits for the 200~Myr
young L0--L7 age limit,
but they extend the plausible intervals almost exclusively toward
higher values of $\alpha$ and lower values of $\beta$. This firmly reinforces
that $\alpha$ is positive and $\beta$ is negative,
pointing to a young and low-mass brown dwarf population in the
Solar neighborhood.

\subsection{The Apparent Peak of the Mass Function}
Many studies of the stellar IMF have found that it reaches a single maximum at
$\approx$0.2--0.3~{\msun} \citep[and references therein]{Bastian:2010ig}.
Our finding of a positive value for $\alpha$, indicating an increase in numbers
toward lower masses over the 0.002--0.085~{\msun} range of our analysis, may
appear to be inconsistent with the oft-cited stellar IMF peak.
We clarify that this peak is seen using log-normal forms of the mass function,
or broken power law forms that parameterize the distribution of log(mass).
Our analysis, and all analyses of the substellar mass function that use the form in
Equation~\ref{eqn.imf}, parameterize the mass distribution in linear space, in
which masses above the substellar regime are far more spread out.
A mass function with $\alpha<1$ would appear in log(mass) space to decline
toward lower masses, even though it rises in linear mass space.
Our result is therefore not inconsistent with the log-normal IMF peak at low
stellar masses.
In the linear-mass power-law parameterization of Equation~\ref{eqn.imf}, the
peak of the mass function may in fact be at its extreme low-mass cutoff, which
we discuss in Section~\ref{compare.kirk.cutoff}.

\section{Comparison With Previous Work}
\label{compare}

\subsection{Nearby Ultracool Dwarfs}
\label{compare.local}
Previous modeling of the local ultracool population has produced relatively
loose constraints on the IMF (Table~\ref{tbl.previous.alpha}).
Our $\alpha=\varalphamedconf$ is consistent with the estimates from
the pioneering works of \citet[$\alpha=0.3\pm0.7$]{Kroupa:2001ki}
and \citet[$\alpha=0.3\pm0.6$]{Allen:2005jf}, although neither of these
works constrain the age distribution.
Our $\alpha$ is notably higher (indicating an IMF with more lower-mass objects)
than the $\alpha\approx0$ estimate of \citet{Metchev:2008gx}, 
based on T~dwarf space density measurements that are consistent with ours
but derived from a less complete sample, in particular for later-T~dwarfs.
Our $\alpha$ disagrees even more strongly with several studies that estimated
$-1<\alpha<0$ by visually comparing modeled space densities or luminosity
functions to the space densities of late-M to late-T dwarfs or subsets thereof
\citep{Pinfield:2008jx,Reyle:2010gq,Kirkpatrick:2012ha,DayJones:2013hm,Burningham:2013gt}.
These studies relied on smaller and usually magnitude-limited samples which
produced larger uncertainties on their space densities and luminosity functions
than we have obtained with our 25~pc parallax-based sample, and did not
fit for the age distribution of sample members.
All of these previous results are consistent with the broad $\alpha$
distribution we found using only the luminosity function of our volume-limited
sample (Figure~\ref{fig.alpha.beta.prob.sm08}); the inclusion of the young
L0--L7~dwarf fraction in our analysis is key to constraining the mass
distribution to positive values of $\alpha$ (Figures
\ref{fig.product.contours.sm08} and~\ref{fig.product.contours.cond}).
We note that if we impose a flat age distribution (i.e., $\beta=0$) on our
analysis, as do two of the above studies, we find a maximum likelihood of
$\alpha=\varalphamedflat^{+\varalphaupperflat}_{\varalphalowerflat}$, which is
consistent with all of the above results for $\alpha$.
However, the flat age distribution is clearly disfavored by our analysis
(Figure~\ref{fig.pdf.imf.sfh}).

\begin{figure}
  \centering
  \includegraphics[width=1\columnwidth, trim = 20mm 0 10mm 0]{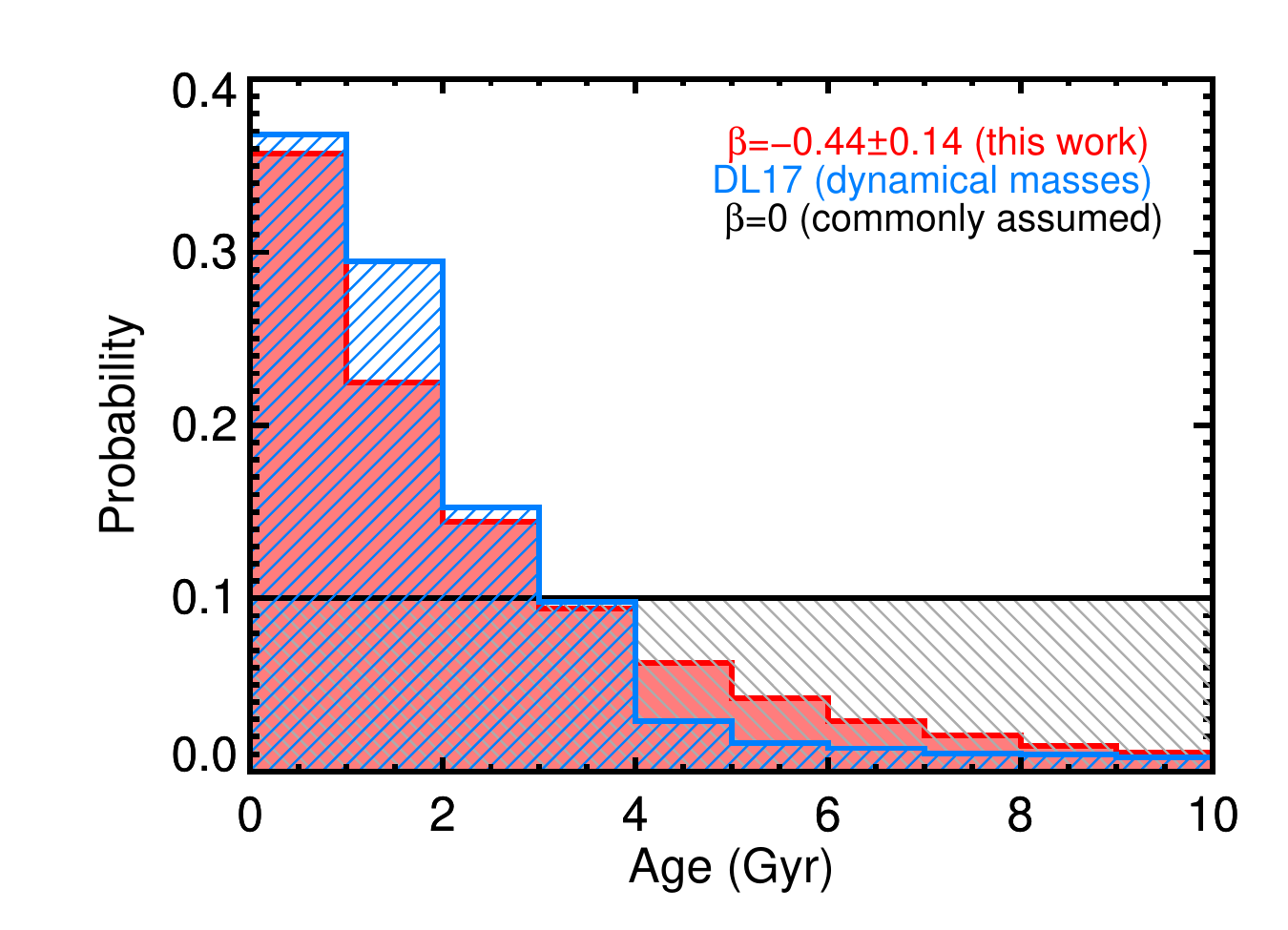}
  \caption{Distribution of ages for our best-fit exponential age distribution
    $\beta=\varbetamedconf$ (red), marginalized over the uncertainty on
    $\beta$. Our age distribution is clearly inconsistent with the uniform age
    distribution (black) assumed by many previous works, including the recent analysis
    by K21. Our result is more consistent with the age distribution found by
    \citet{Dupuy:2017ke} using dynamical masses of late-M to mid-T binaries.}
  \label{beta.compare}
\end{figure}

Previous efforts have produced very few constraints on the age distribution of
ultracool dwarfs (Table~\ref{tbl.previous.alpha}),
with some of those works simply assuming that brown dwarfs have been forming 
at a constant rate since the birth of the Galaxy $\sim$10~Gyr ago.
\citet{Allen:2005jf} demonstrated that evolutionary models predict systematic
differences in the age distributions of ultracool spectral types, with
late-L~dwarfs being the youngest group overall, but found that their observed
luminosity function was unable to sufficiently distinguish between increasing,
constant, and decreasing birthrates for nearby ultracool dwarfs.
\citet{DayJones:2013hm} and \citet{Marocco:2015iz} were able to exclude
age distributions strongly favoring old objects using L0--T8 magnitude-limited
samples, but did not directly fit for $\beta$ or a
similar parameter.
The only previous study that calculated an age distribution,
using the same exponential model that we use,
found $\beta=-0.13\pm0.17$ for masses between 0.072 and 0.1~{\msun}
\citep[early-L and late-M~dwarfs;][]{Deacon:2006ga}, based on a sample of
55~objects.
Our value of $\beta=\varbetamedconf$, based on our much larger volume-limited
sample and --- crucially --- including constraints from the fraction of young
L0--L7~dwarfs, implies a clearly younger population of L and T dwarfs.
Figure~\ref{beta.compare} demonstrates the inconsistency of our result
with a uniform age distribution.

\subsubsection{Galactic Dynamics}
\label{compare.local.dynam}
We note that our age distribution is qualitatively consistent with the age
distribution found by \citet{Dupuy:2017ke} based on dynamical masses of late-M
to mid-T binaries (Figure~\ref{beta.compare}).  That distribution also skews
toward younger objects: They found a median age of 1.3~Gyr, with a 90\%
confidence interval of 0.4--4.2~Gyr.  They highlight the fact that galactic
dynamics excite objects out of the Galactic midplane over time
\citep[e.g.,][]{Robin:2003jk,Mackereth:2019bl} and thereby deplete the solar
neighborhood of older objects.  The skew towards young objects in our
volume-limited ultracool sample may therefore be as much a reflection of this
dynamical heating as it is the underlying birth history of the L and T dwarfs in
the sample.

We note also that as dynamical heating removes objects from the Galactic
midplane, the remaining mass distribution may be impacted as well, a scenario
that has not been addressed to date.
Further modeling of ultracool dwarfs in the Galactic thin disk, thick disk and
halo populations \citep[e.g.,][]{Aganze:2022gf} is needed to establish whether
dynamical heating has altered the ultracool mass distribution of the solar
neighborhood.

\floattable
\begin{deluxetable}{lccCc}
\tablecaption{Previous Estimates of Mass Function Exponent $\alpha$ and Age Distribution
  Exponent $\beta$ \label{tbl.previous.alpha}}
\tabletypesize{\small}
\tablehead{
\colhead{Source} &
\colhead{Region} &
\colhead{Range} &
\colhead{Mass Function} & \colhead{Age Distribution}
}
\startdata
\citet{AlvesdeOliveira:2013dx} & $\rho$ Ophiuchus & Mass: 5--80 {\mjup} & \alpha = 0.7 \pm 0.3 & $\approx$1~Myr \\
\citet{DaRio:2012gs} & Orion Nebula Cluster & Mass: 0.03--0.29 {\msun} & \alpha = 0.6 \pm 0.33 & $\approx$1--3~Myr \\
\citet{Gennaro:2020kz} & Orion Nebula Cluster & Mass: 0.005--0.16 {\msun} & \alpha = 0.58 \pm 0.06 & $\approx$1--3~Myr \\
\citet{AlvesdeOliveira:2013dx} & IC 348 & Mass: 13--80 {\mjup} & \alpha = 1.0 \pm 0.3 & $\approx$3~Myr \\
\citet{PenaRamirez:2012iw} & $\sigma$ Orionis & Mass: 0.006--0.35 {\msun} & \alpha = 0.6 \pm 0.2 & $\approx$3~Myr \\
\citet{Damian:2023wx} & $\sigma$ Orionis & Mass: 0.004--0.19 {\msun} & \alpha = 0.18 \pm 0.19 & $\approx$2--4~Myr \\
\citet{Lodieu:2007if} & Upper Scorpius & Mass: 0.01--0.3 {\msun} & \alpha = 0.6 \pm 0.1 & $\approx$5~Myr \\
\citet{Casewell:2007ij} & Pleiades & Mass: $\approx$0.015--0.065 {\msun} & \alpha = 0.62 \pm 0.14 & $\approx$120~Myr \\
\hline
\citet{Kroupa:2001ki} & Solar neighborhood & Mass: 0.01--0.08 {\msun} & \alpha = 0.3 \pm 0.7 & no constraint \\
\citet{Allen:2005jf} & Solar neighborhood & Mass: 0.04--0.1 {\msun} & \alpha = 0.3\pm 0.6 & no constraint \\
\citet{Deacon:2006ga} & Solar neighborhood & Mass: 0.072--0.1 {\msun} & \alpha = 0.95\pm 1.17 & $\beta=-0.13\pm0.17$ \\
\citet{Pinfield:2008jx} & Solar neighborhood & SpT: T4--T8.5 & -1 < \alpha < -0.5 & no constraint \\
\citet{Metchev:2008gx} & Solar neighborhood & SpT: T0--T8 & \alpha \approx 0 & no constraint \\
\citet{Reyle:2010gq} & Solar neighborhood & SpT: L5--T8 & \alpha \lesssim 0 & no constraint \\
\citet{Kirkpatrick:2012ha} & Solar neighborhood & SpT: T6--Y1 & -0.5 < \alpha < 0 & assumed $\beta=0$ \\
\citet{DayJones:2013hm} & Solar neighborhood & SpT: L4--T4.5 & -1 < \alpha < 0 & $\beta\lesssim0.5$ \\
\citet{Burningham:2013gt} & Solar neighborhood & SpT: T6--T8.5 & -1 < \alpha < -0.5 & assumed $\beta=0$ \\
K21 & Solar neighborhood & SpT: L0--Y2 & \alpha = 0.6 \pm 0.1\tablenotemark{a} & assumed $\beta=0$ \\
\hline
This work (SM08 models)\tablenotemark{b} & Solar neighborhood & SpT: L0--Y2 & \alpha=\varalphamedconf & $\beta=\varbetamedconf$ \\
This work (COND models) & Solar neighborhood & SpT: L0--Y2 & \alpha=\varalphamedcondconf & $\beta=\varbetamedcondconf$ \\
\enddata
 
\tablecomments{This table lists representative constraints on the exponent
  $\alpha$ of a power-law IMF ($\Psi(M)\propto M^{-\alpha}$) and the exponent
  $\beta$ of an exponential age distribution ($b(t) \propto e^{-\beta t}$) from
  the literature, along with our results at the bottom.
  The results from nearby clusters listed at the top show a selection with
  representative values of $\alpha$ from studies that used a power-law IMF.
  For solar neighborhood studies, only \citet{Deacon:2006ga} and this work
  obtain a measurement for the age distribution;
  other work assumes a flat age distribution or obtains no significant constraint.
}
\tablenotetext{a}{Uncertainty is an estimate based on the similarity of {$\chi^2$}
  for $\alpha=\{0.5,0.6,0.7\}$ from a comparison of their {\teff} function to synthetic
  populations.}
\tablenotetext{b}{We adopt these constraints on the mass and age distribution
  for our discussion in Section~\ref{compare}.}
\end{deluxetable}

\subsection{\citet{Kirkpatrick:2021ik}}
\label{compare.kirk}
K21 published the most recent effort to determine the mass
function of ultracool dwarfs in the solar neighborhood.
They based their analysis on a volume-limited sample of L, T, and Y~dwarfs that
they describe as complete for ${\teff}\ge600$~K (spectral types $\lesssim$T8),
drawing upon the same set of parallaxes as our sample.
K21 arrived at $\alpha=0.6\pm0.1$, so in many ways our works appear to be
similar.
However, the K21 sample construction and analysis differ from ours in several
important ways.
We describe these differences in detail.

\begin{itemize}

\item The K21 sample covers the full sky but is volume-limited at 20~pc, making
  theirs a wider but shallower survey that encompasses a 25\% smaller volume of
  space than ours.  In contrast, our sample extends to 25~pc but only includes
  declinations \hbox{$-30^\circ\le\delta\le +60^\circ$} (68\% of the sky).  We
  chose these declination boundaries because this is the part of the sky that
  has been thoroughly searched for all spectral types $\le$T8.  In particular,
  the optical PS1 survey proved to be essential for constructing complete
  samples of L/T transition dwarfs \citep{Best:2015em,Best:2020jr}, and PS1 has
  a southern limit of $\delta=-30^\circ$.  There has been no complete survey for
  L/T~transition objects outside of the PS1 footprint because the $JHK$ bands of
  NIR surveys such as 2MASS cannot distinguish L/T transition dwarfs from
  background M dwarfs \citep[e.g.,][]{Reid:2008fz}, and the $YJHK$ UKIDSS survey
  \citep{Lawrence:2007hu} and optical SDSS survey only observed a subset of the
  PS1 footprint. If the spatial distribution of ultracool dwarfs within 25~pc of
  the Sun is significantly non-uniform (contrary to our assumption), our space
  density measurements could be impacted since our sample does not cover the
  whole sky, but no confirmed evidence of this has been found to date
  \citep[Paper~I]{Kirkpatrick:2019kt}.

\item K21 included the resolved components of known binaries as distinct objects
  in their sample, whereas we excluded binaries in order to avoid mixing the
  potentially different mass distributions and formation histories of binaries
  with those of single ultracool dwarfs (Section~\ref{demo.luminosity}). The K21
  sample therefore contains more objects (525) even though it encompasses a 25\%
  smaller volume of space than our 25~pc sample ({\varnvollim}~objects).

\item K21 demonstrated that 150~K wide {\teff} bins of their sample are
  statistically consistent with completeness for ${\teff}\ge600$~K.  However,
  the {\exvmax} statistic indicates that their sample taken as a whole is
  $\approx$90\% complete for such objects, and only $\approx$80\% complete south
  of declination $-30^{\circ}$, where fewer searches and follow-up observations
  have occurred. K21 also apply 5--13\% corrections for incompleteness in the
  Galactic plane when calculating space densities.  In contrast, we have
  calculated space densities only from volumes within our sample that are
  confirmed to be complete by the {\exvmax} statistic
  (Section~\ref{demo.density}). Our space densities are therefore based
  on smaller but more robustly complete samples.
  
\item K21 constrain the mass function underlying their volume-limited sample
  using a population synthesis approach that is similar to ours, but with a few
  key differences. Notably, K21 draw ages for their synthetic populations from a
  uniform 0--10~Gyr distribution. This has been common practice in
  ultracool population studies, but we show it is inconsistent with the space
  density of young L~dwarfs in the solar neighborhood
  (Section~\ref{popsynth.construct.youngl}; Figure~\ref{fig.youngfrac.chisq})
  and our subsequent finding that the overall age distribution clearly favors
  younger objects (Section~\ref{results}; Figure~\ref{beta.compare}).

\item Rather than using a luminosity function as we do, K21 conduct their
  analysis using the {\teff} function of their sample, with temperatures derived
  mostly from literature studies and $H$-band photometry, and compare this to
  synthetic populations to constrain the IMF.  They argue that determination of
  the bolometric luminosities for their sample should wait until more objects
  have broad spectral energy distribution measurements, in particular for the
  late-T and Y~dwarfs.  However, the {\teff} vs. $M_H$ relation from
  \citet[hereinafter F15]{Filippazzo:2015dv} used by K21 for most objects with
  spectral types $\le$T8 is itself based on {\lbol} values calculated by F15.
  The F15 {\teff} values were also derived from averaging {\teff} at ages 0.5
  and 10~Gyr for all objects except those identified as members of young moving
  groups and clusters, reinforcing K21's assumption of a flat age
  distribution\footnote{Formally, this is different from the 0--10~Gyr
    distribution used by K21 for their synthetic populations, so their analysis
    is not internally consistent, with the largest difference for objects
    younger than 0.5~Gyr.}.  In the end, both K21's and our analyses use
  polynomial relations based on F15's bolometric luminosities for most of the
  objects in our samples to constrain the population properties.  Our choice to
  use bolometric luminosities provides a less biased representation of the
  sample because they do not require an age assumption.  In addition, our
  {\mbol} uncertainties are small ($\approx$0.1--0.2~mag) relative to the 1~mag
  bin size of our luminosity function, whereas the K21 {\teff} uncertainties
  (mostly 79~K or 88~K) are usually $>$50\% of their 150~K bin size.  We also
  recognize that both studies use several alternative methods to determine
  {\lbol} or {\teff} for the small minority of sources that do not have the data
  necessary to use relations based on F15's {\lbol}, which could create
  inhomogeneities in the {\lbol} and {\teff} distributions used in population
  synthesis.  This can best be addressed by more direct measurements of {\lbol}
  and {\teff}.

\item K21 allow the space density of their synthetic populations to be a free
  parameter when fitting their synthetic {\teff} functions to their
  volume-limited sample, whereas we fix our synthetic population space density
  to the value measured from our volume-limited sample. K21's fits therefore
  have an additional degree of freedom, but their best-match synthetic
  population may have a space density that is inconsistent with their
  volume-limited sample.
\end{itemize}

Despite these differences in our sample construction and analyses, K21 derive
space densities for spectral subtype bins with similar precision to ours, and
obtain a similar constraint on the exponent of a power-law IMF:
$\alpha=0.6\pm0.1$, with their quoted uncertainty representing that the
{$\chi^2$} for their {\teff} function fit differs little for
$\alpha\in\{0.5,0.6,0.7\}$.  However, the apparent agreement of our
$\alpha=\varalphamedconf$ with K21's result is in fact misleading given the
clear disagreement in our age distributions.  Formally, the constant age
distribution (i.e., $\beta=0$) assumed by K21 is $\varbetasigmakirk\sigma$
different from our result.  If we were to also assume $\beta=0$, we would find
$\alpha=\varalphamedflat^{+\varalphaupperflat}_{\varalphalowerflat}$ to be most
likely, but this is $\varalphasigmaflat\sigma$ different from the
$\alpha=\varalphamedconf$ and $\beta=\varbetamedconf$ we find in our full
analysis (Figure~\ref{fig.product.contours.sm08}), and
$\varalphasigmaflatkirk\sigma$ different from the K21 value.  It is not clear
what difference(s) between our analyses could lead to such similar values for
$\alpha$ despite the discrepant age distributions. No one difference is an
obvious culprit, so it may be a combination of multiple factors.

K21 favor the SM08 hybrid models over the COND models, largely because they find
an excellent match between their {\teff} distribution and that of SM08, in
particular at the pile-up in {\teff} in the L/T transition which we do not see
reflected in our luminosity function.
This discrepancy would suggest that the SM08 models have correctly predicted a
{\teff} pile-up but incorrectly predicted an {\lbol} pile-up in the L/T
transition.  However, such a situation is physically contradictory for objects
in which cloud-clearing is releasing trapped entropy. If these objects
experience slower declines in {\teff}, causing the pile-up, then we should also
see slower declines in {\lbol}.

\subsubsection{The Low-Mass Cutoff}
\label{compare.kirk.cutoff}
K21 explore one feature of their sample that we do not: the low-mass cutoff of
the local brown dwarf population.
K21 model this as a sharp cutoff at three possible masses: 10, 5, and 1 {\mjup},
but are unable to obtain a constraint, finding that the best fits to their
{\teff} function do not depend strongly on the cutoff value.
They note that more accurate measurements of the space density in their lowest
two {\teff} bins (spanning 150--450~K) will enable a constraint on the low-mass
cutoff.
The IMF constraint that we have found, like that found by K21, predicts an
increase in the number of objects as mass decreases, so a cutoff is needed to
avoid generating populations dominated by extremely low-mass objects.
In our analysis, the minimum mass was implicitly set by the lowest
mass in the SM08 model grid (0.002~{\msun}).  However, since objects at the
faint end of the luminosity range we used for our population synthesis
($\mbol=21$~mag) have higher masses at ${\rm ages}\ge100$~Myr
--- as much as 0.026~{\msun} at 10~Gyr ---
our analysis is not sensitive to the low-mass cutoff.
At present, the best available constraint is that the minimum brown dwarf mass
cannot be greater than the $\approx$5~{\mjup} of the known free-floating
planetary-mass brown dwarfs in nearby star-forming regions
\citep[e.g.,][]{Luhman:2009cn,Best:2017ih,Zhang:2021jq,Damian:2023wx}.

\subsection{Star-forming Regions}
\label{compare.sfr}
Studies of star-forming regions and young clusters that estimate power-law IMFs
have mostly found $0<\alpha<1$, with constraints tending toward the higher end
of that range \citep[Table~\ref{tbl.previous.alpha}; see also,
e.g.,][]{Moraux:2003em,Scholz:2013ga,Gennaro:2020kz}.  Our
$\alpha=\varalphamedconf$ therefore generally agrees with the IMF of nearby
star-forming regions. This is consistent with a low-mass IMF that has not
changed appreciably over the history of our Galaxy, and with galactic dynamics
that have not significantly altered the ultracool mass function of the Solar
neighborhood.

\section{Summary}
\label{summary}
We have updated the volume-limited sample of L and T~dwarfs from
\citet[Paper~I]{Best:2021gm}, expanding it to include L0--Y2 dwarfs and adding
recent discoveries and parallax measurements.  The sample now contains
{\varnvollim} members, covers 68.3\% of the sky ($\delta=-30^\circ$ to
$+60^\circ$), extends to 25~pc, and is defined by parallaxes for 85\% of the
sample, the exception being objects (mostly late-T and Y~dwarfs) identified by
K21 as brown dwarfs having photometric distances within 25~pc.  Our sample is
{\varcompletefull} complete overall but is {\varcompleteearly} complete for
spectral types L0--T4, indicating near-completeness through the L/T transition
out to 25~pc.  We corrected for incompleteness using the {\exvmax} statistic to
identify the maximum distance at which our sample is statistically complete.  We
included an additional correction for Lutz-Kelker bias, although we found it did
not significantly impact our results.  We calculated a space density of
{\vardenspois}~pc$^{-3}$ for our volume-limited sample of L0--Y2~dwarfs,
$\approx$20--80\% larger than many previous estimates but consistent with that
of K21.

We calculated bolometric luminosities and present a completeness-corrected
luminosity function for single objects in our volume-limited sample.
We used our luminosity function in combination with the fraction of young single
L0--L7~dwarfs in our sample and synthetic populations based on SM08 and COND
evolutionary models to simultaneously constrain the mass and age distributions
of single brown dwarfs in the solar neighborhood.
The luminosity function and young L0--L7~dwarf fraction offered complementary
constraints in our analysis, the latter being essential for obtaining a
meaningful age distribution constraint.
For a power-law mass function ($\Psi(M)\propto M^{-\alpha}$) and exponential age
distribution ($b(t) \propto e^{-\beta t}$), we find $\alpha=\varalphamedconf$ and
$\beta=\varbetamedconf$ using the SM08 models, and
we find $\alpha=\varalphamedcondconf$ and
$\beta=\varbetamedcondconf$ using the COND
models.

Our analysis used an age of 0--200~Myr for the young L0--L7~dwarfs, based on
spectroscopic low-gravity features and membership in young moving groups, whose
maximum ages are not precisely known.  Alternate analyses using a wide range of
100 to 300~Myr for the maximum young L0--L7~dwarf age resulted in parameters
spanning $0.36\le\alpha\le0.88$ for the mass function and
$-0.98\le\beta\le-0.26$ for the age distribution, somewhat broader than our 68\%
confidence intervals for a 0--200~Myr age range but clearly confirming that
$\alpha$ is positive and $\beta$ is negative.

Although the SM08 models have been shown to provide a better representation of
individual brown dwarf luminosities due to
their inclusion of clouds for L~dwarfs and subsequent cloud depletion in the
transition to T~dwarfs, we find that synthetic populations from the COND models
provide a better match to our volume-limited sample's luminosity function.
However, the consistency of the constraints on the mass and age distributions
that we find using the SM08 and COND models indicate that the choice of
evolutionary model is not significant in our analysis, and we adopt the
SM08-based constraints for $\alpha$ and $\beta$.
These represent the most precise statistical constraints on both parameters to
date, and the first calculation of the age distribution of brown dwarfs in the
solar neighborhood based on a volume-limited sample.

Our mass distribution indicates an increase in space density toward lower brown
dwarf masses, in tension with many previous estimates for the solar neighborhood
(which favored fewer low-mass objects) but consistent with recent findings in
nearby star-forming regions.  Our age distribution clearly favors younger
ultracool dwarfs rather than the commonly-assumed uniform age distribution,
which may be as much a result of galactic dynamics systematically removing
objects from the midplane over time as it is a result of the historical birth
rate.

\vspace{20 pt}

We thank the anonymous referee for a careful review and helpful comments.
W.~Best acknowledges support from grant HST-GO-15238 provided by STScI and AURA.
We acknowledge the grant provided under the John W. Cox Endowment for Advanced
Studies in Astronomy by the Department of Astronomy at The University of Texas
at Austin that supported A.~Sanghi for the duration of Summer 2020.
This research was funded in part by the Gordon and Betty Moore Foundation
through grant GBMF8550 to M.~Liu.
T.~Dupuy acknowledges support from UKRI STFC AGP grant ST/W001209/1.
This work has benefitted from The UltracoolSheet, maintained by Will Best, Trent
Dupuy, Michael Liu, Aniket Sanghi, Rob Siverd, and Zhoujian Zhang, and developed
from compilations by \citet{Dupuy:2012bp}, \citet{Dupuy:2013ks},
\citet{Liu:2016co}, \citet{Best:2018kw}, \citet{Best:2021gm},
\citet{Sanghi:2023ie}, and \citet{Schneider:2023bv}.
This research has benefitted from the SpeX Prism Library [and the SpeX Prism
Library Analysis Toolkit], maintained by Adam Burgasser at
\url{http://www.browndwarfs.org/spexprism}.
This work has made use of data from the European Space Agency (ESA) mission
{\gaia} (\url{http://www.cosmos.esa.int/gaia}), processed by the \gaia\ Data
Processing and Analysis Consortium (DPAC,
\url{http://www.cosmos.esa.int/web/gaia/dpac/consortium}). Funding for the DPAC
has been provided by national institutions, in particular the institutions
participating in the \gaia\ Multilateral Agreement.
This research has made use of NASA's Astrophysical Data System and 
the SIMBAD and Vizier databases operated at CDS, Strasbourg, France.
For the purpose of open access, the author has applied a Creative Commons
Attribution (CC BY) license to any Author Accepted Manuscript version arising
from this submission.

\appendix

\section{An Updated ``Super-magnitude'' Method for Calculating Bolometric Luminosities}
\label{appendix.supermag}

The standard method for determining the bolometric luminosity of a source is to
integrate its spectral energy distribution (SED) as a function of wavelength or
frequency.  Ideally this means integrating a broad spectrum or multi-band
photometry spanning most of the SED \citep[e.g., F15;][]{Sanghi:2023ie}.  For
cold brown dwarfs, however, this method is currently not useful as most of the
flux emerges in the mid-infrared where spectra and broad-band photometry are
difficult to obtain from the ground, especially for fainter objects, and
\textit{JWST} has not yet observed a large sample of brown dwarfs in the
mid-infrared.  Empirical relations in the literature mapping ultracool dwarf
spectral types and broadband photometry to bolometric luminosities
\citep[e.g.,][F15]{Liu:2010cw,Dupuy:2017ke} are not valid for spectral types
$\gtrsim$T8 due to the lack of such cold objects with independently-determined
bolometric luminosities.

To overcome this barrier, \citet[hereinafter DK13]{Dupuy:2013ks} added the
absolute fluxes in several broadband NIR and MIR bandpasses to obtain a
``super-magnitude'' for each object.  They used the {\jmko}, {\hmko}, and
{\spitzer}/IRAC [3.6] and [4.5] bands to define the $m_{JH12}$ super-magnitude
whenever such photometry was available in all four bands for an object, since
these bands capture $\gtrsim$50\% of the total flux for late-T and Y~dwarfs.
When those four bands were not all available, they used super-magnitude from
other combinations of bandpasses to match the available photometry.  DK13 then
calculated the super-magnitudes using the same bandpasses for model spectra
\citep{Morley:2012io} spanning appropriate ranges of temperature, surface
gravity, and cloud thickness.  They used the mean super-magnitudes and the
corresponding model bolometric luminosities to derive bolometric corrections for
the super-magnitudes, enabling bolometric magnitude determinations for T8 and
later-type dwarfs.

We have updated this method in several ways:
\begin{enumerate}
\item We now use the flux tables for the Sonora-Bobcat cloudless atmosphere
  models \citep{Marley:2021ba} to calculate model super-magnitudes using
  combinations of the {\jmko}, {\hmko}, [3.6], [4.5], and AllWISE $W1$ and $W2$
  bands. The zero-points we used to convert {$J$, $H$, [3.6], [4.5], $W1$, $W2$}
  fluxes into Vega magnitudes were, respectively, {9.31, 8.52, 2.13, 1.26, 4.94,
    1.90}$\times10^{-11}$~\wmss. We calculated these by direct integration of
  the high-resolution model Vega spectrum from SpeXtool \citep{Cushing:2004bq}
  over each bandpass. The zero-points for our super-magnitudes are the sum of
  the relevant zero-point fluxes, e.g., for $m_{JH12}$ the zero-point is
  $2.12\times10^{-10}$~\wmss.

\item Rather than limiting our super-magnitude calculations to ranges of {\teff}
  and surface gravity appropriate only for late-T and Y dwarfs, we have
  calculated super-magnitudes using all of the Sonora-Bobcat models
  ($200\le\teff\le2400$~K and $3.0\le\logg\le5.5$).

\item Using the super-magnitudes and luminosities from the Sonora-Bobcat models,
  we fit 5th-order polynomials to convert the super-magnitudes directly into
  bolometric luminosities, eliminating the intermediate calculation of
  bolometric corrections.  This change means that the polynomials require {\it
    absolute} super-magnitudes to accurately determine the bolometric
  luminosities.
  
\item As the Sonora-Bobcat models are computed for three metallicities,
  $\feh=\{-0.5,0.0,0.5\}$, we derived separate polynomials for these three
  metallicities.
\end{enumerate}

The polynomials converting super-magnitudes to bolometric luminosities are
presented in Table~\ref{tbl.supermag.poly}, along with the RMS of the residuals
from the polynomial fits in super-magnitude bins.  The residuals for the
polynomial fits are shown in Figure~\ref{fig.supermag.resid}.  The residuals for
each super-magnitude show little variation over the full range of metallicities.
The polynomials give exceptionally tight relations ($\sigma<0.02$~dex) between
$m_{JH12}$ and {\lbol} for $m_{JH12}<17$, corresponding to $\teff\gtrsim400$~K.
At lower {\teff} there is an expanding envelope of uncertainty about the
polynomial relation.  The residuals for the $m_{JHW1W2}$ polynomials (replacing
the {\spitzer}/IRAC [3.6] and [4.5] bands with the comparable WISE $W1$ and
$W2$~bands) are very similar.  For objects lacking photometry in the
near-infrared $J$ and $H$ bands (which is the case for some known Y~dwarfs), the
$m_{12}$ and $m_{W1W2}$ super-magnitude polynomials have residuals with a spread
of $\approx$0.10~mag, tightening down to $\approx$0.03~mag at
$m_{12}\approx15$~mag and $m_{W1W2}\approx16$~mag ($\teff\sim500$~K).

\floattable
\begin{splitdeluxetable}{LccccccCBLcccccccccc}
\centering
\tablecaption{ Polynomials Converting Super-magnitude to {\mbol} \label{tbl.supermag.poly} }
\tabletypesize{\scriptsize}
\tablewidth{0pt}
\tablehead{   
  \multicolumn{9}{c}{} &
  \multicolumn{10}{c}{RMS of polynomial over magnitude bins} \\
  \cline{10-19}
  \colhead{Super-magnitude} &
  \colhead{$c_0$} &
  \colhead{$c_1$} &
  \colhead{$c_2$} &
  \colhead{$c_3$} &
  \colhead{$c_4$} &
  \colhead{$c_5$} &
  \colhead{Valid $M$ range} & 
  \colhead{Super-magnitude} &
  \colhead{12--13} &
  \colhead{13--14} &
  \colhead{14--15} &
  \colhead{15--16} &
  \colhead{16--17} &
  \colhead{17--18} &
  \colhead{18--19} &
  \colhead{19--20} &
  \colhead{20--21} &
  \colhead{21--22} \\
  \colhead{} &
  \colhead{} &
  \colhead{} &
  \colhead{} &
  \colhead{} &
  \colhead{} &
  \colhead{} &
  \colhead{(mag)} &
  \colhead{} &
  \colhead{(mag)} &
  \colhead{(mag)} &
  \colhead{(mag)} &
  \colhead{(mag)} &
  \colhead{(mag)} &
  \colhead{(mag)} &
  \colhead{(mag)} &
  \colhead{(mag)} &
  \colhead{(mag)} &
  \colhead{(mag)}
}
\startdata
\multicolumn{8}{c}{$\feh=-0.5$} &
\multicolumn{11}{c}{$\feh=-0.5$} \\
\hline
M_{JH12} & $-$6.4398e+00 & $-$1.4027e$-$01 & \phm{$-$}2.3226e$-$01 & $-$3.0873e$-$02 & \phm{$-$}1.4257e$-$03 & $-$2.2589e$-$05 & 12-22 & M_{JH12} & 0.011 & 0.009 & 0.012 & 0.016 & 0.012 & 0.025 & 0.038 & 0.049 & 0.065 & 0.026 \\
M_{JHW1W2} & $-$2.7293e+01 & \phm{$-$}6.0786e+00 & $-$4.9259e$-$01 & \phm{$-$}1.0392e$-$02 & \phm{$-$}2.7620e$-$04 & $-$1.0041e$-$05 & 12-22 & M_{JHW1W2} & 0.017 & 0.019 & 0.026 & 0.033 & 0.017 & 0.015 & 0.026 & 0.040 & 0.052 & 0.003 \\
M_{12} & $-$1.5231e+02 & \phm{$-$}5.0524e+01 & $-$6.5969e+00 & \phm{$-$}4.1605e$-$01 & $-$1.2839e$-$02 & \phm{$-$}1.5571e$-$04 & 12-20 & M_{12} & 0.080 & 0.069 & 0.030 & 0.020 & 0.036 & 0.049 & 0.065 & 0.028 & \nodata & \nodata \\
M_{W1W2} & $-$1.5505e+02 & \phm{$-$}4.9420e+01 & $-$6.2087e+00 & \phm{$-$}3.7708e$-$01 & $-$1.1211e$-$02 & \phm{$-$}1.3099e$-$04 & 12-20 & M_{W1W2} & 0.086 & 0.082 & 0.060 & 0.022 & 0.018 & 0.031 & 0.042 & 0.052 & \nodata & \nodata \\
\hline
\multicolumn{8}{c}{$\feh=0.0$} &
\multicolumn{11}{c}{$\feh=0.0$} \\
\hline
M_{JH12} & $-$2.7157e$-$01 & $-$2.6723e+00 & \phm{$-$}6.1027e$-$01 & $-$5.7324e$-$02 & \phm{$-$}2.3075e$-$03 & $-$3.3885e$-$05 & 12-22 & M_{JH12} & 0.010 & 0.012 & 0.012 & 0.010 & 0.010 & 0.027 & 0.046 & 0.062 & 0.073 & 0.086 \\
M_{JHW1W2} & $-$1.0493e+01 & \phm{$-$}5.6039e$-$02 & \phm{$-$}3.3836e$-$01 & $-$4.5086e$-$02 & \phm{$-$}2.0762e$-$03 & $-$3.2808e$-$05 & 12-22 & M_{JHW1W2} & 0.012 & 0.010 & 0.013 & 0.022 & 0.014 & 0.017 & 0.040 & 0.055 & 0.078 & 0.069 \\
M_{12} & $-$2.4219e+02 & \phm{$-$}7.7727e+01 & $-$9.8550e+00 & \phm{$-$}6.0935e$-$01 & $-$1.8529e$-$02 & \phm{$-$}2.2228e$-$04 & 12-20 & M_{12} & 0.068 & 0.063 & 0.033 & 0.019 & 0.043 & 0.062 & 0.073 & 0.089 & \nodata & \nodata \\
M_{W1W2} & $-$2.8736e+02 & \phm{$-$}8.9031e+01 & $-$1.0910e+01 & \phm{$-$}6.5390e$-$01 & $-$1.9305e$-$02 & \phm{$-$}2.2513e$-$04 & 12-20 & M_{W1W2} & 0.071 & 0.075 & 0.058 & 0.026 & 0.023 & 0.046 & 0.063 & 0.081 & \nodata & \nodata \\
\hline
\multicolumn{8}{c}{$\feh=+0.5$} &
\multicolumn{11}{c}{$\feh=+0.5$} \\
\hline
M_{JH12} & $-$1.9805e+00 & $-$2.6266e+00 & \phm{$-$}6.5022e$-$01 & $-$6.1584e$-$02 & \phm{$-$}2.4614e$-$03 & $-$3.5652e$-$05 & 12-22 & M_{JH12} & 0.019 & 0.020 & 0.018 & 0.014 & 0.011 & 0.022 & 0.045 & 0.064 & 0.080 & 0.094 \\
M_{JHW1W2} & $-$2.3089e+01 & \phm{$-$}3.4414e+00 & $-$2.6539e$-$02 & $-$2.5080e$-$02 & \phm{$-$}1.5099e$-$03 & $-$2.6089e$-$05 & 12-22 & M_{JHW1W2} & 0.015 & 0.007 & 0.010 & 0.018 & 0.014 & 0.017 & 0.042 & 0.063 & 0.075 & 0.095 \\
M_{12} & $-$3.6928e+02 & \phm{$-$}1.1681e+02 & $-$1.4623e+01 & \phm{$-$}8.9825e$-$01 & $-$2.7232e$-$02 & \phm{$-$}3.2676e$-$04 & 12-20 & M_{12} & 0.059 & 0.067 & 0.042 & 0.016 & 0.040 & 0.063 & 0.080 & 0.098 & \nodata & \nodata \\
M_{W1W2} & $-$2.7307e+02 & \phm{$-$}8.3427e+01 & $-$1.0076e+01 & \phm{$-$}5.9445e$-$01 & $-$1.7261e$-$02 & \phm{$-$}1.9793e$-$04 & 12-21 & M_{W1W2} & 0.068 & 0.074 & 0.066 & 0.030 & 0.023 & 0.050 & 0.068 & 0.085 & 0.097 & \nodata \\
\enddata
\tablecomments{
  The polynomials are defined as ${\rm log}(\lbol/\lsun) = \sum_{i=0} c_i M^i$,
  where $M$ is the absolute super-magnitude of the object.  The rightmost column in the top section gives
  the valid range for the absolute super-magnitude (not bolometric magnitude).
  The super-magnitudes used to derive these polynomials were calculated from
  Sonora-Bobcat cloudless atmospheric models.
}
\end{splitdeluxetable}

\begin{figure*}
  \gridline{\fig{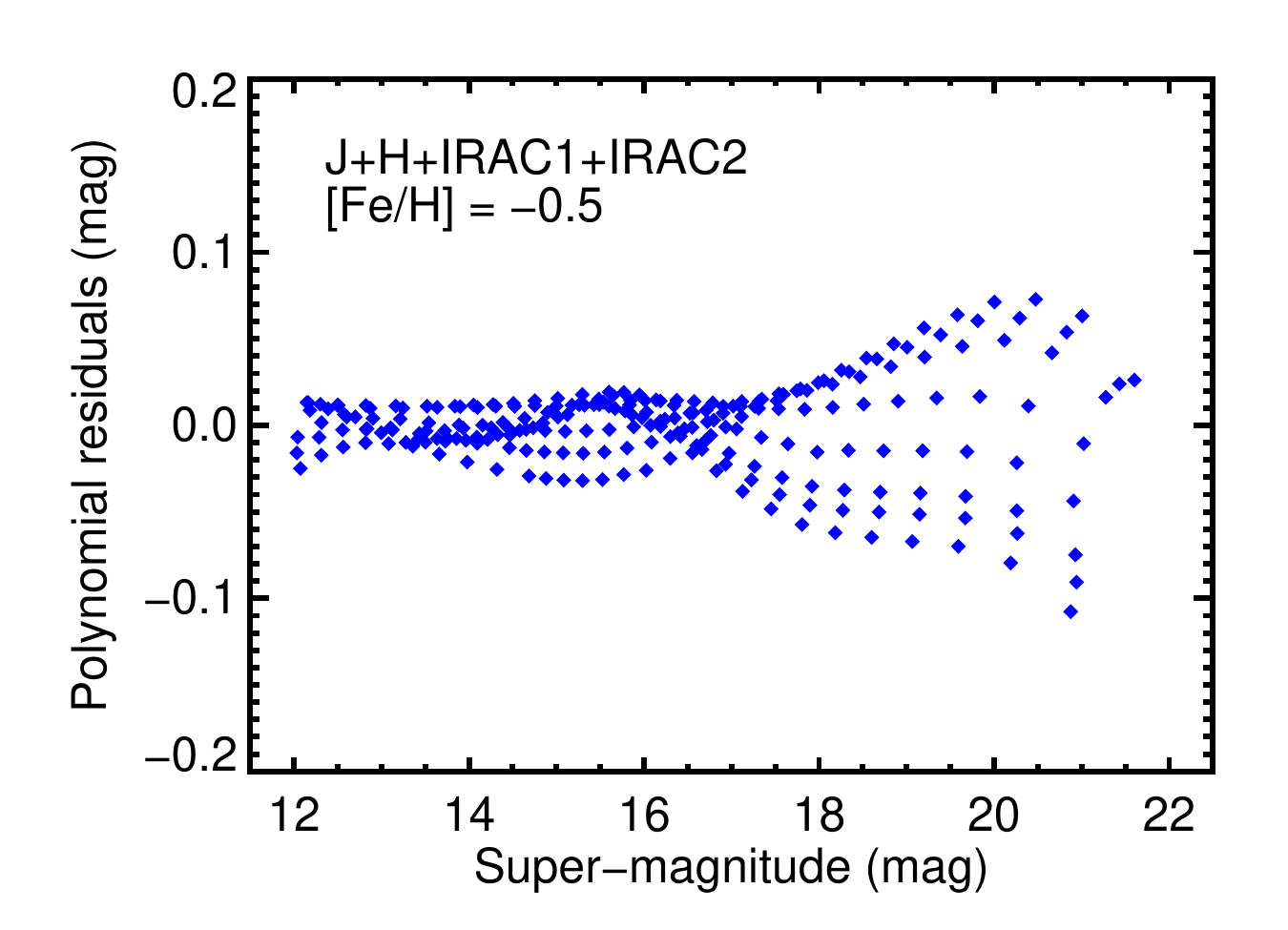}{0.5\textwidth}{}
    \fig{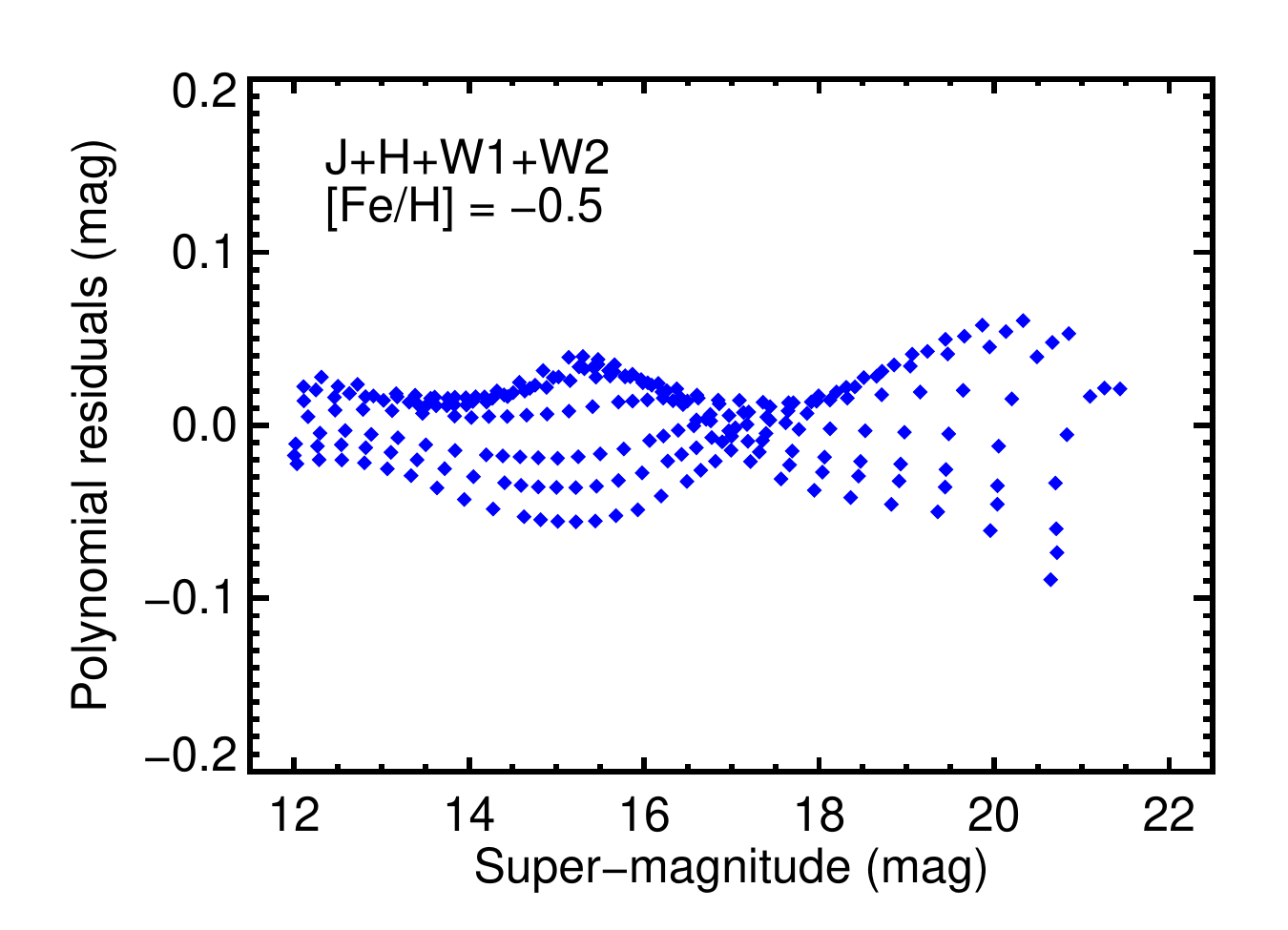}{0.5\textwidth}{}
  }
  \gridline{\fig{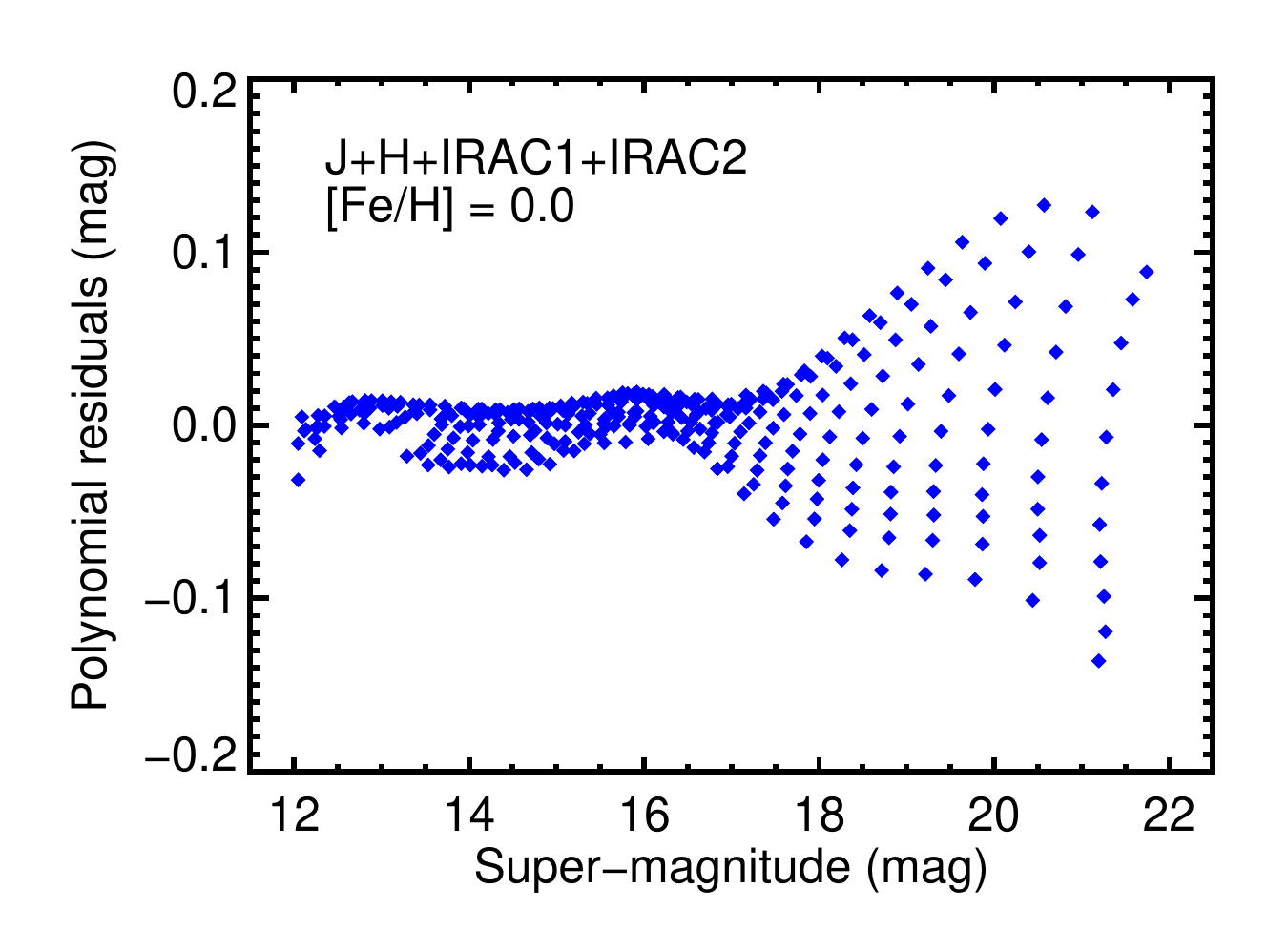}{0.5\textwidth}{}
    \fig{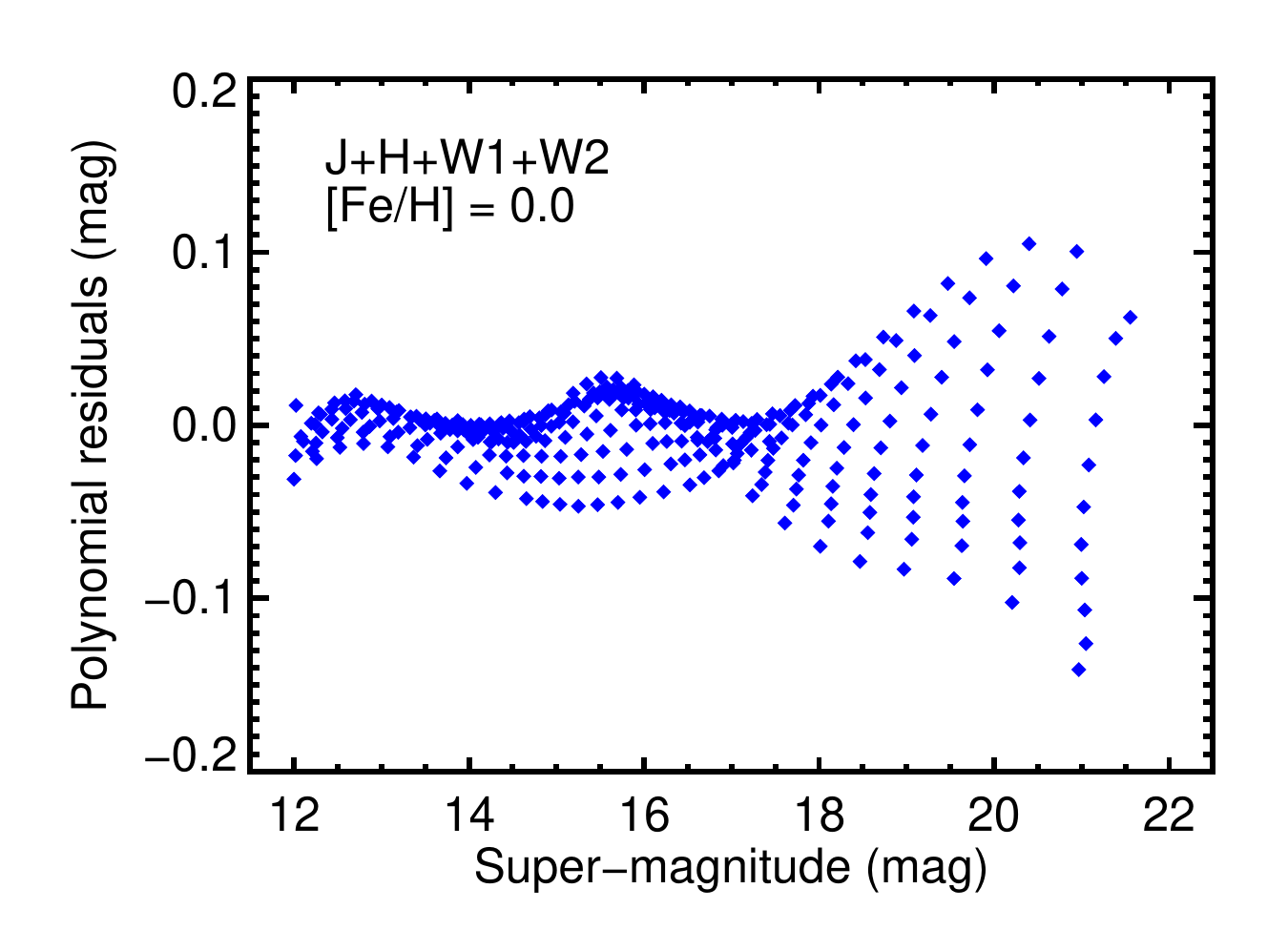}{0.5\textwidth}{}
  }
  \gridline{\fig{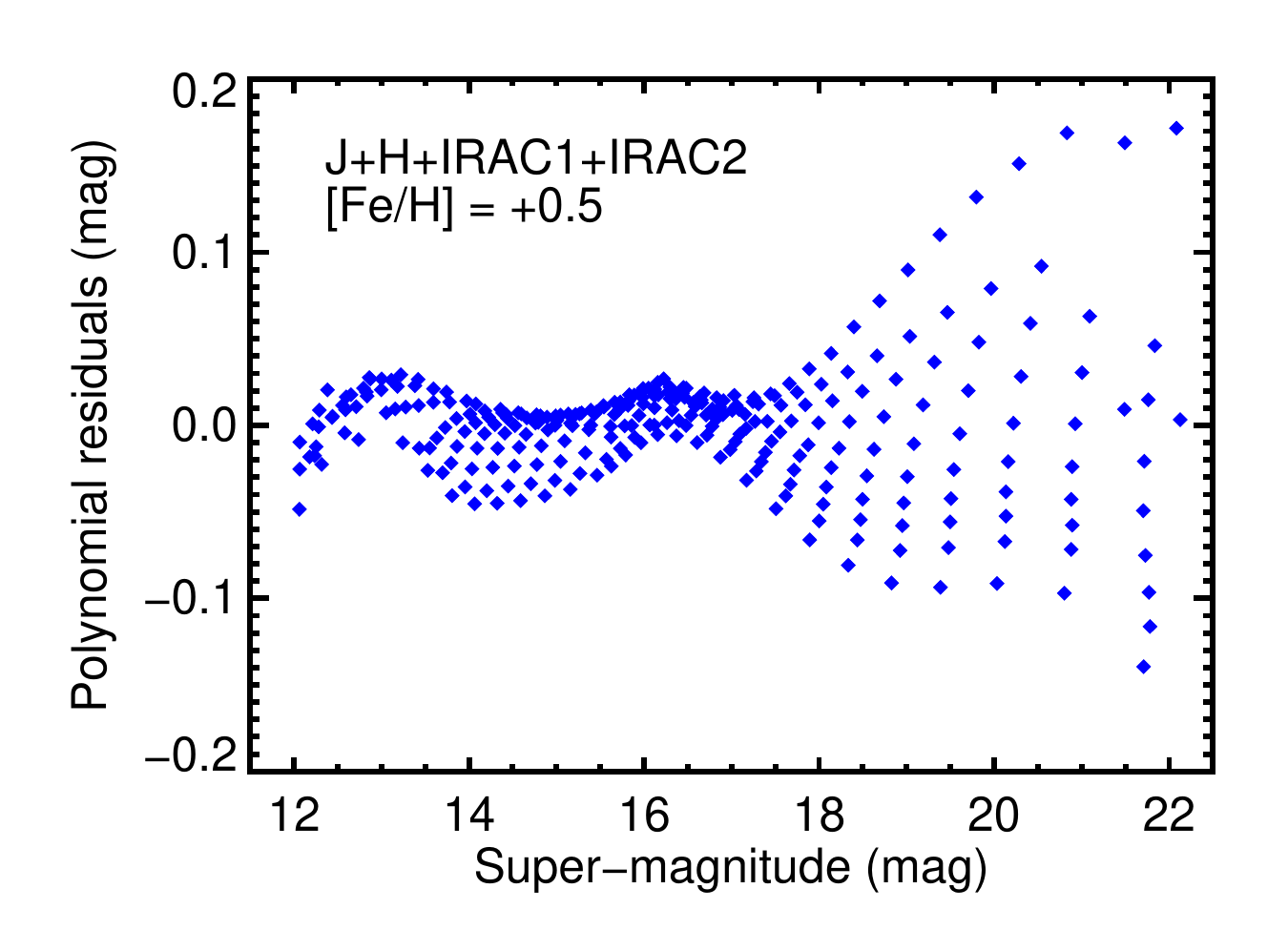}{0.5\textwidth}{}
    \fig{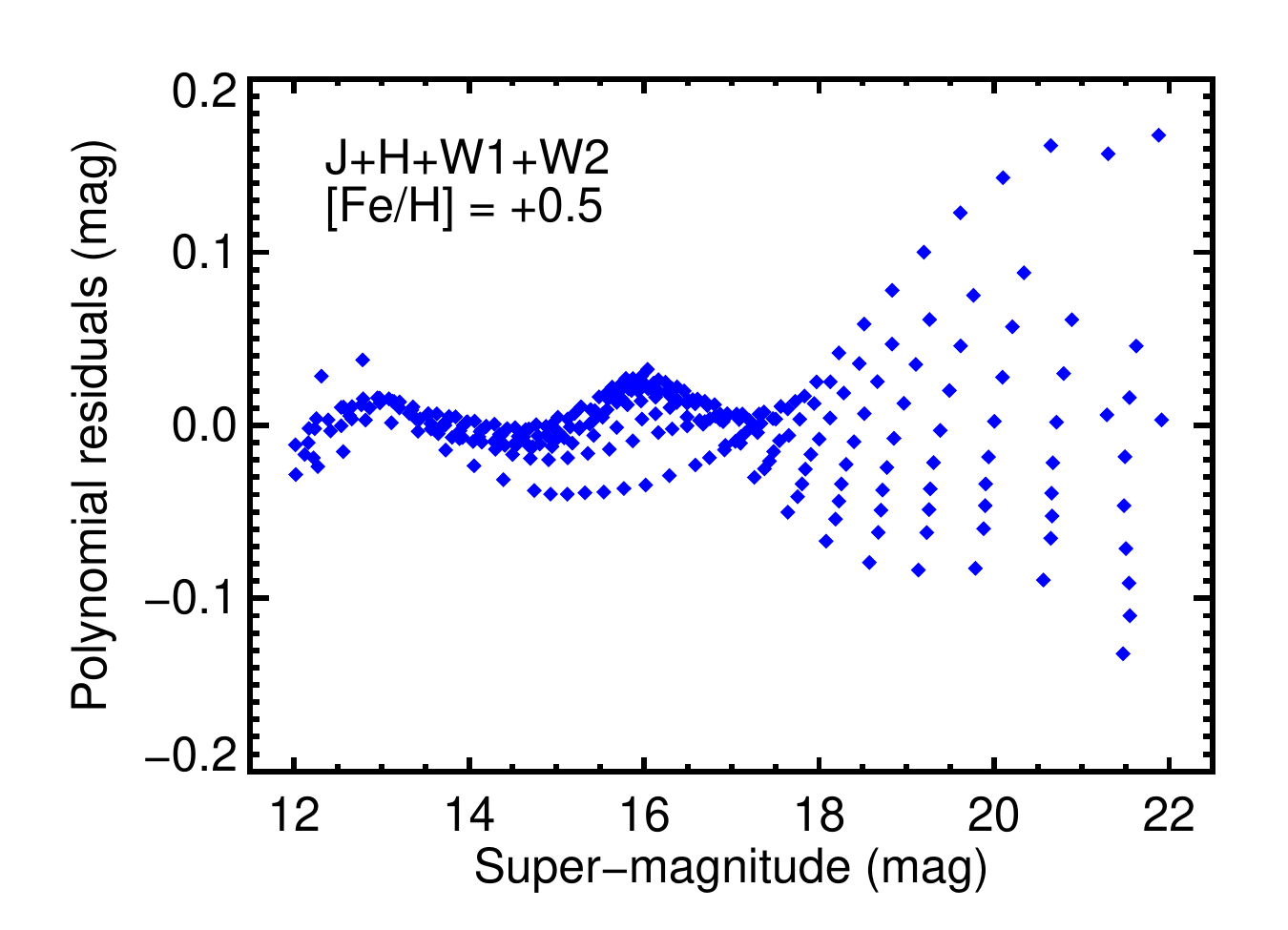}{0.5\textwidth}{}
  }
  \caption{Residuals for the 5th-order polynomial fits converting
    super-magnitudes to bolometric luminosities, using Sonora-Bobcat cloudless
    atmosphere models. }
  \figurenum{fig.supermag.resid.1}
\end{figure*}

\begin{figure*}
  \gridline{\fig{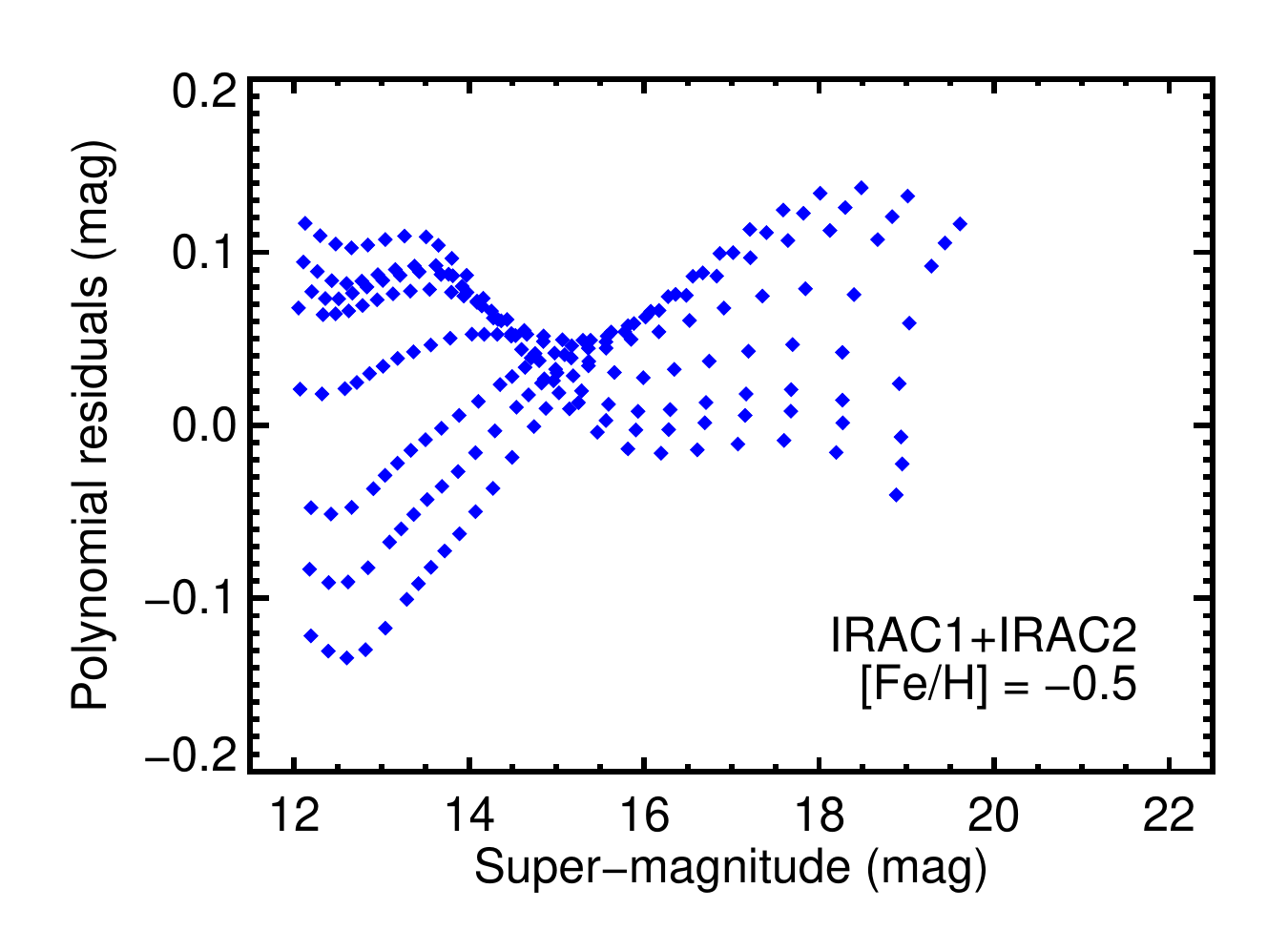}{0.5\textwidth}{}
    \fig{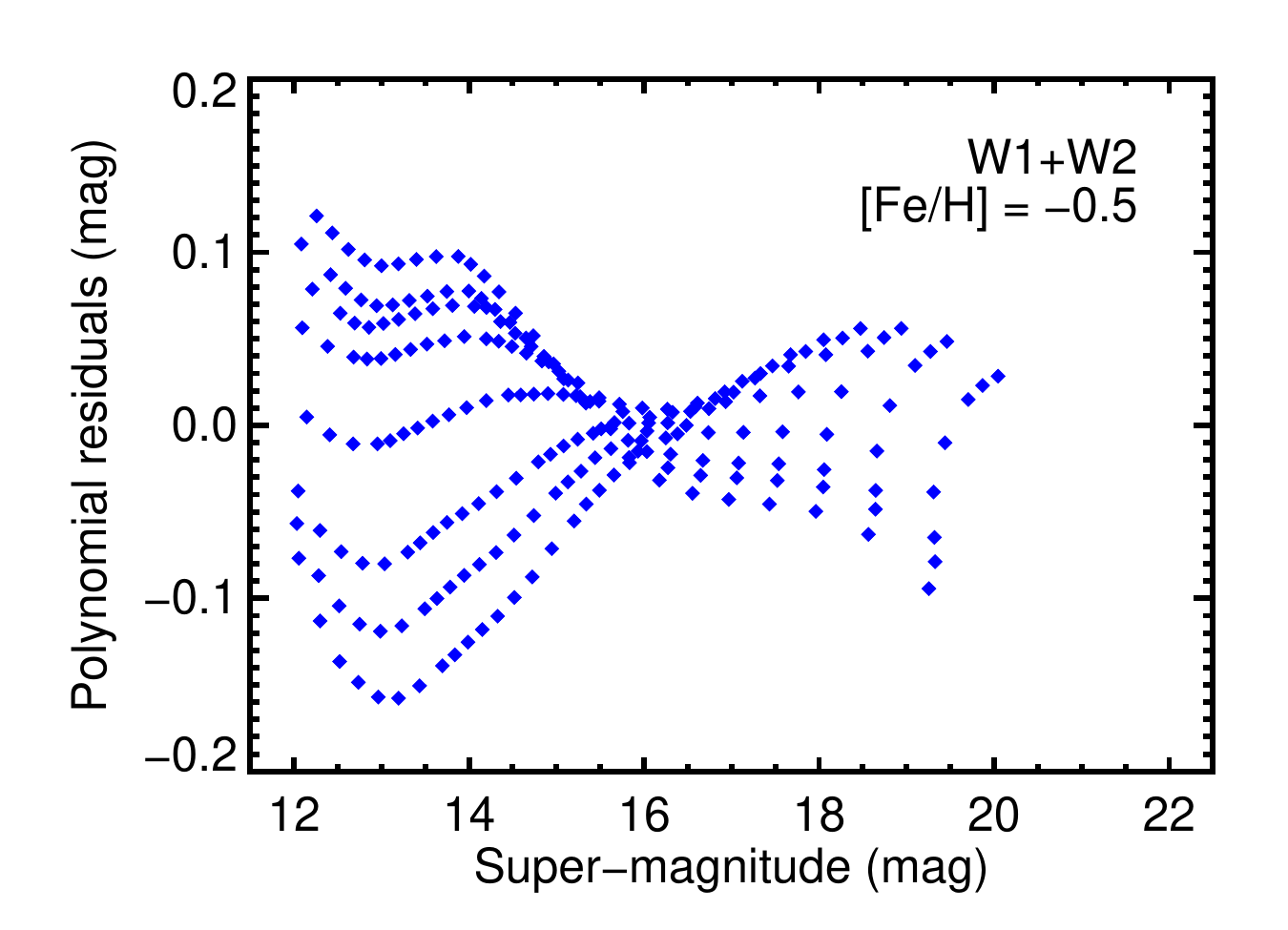}{0.5\textwidth}{}
  }
  \gridline{\fig{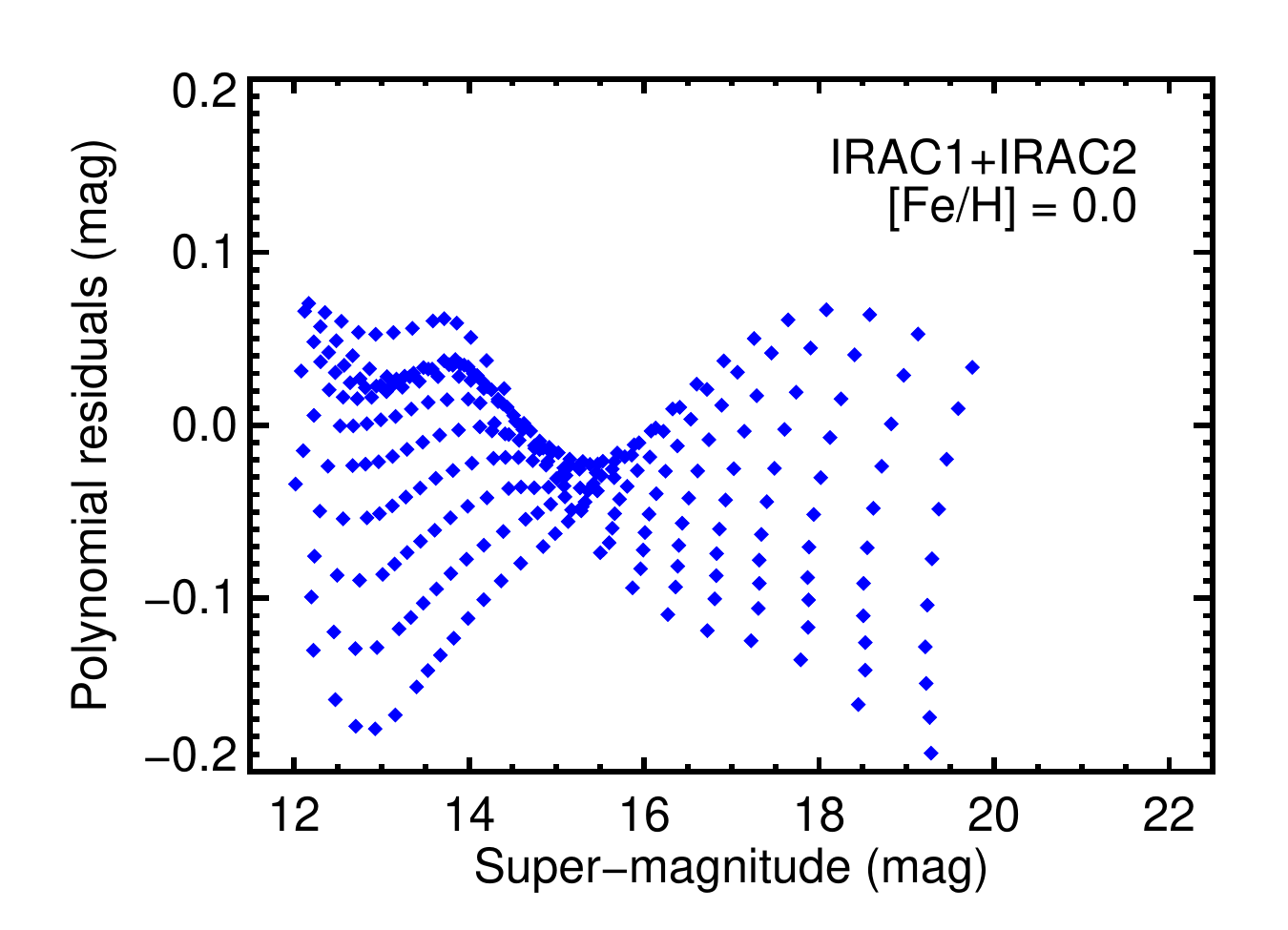}{0.5\textwidth}{}
    \fig{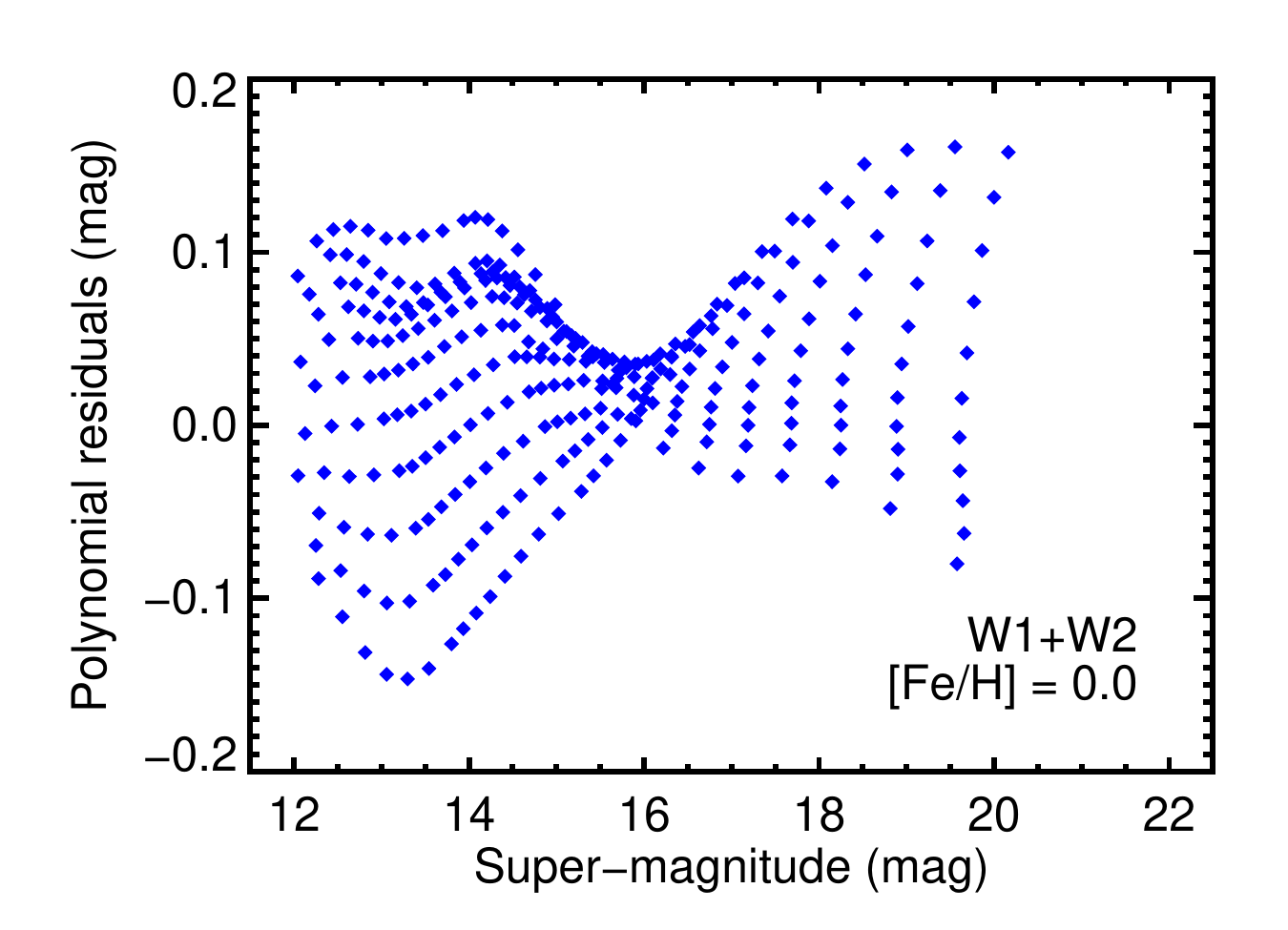}{0.5\textwidth}{}
  }
  \gridline{\fig{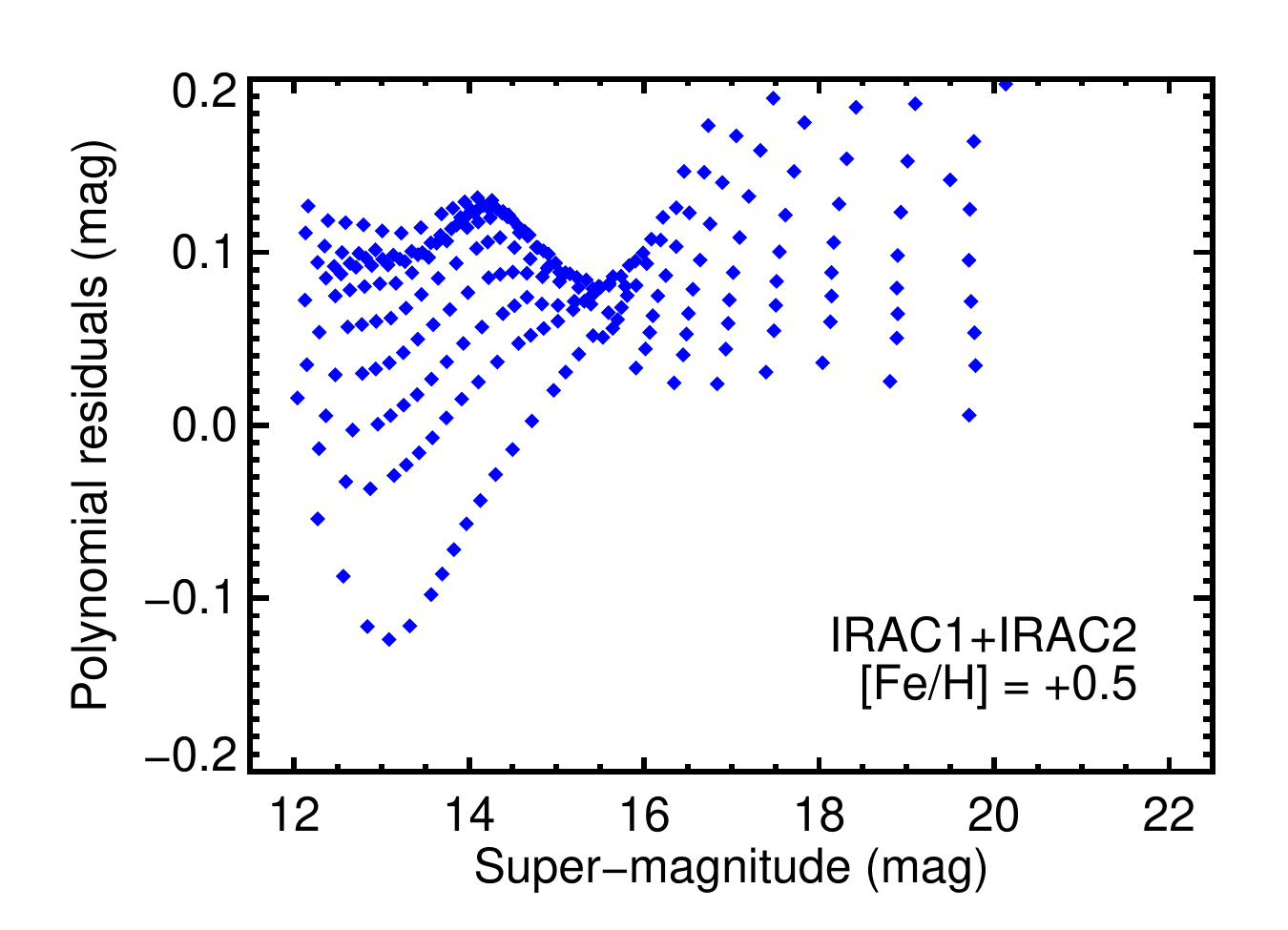}{0.5\textwidth}{}
    \fig{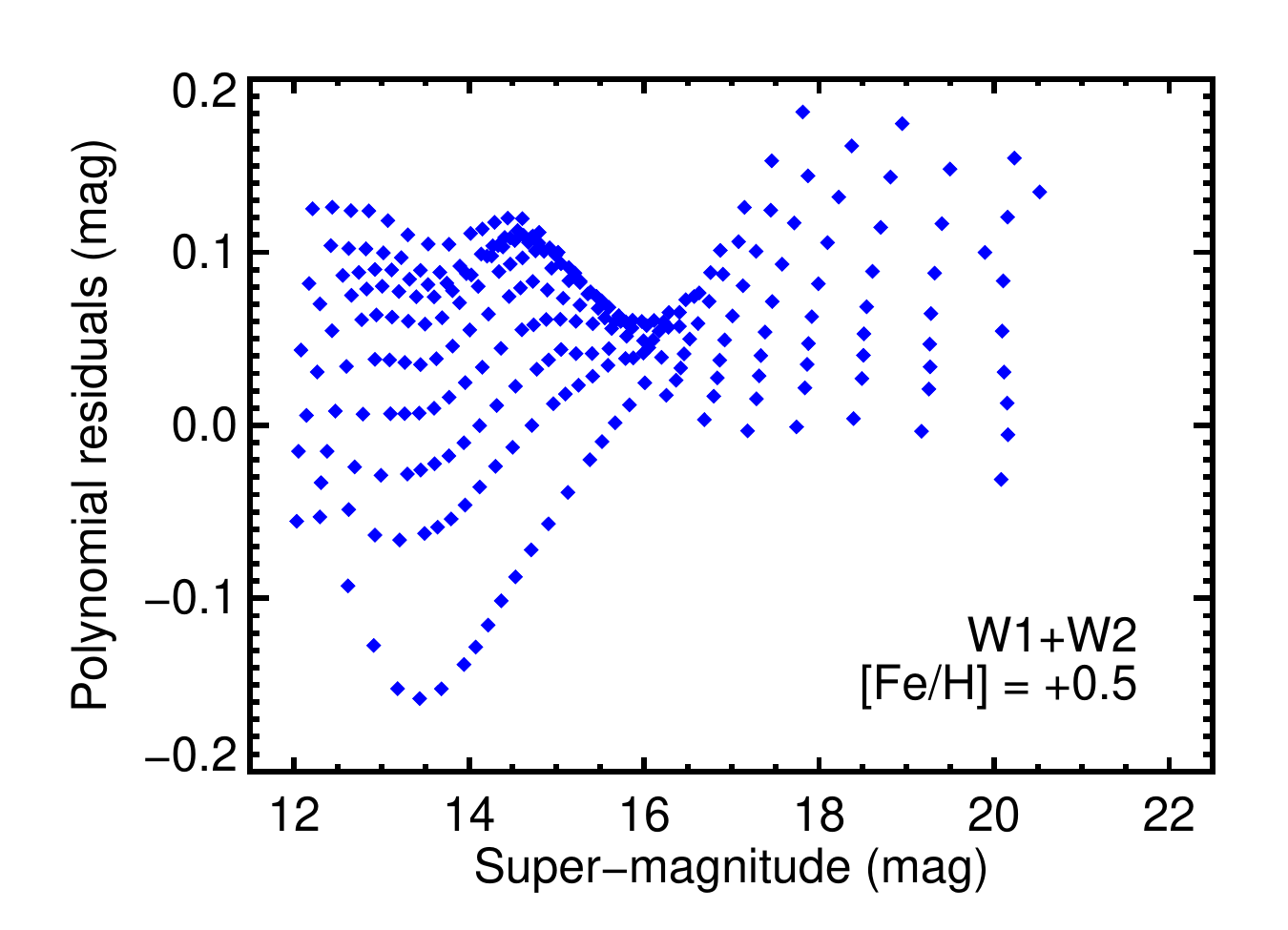}{0.5\textwidth}{}
  }
  \caption{Continued.}
  \label{fig.supermag.resid}
\end{figure*}

\begin{figure*}
  \centering
  \begin{minipage}[t]{0.48\textwidth}
    \includegraphics[width=1\columnwidth, trim = 20mm 0 10mm 0]{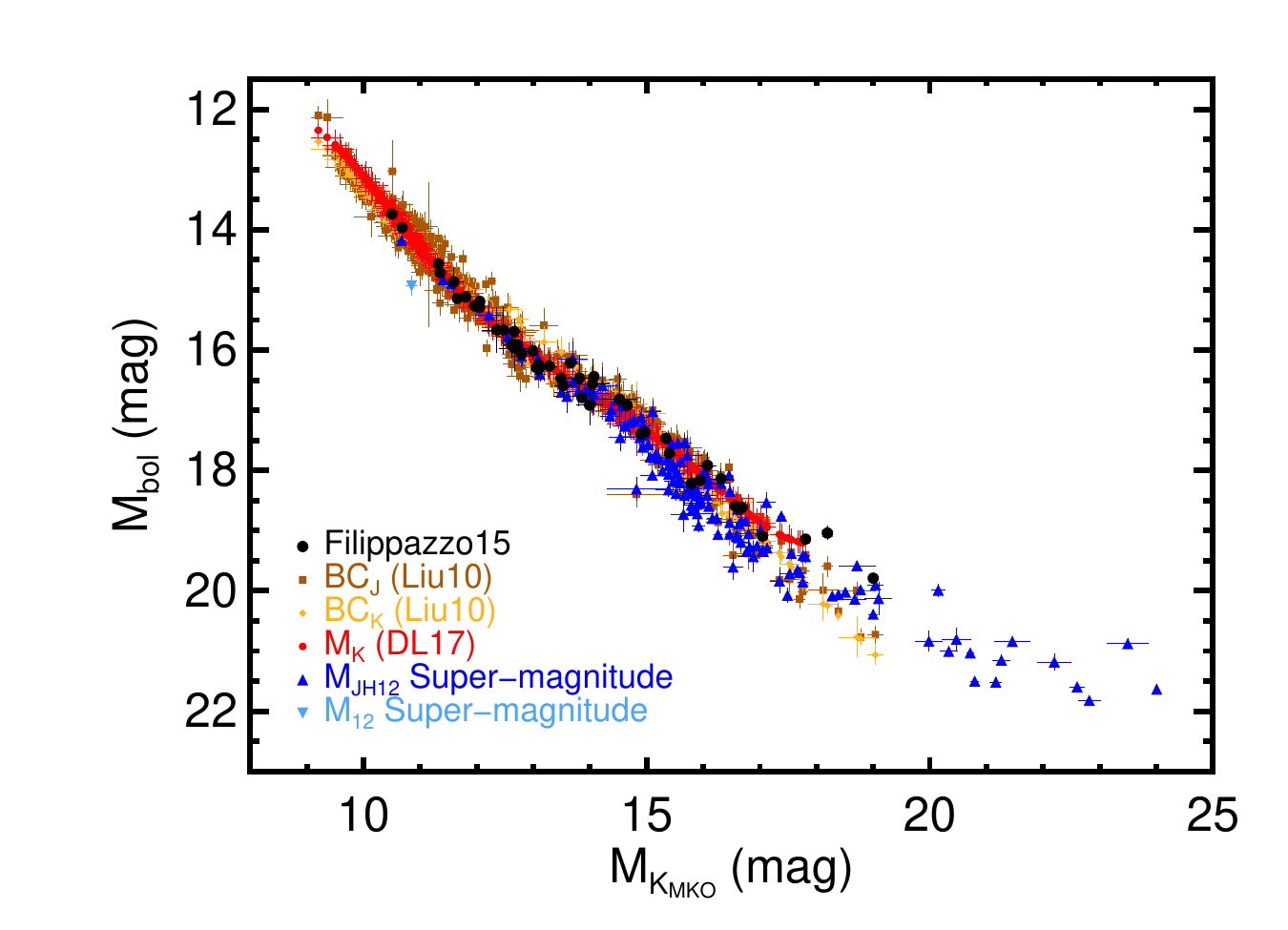}
  \end{minipage}
  \hfill
  \begin{minipage}[t]{0.48\textwidth}
    \includegraphics[width=1\columnwidth, trim = 20mm 0 10mm 0]{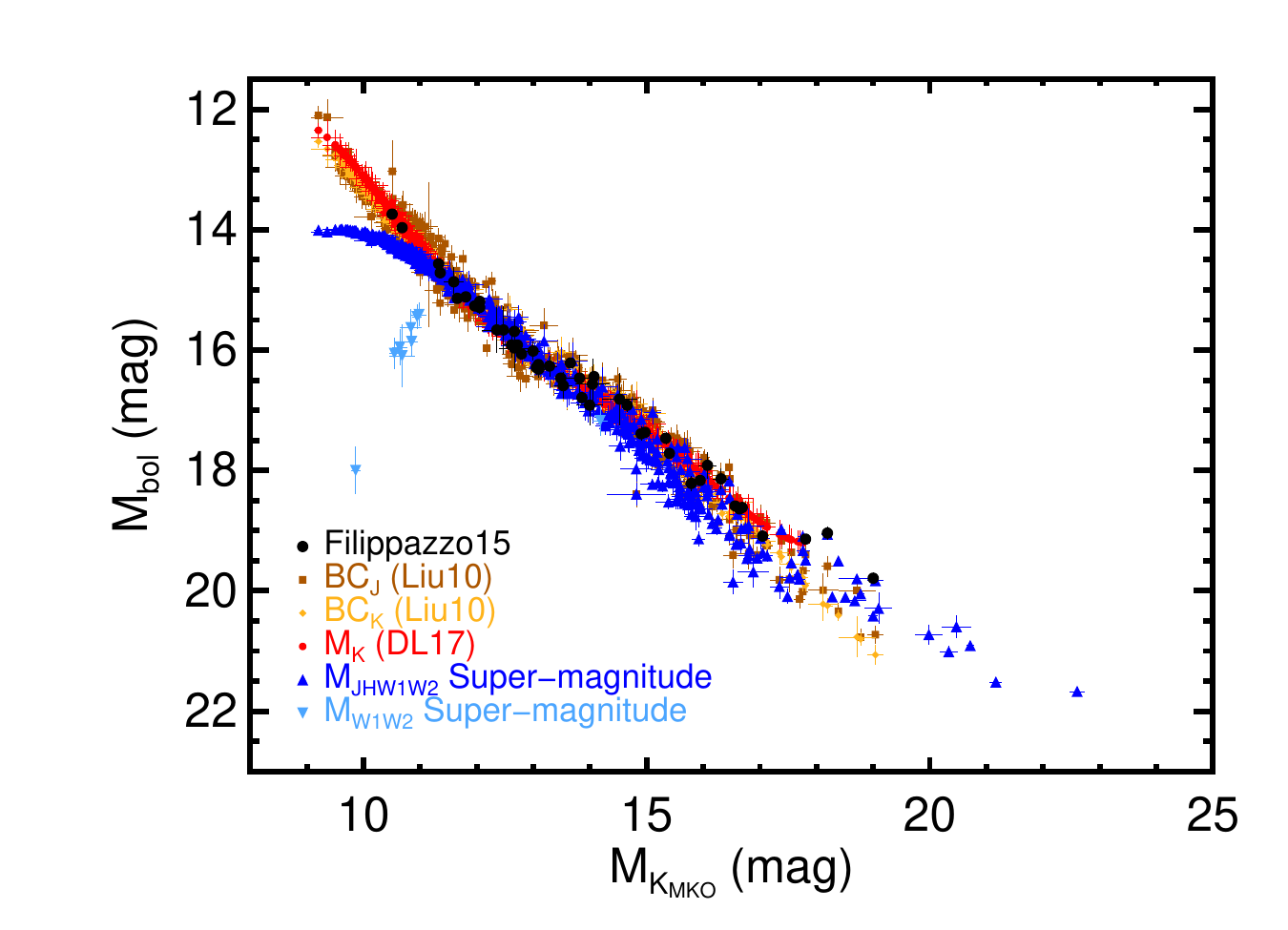}
  \end{minipage}
  \begin{minipage}[t]{0.48\textwidth}
    \includegraphics[width=1\columnwidth, trim = 20mm 0 10mm 0]{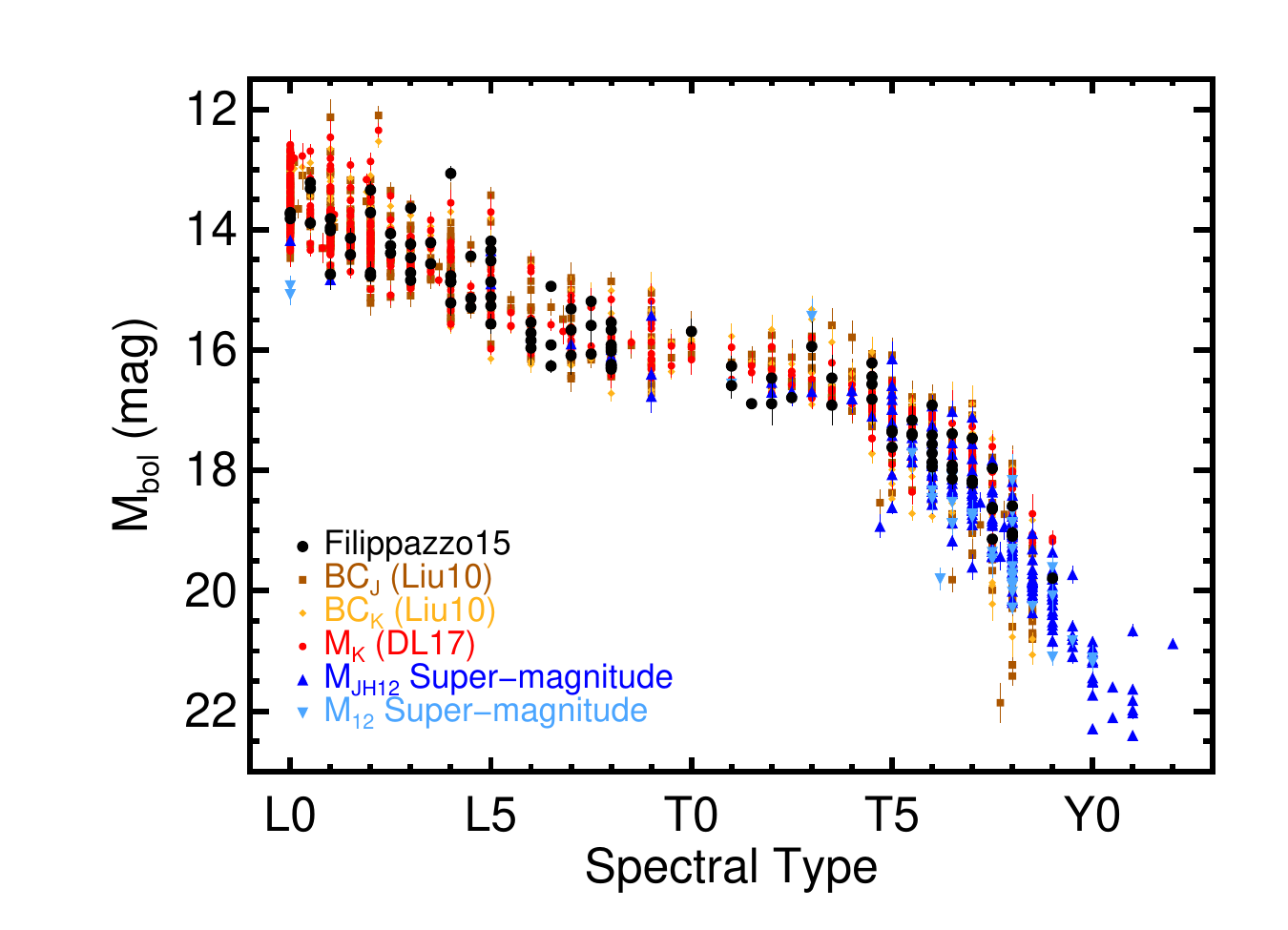}
  \end{minipage}
  \hfill
  \begin{minipage}[t]{0.48\textwidth}
    \includegraphics[width=1\columnwidth, trim = 20mm 0 10mm 0]{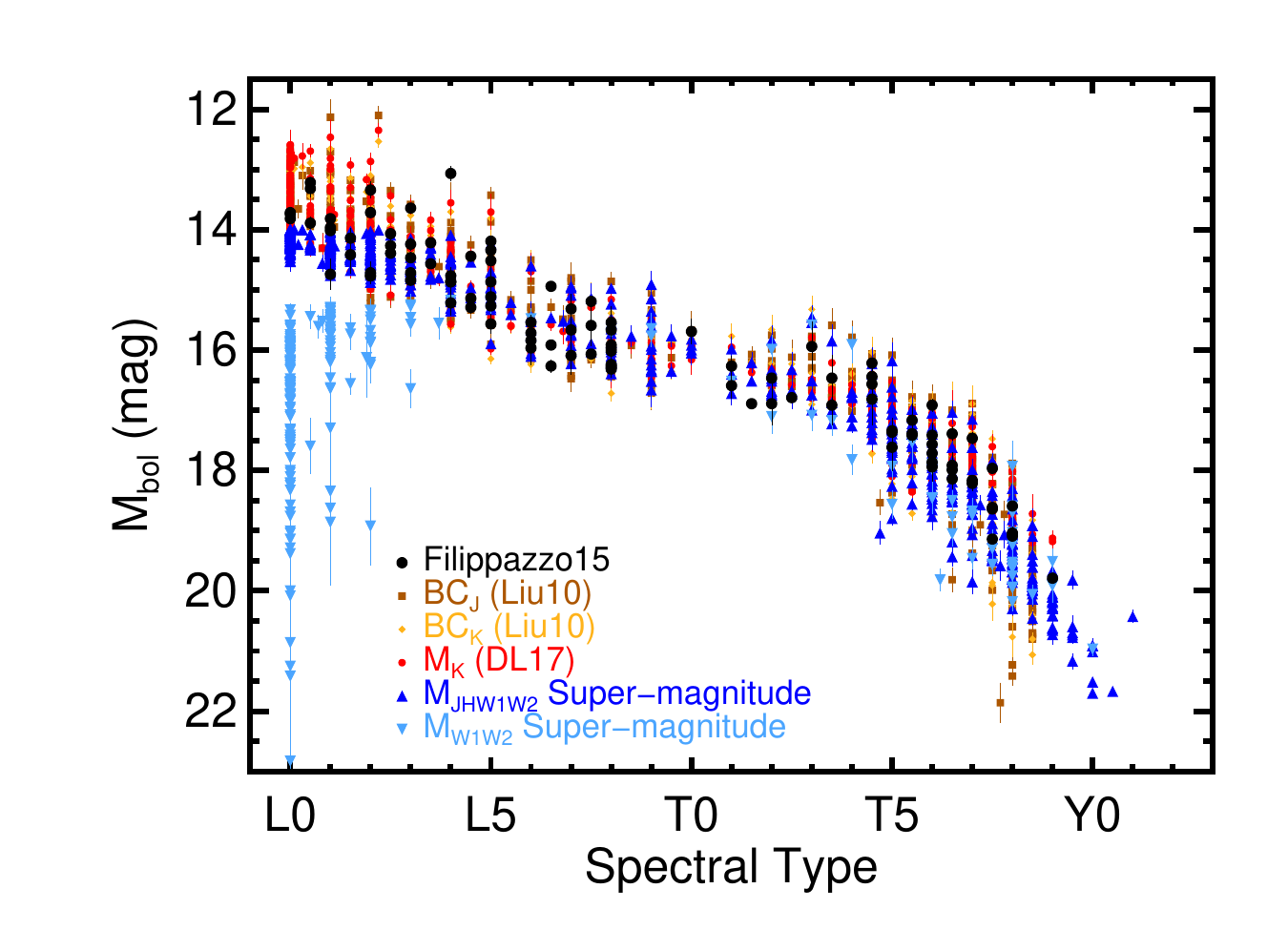}
  \end{minipage}
  \caption{{\it Left}: Bolometric magnitudes calculated with our updated
    super-magnitude method for L0--Y2 dwarfs, using $m_{JH12}$ (dark blue) or
    $m_{12}$ (light blue, when {\jmko} or {\hmko} photometry were not
    available), shown as a function of {\mkmko} (top) and of spectral type
    (bottom).  Included for comparison are bolometric magnitudes determined
    (F15; black circles), or calculated using the {\jmko} and {\kmko} bolometric
    corrections of \citet[brown squares and orange diamonds,
    respectivley]{Liu:2010cw}, or the {\loglbol} vs. {\mkmko} polynomial of
    \citet[red circles]{Dupuy:2017ke}.  {\it Right}: Same plots but showing
    {\mbol} calculated using the $m_{JHW1W2}$ (dark blue) or $m_{W1W2}$
    super-magnitudes.  All methods generally agree, but the
    super-magnitude-based {\mbol} are $\approx$0.3~mag fainter for
    $\mkmko\approx15-17$~mag (mid-T spectral types) and diverge sharply from
    other methods at spectral types earlier than $\approx$L4.  The
    super-magnitude method extends the {\mbol} sequence to fainter and cooler
    objects than previous methods are able to reach and should be used to
    determine {\mbol} for spectral types $\approx$T8 and later.  }
  \label{fig.mbol.compare}
\end{figure*}

Figure~\ref{fig.mbol.compare} compares our bolometric magnitudes derived from
super-magnitudes to the {\mbol} from other sources and methods in the
literature, presented as a function of {\mkmko} and of spectral type for all
L0--Y2~dwarfs with parallaxes and appropriate photometry.  All underlying data
(photometry and spectral types) used to compute the {\mbol} were taken from the
UltracoolSheet.  We show one set of {\mbol} derived from the $m_{JH12}$
super-magnitude, using $m_{12}$ in cases where {\jmko} or {\hmko} were
unavailable; and a second set derived from $m_{JHW1W2}$ and $m_{W1W2}$.
Literature sources include the {\lbol} computed from low-resolution spectra and
broadband photometry by F15, the spectral-type based {\jmko} and {\kmko}
bolometric corrections of \citet{Liu:2010cw}, and the {\loglbol} vs. {\mkmko}
polynomial of \citet{Dupuy:2017ke}.

In general all methods for computing {\mbol} agree well for most of the L, T,
and Y dwarf regime.  The {\mbol} derived from $m_{JHW1W2}$ and $m_{W1W2}$
diverges sharply from other methods for $\mkmko<11.5$ and spectral types earlier
than L4. (Insufficient [3.6] and [4.5] photometry exists for such objects to
corroborate this trend using $m_{JH12}$ and $m_{12}$.)  The
super-magnitude-based {\mbol} are $\approx$0.3~mag fainter for
$\mkmko\approx15-17$~mag (mid-T spectral types), suggesting that the atmosphere
models are under-predicting the bolometric luminosities relative to the flux in
passbands used in our super-magnitudes for these objects.  The super-magnitudes
clearly extend the {\mbol} sequence to fainter and cooler objects than previous
methods are able to reach.  The steady decline of bolometric luminosity as a
function of {\mkmko} appears to flatten significantly at $\mkmko\gtrsim19$~mag,
suggesting a more rapid decline in $K$-band flux as these cool objects become
colder.  Empirical calibration for these objects must await mid-infrared spectra
from the James Webb Space Telescope.  Until then, we present our updated
super-magnitude method as the best way to determine bolometric luminosities for
objects with $\mkmko\gtrsim18$ or spectral types T8 and later.  For warmer
objects, we note that the spectral type-based {\kmko} bolometric corrections
from \citet{Liu:2010cw} and the {\mkmko} polynomial of \citet{Dupuy:2017ke} give
very similar results, and are both consistent with the calculations of F15.  If
direct calculation of {\mbol} from spectra and/or broadband photometry is not
available or feasible, these methods or other similar conversions from the
literature \citep[e.g., F15;][]{Faherty:2016fx} should be sufficient.

\end{document}